\documentclass[a4paper,12pt,twoside]{article}
\usepackage{latexsym}
\usepackage{amsfonts}
\usepackage{epsfig}

\else\oddsidemargin25mm\evensidemargin25mm\marginparwidth25mm\fi
\begin{document}
\def\bfone{\relax{\rm 1\kern-.35em 1}}
\def\dop{{\rm d}\hskip -1pt}
\def\a{\alpha}
\def\b{\beta}
\def\g{\gamma}
\def\d{\delta}
\def\e{\epsilon}
\def\ve{\varepsilon}
\def\t{\theta}
\def\l{\lambda}
\def\m{\mu}
\def\n{\nu}
\def\pg{\pi}
\def\r{\rho}
\def\s{\sigma}
\def\t{\tau}
\def\c{\chi}
\def\p{\psi}
\def\o{\omega}
\def\G{\Gamma}
\def\D{\Delta}
\def\T{\Theta}
\def\L{\Lambda}
\def\Pg{\Pi}
\def\S{\Sigma}
\def\O{\Omega}
\def\pb{\bar{\psi}}
\def\cb{\bar{\chi}}
\def\lb{\bar{\lambda}}
\def\i{\imath}
\def\oL{\overline{L}}
\def\eq#1{(\ref{#1})}
\newcommand{\be}{\begin{equation}}
\newcommand{\ee}{\end{equation}}
\newcommand{\ba}{\begin{eqnarray}}
\newcommand{\ea}{\end{eqnarray}}
\newcommand{\ban}{\begin{eqnarray*}}
\newcommand{\ean}{\end{eqnarray*}}
\newcommand{\nn}{\nonumber}
\newcommand{\nin}{\noindent}
\newcommand{\fgl}{\mathfrak{gl}}
\newcommand{\fu}{\mathfrak{u}}
\newcommand{\fsl}{\mathfrak{sl}}
\newcommand{\fsp}{\mathfrak{sp}}
\newcommand{\fusp}{\mathfrak{usp}}
\newcommand{\fsu}{\mathfrak{su}}
\newcommand{\fp}{\mathfrak{p}}
\newcommand{\fso}{\mathfrak{so}}
\newcommand{\fl}{\mathfrak{l}}
\newcommand{\fg}{\mathfrak{g}}
\newcommand{\fr}{\mathfrak{r}}
\newcommand{\fe}{\mathfrak{e}}
\newcommand{\rE}{\mathrm{E}}
\newcommand{\rSp}{\mathrm{Sp}}
\newcommand{\rSO}{\mathrm{SO}}
\newcommand{\rSL}{\mathrm{SL}}
\newcommand{\rSU}{\mathrm{SU}}
\newcommand{\rUSp}{\mathrm{USp}}
\newcommand{\rU}{\mathrm{U}}
\newcommand{\rF}{\mathrm{F}}
\newcommand{\R}{\mathbb{R}}
\newcommand{\C}{\mathbb{C}}
\newcommand{\Z}{\mathbb{Z}}
\newcommand{\Hb}{\mathbb{H}}
\def\oL{\overline{L}}


\begin{titlepage}
\begin{flushright}
\hskip 5.5cm \hskip 1.5cm
\vbox{\hbox{CERN-TH/2003-046}\hbox{DFTT3/2003}\hbox{SPIN-03/03}\hbox{ITP-UU-03/03
}\hbox{March, 2003}} \end{flushright} \vskip 1cm
\begin{center}
{\LARGE {$N=4$ Supergravity Lagrangian for Type IIB\\ on $T^6/\mathbb{Z}_2$ Orientifold\\ in Presence of Fluxes and $D3$-Branes }}\\
\vskip 1.5cm
  {\bf R. D'Auria$^{1,2,\star}$, S. Ferrara$^{3\ast}$, F.Gargiulo$^{2,4,\sharp}$, M. Trigiante$^{5,\flat}$ and S. Vaul\`a$^{1,2,\circ}$} \\
\vskip 0.5cm
\end{center}
{\small $^1$ Dipartimento di Fisica, Politecnico di
Torino, Corso Duca degli Abruzzi 24, I-10129 Torino, Italy}\\
{\small $^2$ Istituto Nazionale di Fisica Nucleare (INFN) -
Sezione di
Torino, Via P. Giuria 1, I-10125 Torino, Italy}\\
{\small $^3$ CERN, Theory Division, CH 1211 Geneva 23, Switzerland
and INFN, Laboratori Nazionali di Frascati, Italy.}\\
{\small $^4$ Dipartimento di Fisica Teorica, Universit\`a degli
Studi di Torino, Via P. Giuria 1, I-10125 Torino, Italy.}\\
{\small $^5$ Spinoza Institute, Leuvenlaan 4 NL-3508, Utrecht, The
Netherlands.} \vskip 0.5cm
\begin{center}
e-mail: $^{\star}$ riccardo.dauria@polito.it, $^{\ast}$
sergio.ferrara@cern.ch, $^\sharp$ gargiulo@to.infn.it,\\ $^\flat$
M.Trigiante@phys.uu.nl, $^{\circ}$ silvia.vaula@polito.it
\end{center}
\vskip 1cm
\begin{abstract}
We derive the Lagrangian and the transformation laws of $N=4$
gauged supergravity coupled to matter multiplets whose
$\sigma$-model of the scalars is ${\rm SU}(1,1)/{\rm U}(1)\otimes
{\rm SO}(6,6+n)/{\rm SO}(6)\otimes {\rm SO}(6+n)$ and which
corresponds to the effective Lagrangian of the Type IIB string
compactified on the $T^6/\mathbb{Z}_2$ orientifold with fluxes
turned on and in presence of $n$ $D3$--branes. The gauge group is
$T^{12}\otimes G$ where $G$ is the gauge group on the brane and
$T^{12}$ is the gauge group
on the bulk corresponding to the gauged translations of the R-R scalars coming from the R-R four--form.\\
The $N=4$ bulk sector of this theory can be obtained as a
truncation of the Scherk--Schwarz spontaneously broken $N=8$
supergravity. Consequently the full bulk spectrum satisfies
quadratic and quartic mass sum rules,
identical to those encountered in Scherk--Schwarz reduction gauging a flat group.\\
This theory gives rise to a no scale supergravity extended with
partial super--Higgs mechanism.

\end{abstract}

\end{titlepage}

\section{Introduction}
In recent time, compactification of higher dimensional theories in presence of $p$--form fluxes \cite{ps}--\cite{Tripathy:2003qw} has given origin to new four--dimensional vacua with spontaneously broken supersymmetry and with vanishing vacuum energy. These models realize, at least at the classical level, the no--scale structure \cite{Cremmer:1983bf}--\cite{Cremmer:1984hj} of extended supergravities in an $M$ or String theory setting \cite{Ferrara:1987jq}--\cite{Bagger:2001ep} .\\
No--scale supergravities also arise from Scherk--Schwarz generalized dimensional reduction \cite{Scherk:1979zr}--\cite{Sezgin:ac}, where a flat group is gauged.\\
>From a pure four--dimensional point of view all these models can be viewed as particular cases of gauged--extended supergravities (for recent reviews see \cite{Andrianopoli:2002vy}--\cite{VanProeyen:2003zj}). The gauge couplings correspond to fluxes turned on\footnote{Note that in String and $M$ theories the fluxes satisfy some quantization conditions \cite{ps}--\cite{Tripathy:2003qw}}. This is so because for $N>1$ supersymmetry a scalar potential is necessarily due to the presence of gauge symmetries. It has been shown  that a common feature of all no--scale structures is that the complete gauge group of the theory contains a sector where ``axionic'' symmetries are gauged \cite{Cecotti:1985sf} --\cite{deWit:2002vt}.\\
The Higgs effect in this sector is then tightly connected to the super--Higgs mechanism \cite{Andrianopoli:2002rm},\cite{Louis:2002vy}. The complete gauge group is usually larger than this sector and the additional gauge bosons are frequently associated to central charges of the supersymmetry algebra.\\
For instance, in $N=8$ spontaneously broken supergravity \`a la
Scherk--Schwarz, the translational part $T_{\L}$ is
27--dimensional and the extra sector $T_0$ is one--dimensional,
then completing a 28--dimensional flat group
\cite{Scherk:1979zr}, \cite{Andrianopoli:2002mf} \be
[T_{\L},T_0]=f_{\L 0}^{\D}T_{\D};\quad [T_{\L},T_{\S}]=0;\quad
\L,\S=1\dots 27\ee This algebra is a 28--dimensional subalgebra
of $\mathfrak{e}_{7,7}$. In the case of the IIB orientifold
$T^6/\mathbb{Z}_2$, the translational part $T_{\L}$ is
12--dimensional, while the extra sector $T_i$ are the Yang--Mills
generators on the brane \cite{fp}, \cite{kst},
\cite{D'Auria:2002th} \be [T_{\L},T_i]=0;\quad
[T_{\L},T_{\S}]=0;\quad
[T_i,T_j]=c_{ij}^{\phantom{q}\phantom{q}k}T_k;\quad \L\S=1\dots
12;\ i,j,k=1\dots dim\,G\ee
What is common to these groups is that they must have a symplectic action on the vector field strengths and their dual \cite{Gaillard:1981rj}. This implies that they must be embedded in ${\rm Sp}(2n,\mathbb{R})$, where $n=12+{\rm dim}\,G$ in the orientifold case.\\
The particular choice of the embedding determines the structure of the gauged supergravity. In the case of the type IIB supergravity in presence of $D3$--branes, the strong requirement is that the original ${\rm SL}(2,\mathbb{R})$ symmetry acts linearly on the twelve bulk vectors $(B_{\m \L},C_{\m L})$, $\L=1\dots 6$, but acts as an electric magnetic duality on the vectors $A_{\m}^i$, $i=1\dots n$ living on the $D3$--branes\footnote {$(B_{\m I},C_{\m I})$ are the  ${\rm SL}(2,\mathbb{R})$ doublet N-S and R-R two--forms with one leg on space--time and one leg on the torus}.\\
Mathematically this corresponds to a very particular embedding of ${\rm SL}(2,\mathbb{R})\times {\rm SO}(6,6+n)$ into ${\rm Sp}(24+2n,\mathbb{R})$.\\
The relevant decomposition is
\begin{eqnarray} \mathfrak{ so}(6,6+n)&=&\mathfrak{
sl}(6,\mathbb{R})^0+\mathfrak{ so}(1,1)^0+ \mathfrak{
so}(n)^0+{\bf (15^\prime,1)}^{+2}+{\bf
(15,1)}^{-2}+\nonumber\\&&{\bf (6^\prime,n)}^{+1}+{\bf (6,n)}^{-1}
\end{eqnarray}
where $\mathfrak{so}(n)\supset Adj\,G_{({\rm dim}\, n)}$ (note
that if $G={\rm U}(N)$, then $n=N^2$). The symplectic embedding
of the $12+n$ vectors such that ${\rm SL}(2,\mathbb{R})$ is
diagonal on 12 vectors and off diagonal in the remaining
Yang--Mills vectors on the branes, is performed in section 3.\\
Interestingly, the full bulk sector of the $T^6/\mathbb{Z}_2$ Type
IIB orientifold can be related to a $N=4$ truncation of the $N=8$
spontaneously broken supergravity \`a la Scherk--Schwarz
\cite{Scherk:1979zr}, \cite{Cremmer:1979uq}. This
will be proven in detail in Section 8.\\
The ${\rm U}(4)$ $R$--symmetry of the Type IIB theory is
identified with the ${\rm U}(4)\subset {\rm USp}(8)$ of the $N=8$
theory, while ${\rm SL}(2,\mathbb{R})\times {\rm GL}(6)$ is
related to the subgroup of ${\rm E}_{6(6)}\times {\rm
SO}(1,1)\subset {\rm E}_{7(7)}$. The $N=4$ truncation is obtained
by deleting the left--handed gravitino in the $\overline{\bf 4}^{\
-\frac{1}{2}}$ and keeping the ${\bf 4}^{+\frac{1}{2}}$ in the
decomposition of the ${\bf 8}$ of ${\rm USp}(8)$ into ${\rm
U}(4)$ irreducible representations: ${\bf 8}\longrightarrow
\overline{\bf 4}^{\ -\frac{1}{2}}+{\bf
4}^{+\frac{1}{2}}$.\\
The $N=8$ gravitino mass matrix (the ${\bf 36}$ of ${\rm USp}(8)$)
decomposes as follows \be {\bf 36}\longrightarrow {\bf 1}^0+{\bf
15}^0+{\bf 10}^{+1}+\overline{\bf 10}^{\ -1} \ee and the
representation ${\bf 10}^{+1}$ corresponds to the $N=4$ gravitino
mass matrix of the orientifold theory \cite{D'Auria:2002th}, \cite{D'Auria:2002tc}.\\
The vacuum condition of the $N=8$ Scherk--Schwarz model
corresponds to the vanishing of a certain representation ${\bf
42}$ of ${\rm USp}(8)$ \cite{Sezgin:ac},
\cite{Andrianopoli:2002mf}. Its $N=4$ decomposition is \be {\bf
42}\longrightarrow {\bf 20}^0+{\bf 1}^{+2}+{\bf
1}^{-2}+\overline{\bf 10}^{\ +1}+{\bf 10}^{-1} \ee and the vacuum
condition of the $N=4$ orientifold theory corresponds to setting
to zero \cite{D'Auria:2002th}, \cite{D'Auria:2002tc} the
representation ${\bf 10}^{-1}$ (the other representations being
deleted in the truncation).\\
This theory has a six--dimensional moduli space ($6+6N$, $N$
being the dimensional of the Cartan subalgebra of $G$,
 if the $D3$--brane coordinates are added) which is locally three copies of ${\rm SU}(1,1)/{\rm U}(1)$ \cite{Andrianopoli:2002aq},
 \cite{Ferrara:2002bt}, \cite{D'Auria:2002th}. The spectrum depends on the overall scale
 $\g=(R_1R_2R_3)^{-1}=e^{\frac{K}{2}}$, where $K$ is the K\"ahler potential of the moduli space.
 In units of this scale, if we call $m_i$ $(i=1,2,3,4)$ the four gravitino masses, the overall
 mass spectrum has a surprisingly simple form, and in fact it coincides with a particular truncation (to half of the states) of the $N=8$ spectrum of Scherk--Schwarz spontaneously broken supergravity \cite{Cremmer:1979uq}.\\
The mass spectrum satisfies the quadratic and quartic relations:
\ba
&&\sum_J (2J+1)(-1)^{2J}m_J^2=0\nn\\
&&\sum_J (2J+1)(-1)^{2J}m_J^4=0 \ea These relations imply that
the one--loop divergent contribution to the vacuum energy is
absent, in the field theory approximation \cite{Zumino:1974bg},
\cite{dew}. In the present investigation we complete the analysis
performed in reference \cite{D'Auria:2002tc},
\cite{D'Auria:2002th}. In these previous works the part referring
to the bulk sector of the theory and the vacua in presence of
$D3$--branes degrees of freedom were obtained.

The paper is organized as follows:

In Section 2 we describe the $N=4$ $\s$--model geometry of the
bulk sector coupled to $n$ $D3$--branes.

In Section 3 we give in detail the symplectic embedding which
describes the bulk IIB theory coupled to $D3$--brane gauge fields.

In Section 4 the gauging of the $N=4$ theory is given.

In Section 5 the Lagrangian (up to four fermions terms) and the
supersymmetry transformation laws (up to three fermions terms) are
obtained.

In Section 6 the potential and its extrema are discussed.

In Section 7 the mass spectrum is given.

In section 8 we describe the embedding of our model in the $N=8$
supergravity and its relation with the Scherk--Schwarz
compactification.

In Appendix A we describe the geometric method of the Bianchi
identities in superspace in order to find the supersymmetry
transformation laws on space--time.

In Appendix B we use the geometric method (rheonomic approach)in
order to find a superspace Lagrangian which reduces to the
space--time Lagrangian after suitable projection on the
space--time.

In Appendix C we give a more detailed discussion of the freezing
of the moduli when we reduce in steps
$N=4\longrightarrow\,3,\,2,\,1,\,0$ using holomorphic coordinates
on the $T^6$ torus.

In Appendix D we give some conventions.

\section{The Geometry of the scalar sector of the $T^6/\mathbb{Z}_2$\\ orientifold in presence of $D3$--branes}
\subsection{The $\s$--model of the bulk supergravity sector}
\setcounter {equation}{0} \addtocounter{section}{0}
For the sake of establishing notations, let us first recall the
physical content of the $N=4$ matter coupled supergravity
theory.\\ The gravitational multiplet is
\be\{V^a_\m;\,\p_{A\m};\,\p^A_\m;\,A_{1\m}^{I};\,\c^A;\,\c_A;\,\phi_1;\,\phi_2\}\ee
where $\p_{A\m}$ and $\p^{A}_{\m}$ are chiral and antichiral
gravitini, while $\c^A$ and $\c_A$ are chiral and antichiral
dilatini; $V^a_\m$ is the vierbein, $A_{1\m}^I$, $I=1,\dots 6$ are
the graviphotons and the complex scalar fields $\phi_1,\,\phi_2$
satisfy the
constraint $\phi_1\overline{\phi}_1-\phi_2\overline{\phi}_2=1$.\\
We also introduce $6+n$ Yang--Mills vector multiplets, from which
$6$ will be considered as vector multiplets of the bulk, namely
\be\{A_{2\m}^{I};\,\l^I_A;\,\l^{IA};\,s^r\}\ee where $\l^I_A$ and
$\l^{IA}$ are respectively chiral and antichiral gaugini,
$A_{2\m}^I$ are matter vectors and $s^r$, $r=1,\dots 36$ are real
scalar fields.\\
Correspondingly we denote the $n$ vector multiplets, which
microscopically live on the $D3$--branes, as
\be\{A_{\m}^i;\,\l^i_A;\,\l^{iA};\,q^I_i\}\ee where $i=1,\dots n$.
\vskip 0.5cm It is well known that the scalar manifold of the
$N=4$ supergravity coupled to $6+n$ vector multiplets is given by
the coset space \cite{Bergshoeff:1985ms},\cite{deRoo:1985jh}\be
\frac{{\rm SU}(1,1)}{{\rm U}(1)}\otimes\frac{{\rm SO}(6,6+n)}{{\rm
SO}(6)\times {\rm SO}(6+n)}\ee Denoting by $w[\,\,]$, the weights
of the fields under the ${\rm U}(1)$ factor of the ${\rm U}(4)$
$R$--symmetry, the weights of the chiral spinors
are\footnote{Throughout the paper lower ${\rm SU}(4)$ indices
belong to the fundamental representation, while upper ${\rm
SU}(4)$ indices belong to its complex conjugate}
\begin{equation}
\label{pesos}w[\p_A]=\frac{1}{2};\quad w[\c^A]=\frac{3}{2};\quad
w[\l_{IA}]=-\frac{1}{2};\quad w[\l_{iA}]=-\frac{1}{2}
\end{equation}
and for the ${\rm SU}(1,1)/{\rm U}(1)$ scalars we have \be
w[\phi_1]=w[\phi_2]=-1;\quad
w[\overline{\phi}_1]=w[\overline{\phi}_2]=1\ee Let us now
describe the geometry of the coset $\s$--model.

For the ${\rm SU}(1,1)/{\rm U}(1)$ factor of the $N=4$
$\s$--model we use the following parameterization
\cite{D'Auria:2002tc}:
\begin{equation}\label{cososcal}
 S_{{\rm SU}(1,1)}=
\begin{pmatrix}
{\phi_1&\overline{\phi}_2\cr \phi_2&\overline{\phi}_1\cr}
\end{pmatrix}  \ \ \ \ \ \
(\phi_1\overline{\phi}_1-\phi_2\overline{\phi}_2=1)
\end{equation}
\nin Introducing the 2-vector
\begin{equation}
\begin{pmatrix}{L^1\cr L^2\cr}
\end{pmatrix}
=\frac{1}{\sqrt{2}}
\begin{pmatrix}{\phi_1+\phi_2\cr-i(\phi_1-\phi_2)\cr}
\end{pmatrix}\end{equation}
\be w[L^\a]=-1;\quad w[\oL^\a]=1\ee
   the
identity $\phi_1\overline{\phi}_1-\phi_2\overline{\phi}_2=1$
becomes:
\begin{equation} \label{procione}
L^{\a}\overline{L}^{\b}-\overline{L}^{\a}L^{\b}=i\e^{\a\b}
\end{equation}
\nin The indices $\a=1,2$ are lowered by the Ricci tensor
$\epsilon_{\a\b}$, namely:
\begin{equation}L_{\a}\equiv\epsilon_{\a\b}L^{\b}\end{equation}
A useful parametrization of the ${\rm SU}(1,1)/{\rm U}(1)$ coset
is in terms of the N--S, R--R string dilatons of Type IIB theory
\cite{D'Auria:2002th} \be\label{dilaton0}
\frac{\phi_2}{\phi_1}=\frac{i-S}{i+S}\ee with $S=ie^{\varphi}+C$,
from which follows, fixing an arbitrary $U(1)$ phase:
\ba\label{dilaton3}&&S=-\frac{L^2}{L^1}\\
&&\label{dilaton1}\phi_1=-\frac{1}{2}[i(e^{\varphi}+1)+C]e^{-\frac{\varphi}{2}}\\
&&\label{dilaton2}\phi_2=\frac{1}{2}[i(e^{\varphi}-1)+C]e^{-\frac{\varphi}{2}}\\
&&\label{L1}L^1=-\frac{i}{\sqrt{2}}e^{-\frac{\varphi}{2}}\\
&&\label{L2}L^2=-\frac{1}{\sqrt{2}}(e^{\frac{\varphi}{2}}-iCe^{-\frac{\varphi}{2}})\ea
Note that the physical complex dilaton $S$ is ${\rm U}(1)$ independent.\\
We will also use the isomorphism ${\rm SU}(1,1)\sim {\rm
SL}(2,\mathbb{R})$ realized with the Cayley matrix $\mathcal{C}$
\begin{equation} \label{cayley}\mathcal{C}=\frac{1}{\sqrt{2}}
\begin{pmatrix}
{1&1\cr-i&i\cr}
\end{pmatrix}
\end{equation}

\begin{equation}
S_{{\rm SL}(2,R)}=\mathcal{C}S_{{\rm
SU}(1,1)}\mathcal{C}^{-1}=\frac{1}{\sqrt{2}}
\begin{pmatrix}
{ L^1+\oL^1&i(L^1-\oL^1)\cr L^2+\oL^2&i(L^2-\oL^2)\cr}
\end{pmatrix}
\equiv
\begin{pmatrix}
{\a&\b\cr \g&\d\cr}
\end{pmatrix}
\end{equation}

\nin We note that the 2--vector $(L^\a$ $\oL^\a)$ transform as a
vector of ${\rm SL}(2,\mathbb{R})$ on the left and ${\rm
SU}(1,1)$ on the right. Indeed:
\begin{equation}
\widetilde{S}=\mathcal{C}S_{{\rm SU}(1,1)}=
\begin{pmatrix}{L^1&\oL^1\cr L^2&\oL^2\cr}
\end{pmatrix}
\end{equation}

The left-invariant Lie algebra valued 1-form of ${\rm SU}(1,1)$is
defined by:

\begin{equation} \theta\equiv S^{-1}dS=
\begin{pmatrix}{q&\overline{p}\cr
p&-q\cr}\end{pmatrix}\end{equation}

\nin where the coset connection 1-form $q$ and the vielbein 1-form
$p$ are given by:
\begin{eqnarray} &&\label{q0}q=i\e_{\a\b}L^{\a}d\overline{L}^{\b}\\
&&\label{p0}p=-i\e_{\a\b}L^{\a}dL^{\b}
\end{eqnarray}
 Note that we have the
following relations
\begin{eqnarray}
&&\nabla L^{\a}\equiv dL^{\a}+q
L^{\a}=-\overline{L}^{\a}p\\
&&\nabla\overline{L}^{\a}\equiv
d\overline{L}^{\a}-q\overline{L}^{\a}=-L^{\a}\overline{p}
\end{eqnarray}

To discuss the geometry of the ${\rm SO}(6,n)/{\rm SO}(6)\times
{\rm SO}(6+n)$ $\sigma$--model, it is convenient to consider
first the case $n=0$, that is the case when only six out the
$6+n$ vector multiplets are present (no $D3$--branes). This case
was studied in reference \cite{D'Auria:2002tc}.

In this case the coset reduces to $\frac{{\rm SO}(6,6)}{{\rm
SO}(6)\times {\rm SO}(6)}$ ; with respect to the subgroup $ {\rm
SL}(6,\mathbb{R})\times {\rm SO}(1,1)$ the ${\rm SO}(6,6)$
generators decompose as follows:
\be\label{dec0}\fso(6,6)=\fsl(6,\R)^0+\fso(1,1)^0+(\mathbf{15}',\mathbf{1})^{+2}+(\mathbf{15},\mathbf{1})^{-2}\ee
where the superscripts refer to the ${\rm SO}(1,1)$ grading. We
work in the basis where the ${\rm SO}(6,6)$ invariant metric has
the following form

\begin{equation}
\label{eta6} \eta=
\begin{pmatrix}{{\bf0}_{6\times 6}&{\bf{1}}_{6\times 6}\cr {\bf{1}}_{6\times 6}&{\bf{0}}_{6\times 6}\cr}
\end{pmatrix}
\end{equation}
Thus, the generators in in the right hand side of (\ref{dec0})
are:
\begin{eqnarray}
\mathfrak{sl}(6,\mathbb{R})&:& \left(\matrix{A & 0\cr 0 & - A^T
\cr }\right)\,\,\,\quad\qquad\qquad\,\,\,;\,\,\,\mathfrak{
so}(1,1)\,:\,\left(\matrix{\bfone & 0 \cr 0
& -\bfone  }\right)\nonumber\\
{\bf (15^\prime,1)}^{+2}&:& T_{[\Lambda\Sigma]}=\left(\matrix{0 &
t_{[\Lambda\Sigma]} \cr 0 & 0 } \right)\,\,\qquad\,;\,\,\,{\bf
(15,1)}^{-2}\,:\,(T_{[\Lambda\Sigma]})^T
\end{eqnarray}
where we have defined:
\begin{eqnarray}
t_{[\Lambda\Sigma]}{}^{\Gamma\Delta} &=&
\delta^{\Gamma\Delta}_{\Lambda\Sigma};\quad\quad
\Lambda,\,\Sigma=1,\dots,6
\end{eqnarray}
and $A$ are the ${\rm SL}(6,\mathbb{R})$ generators. It is useful
to split the scalar fields $s^r$ into those which span the ${\rm
GL}(6,\mathbb{R})/{\rm SO}(6)_d$ submanifold and which
parametrize the corresponding coset representative $L_{{\rm
GL}(6,\mathbb{R})}$ from the axions parametrizing the $15'^{+2}$
translations. We indicate them respectively with $E=E^T\equiv
E^I_{\L}$, $E^{-1}\equiv (E^{-1})_I^{\L}$ symmetric $6\times6$
matrices
 and with $B=-B^T\equiv
B^{\Lambda\Sigma}$, $\Lambda,\,\Sigma=1,\dots,6$, $I= 1,\dots,6$.
Note that the capital Greek indices refer to global ${\rm GL}(6)$
while the capital Latin indices refer to local ${\rm SO}(6)_{(d)}$
transformations.
 The coset
representatives $L_{{\rm GL}(6,\mathbb{R})}$ and the full coset
representative $L$ can thus be constructed as follows:
\begin{eqnarray}
L=\exp{\left(-B^{\Lambda\Sigma}T_{[\Lambda\Sigma]}\right)}L_{{\rm
GL}(6,\mathbb{R})}&=& \left(\matrix{E^{-1} & -B E \cr 0 & E
}\right)\nonumber\\&&\nonumber\\L_{{ {\rm
GL}}(6,\mathbb{R})}&=&\left(\matrix{E^{-1} & 0 \cr 0 & E}\right)
\end{eqnarray}
Note that the coset representatives $L$ are orthogonal with
respect to the metric \eq{eta6}, namely $L^T\eta L=\eta$.

The left invariant 1--form $L^{-1}dL\equiv\G$ satisfying
$d\G+\G\wedge\G=0$ turns out to be
\begin{equation}
 \G =
\begin{pmatrix}
{EdE^{-1}&-EdBE\cr 0&E^{-1}dE\cr}
\end{pmatrix}
\end{equation}
\nin As usual we can decompose the left invariant 1-form into the
connection $\O$, plus the vielbein $\mathcal{P}$: \be
\label{decomp} \G=\O^HT_H+\mathcal{P}^KT_K
\end{equation}
\nin The matrices $T_H$ are the generators of the isotropy group
${\rm SO}(6)_1\times {\rm SO}(6)_2$,
 where we have indicated with ${\rm SO}(6)_1\sim {\rm SU}(4)$ the semisimple part
 of the $R$--symmetry group ${\rm U}(4)$ and with  ${\rm SO}(6)_2$ the "matter group".\\
Since we are also interested in the connection of the diagonal
subgroup ${\rm SO}(6)_{(d)}$, we will use in the following two
different basis for the generators, the first one that makes
explicit the direct product structure of the isotropy group
(Cartan basis) and the latter in which we identify the diagonal
subgroup of the two factors (diagonal basis). We have
respectively:
\begin{equation}
T_H=
\begin{pmatrix}{T_1&0\cr
0&T_2\cr}
\end{pmatrix}\quad\quad
T_H'=
\begin{pmatrix}{T_{(v)}&T_{(a)}\cr
T_{(a)}&T_{(v)}\cr}
\end{pmatrix}
\end{equation}
where $T_{(v)}$ is the generator of the diagonal ${\rm
SO}(6)_{(d)}$ of ${\rm SO}(6)_1\times {\rm SO}(6)_2$ and
$T_{(a)}$ is the generator of the orthogonal complement. The two
basis are related by
\begin{equation}
T_H'=D^{-1}T_HD
\end{equation} where $D$ is the matrix:
\begin{equation}\label{matD}D=\frac{1}{\sqrt{2}}
\begin{pmatrix}
{1&1\cr 1&-1\cr}
\end{pmatrix}
\end{equation}
In the diagonal basis we can extract the connections $\o^{(d)}$
and $\widehat{\o}$ of the diagonal ${\rm SO}(6)_{(d)}$ subgroup
and of its orthogonal part by tracing with the $T_H^{'}$
generators or, more simply, by decomposing $L^{-1}dL$ into its
antisymmetric part, giving the connection, and its symmetric part
giving the vielbein. \nin In the following we will write $\O$ and
$\mathcal{P}$ as follows: \be\O=\o^{(d)}+\widehat{\o}\ee
\be\o^{(d)}=\frac{1}{2}
\begin{pmatrix}{EdE^{-1}-dE^{-1}\,E&0\cr
0&EdE^{-1}-dE^{-1}\,E\cr}
\end{pmatrix}
\end{equation}
\begin{equation}
\widehat{\o} =\frac{1}{2}
\begin{pmatrix}
{0&-EdBE\cr -EdBE&0\cr}
\end{pmatrix}
\end{equation}
\nin The vielbein $\mathcal{P}$ is, by definition $
\mathcal{P}=\G-\O$ so that we get

\begin{equation}\O=
\begin{pmatrix}{\o^{IJ}&-P^{[IJ]}\cr
-P^{[IJ]}&\o^{IJ}\cr}\end{pmatrix};\quad\quad\mathcal{P}=
\begin{pmatrix}{P^{(IJ)}&-P^{[IJ]}\cr
P^{[IJ]}&-P^{(IJ)}\cr}\end{pmatrix}
\end{equation}
\nin where
 \begin{eqnarray} &&\o^{IJ}=\frac{1}{2}(EdE^{-1}-dE^{-1}\,E)^{IJ}\\
&&P^{(IJ)}=\frac{1}{2}(EdE^{-1}+dE^{-1}\,E)^{IJ}\\
&&P^{[IJ]}=\frac{1}{2}(EdBE)^{IJ}
\end{eqnarray}
In particular:
\begin{equation}
\nabla^{(d)}E^{I}_{\,\,\L}\equiv
dE^{I}_{\,\,\L}-E^I_{\L}\o_I\,\,^J =-E^{J}_{\,\,\L}P^{(JI)}.
\end{equation}
In this basis the Maurer-Cartan equation
\begin{equation}
d\G+\G\wedge\G=0
\end{equation}
\nin take the form:
 \begin{eqnarray}
\label{mauca1}&&R^{(d)IJ}=-P^{(IK)}\wedge P^{(KJ)}\\
\label{mauca2}&&\nabla^{(d)}P^{[IJ]}=-P^{(IK)}\wedge P^{[KJ]}+P^{[IK]}\wedge P^{(KJ)}\\
\label{mauca3}&&\nabla^{(d)}P^{(IJ)}=0
\end{eqnarray}
 where $\nabla^{(d)}$ is the ${\rm SO}(6)_d$  covariant
derivative and $R^{(d)}$ is the ${\rm SO}(6)_d$ curvature: \be
R^{(d)\,IJ}=d\o^{IJ}+\o^{IK}\wedge\o_K^{\,\,\,J}
\end{equation}
The usual Cartan basis ($T_H$--basis) where the connection is
block-diagonal and the vielbein is block off--diagonal is
obtained by rotating $\G$ with the matrix $D$. We find: \be\G
=\begin{pmatrix}{\o_1^{IJ}&-(P^{IJ})^T\cr
-P^{IJ}&\o_2^{IJ}\cr}\end{pmatrix}=
\begin{pmatrix}{\o^{IJ}-P^{[IJ]}&P^{(IJ)}+P^{[IJ]} \cr
P^{(IJ)}-P^{[IJ]}&\o^{IJ}+P^{[IJ]}\cr}\end{pmatrix}
\end{equation}
where  $\o_1$ and $\o_2$ are the connections of ${\rm SO}(6)_1$
and ${\rm SO}(6)_2$ respectively, while $
-(P^{IJ})^T=P^{(IJ)}+P^{[IJ]}$  is the vielbein. In this case the
the curvature of the ${\rm SO}(6,6)/{\rm SO}(6)\otimes {\rm
SO}(6)$ manifold takes the form: \be \mathcal{R}=
\begin{pmatrix}{R_1&0\cr 0&R_2\cr}
\end{pmatrix} \end{equation} where:
\begin{eqnarray}
&&R_1^{IJ}\equiv d\o_1^{IJ}+\o_1^{IK}\wedge\o_{1\,K}^{\ \ \ \ \,J}=-P^{KI}\wedge P^{\ \ \,J}_K\\
&&R_2^{IJ}\equiv d\o_2^{IJ}+\o_2^{IK}\wedge\o_{2\,K}^{\ \ \ \
\,J}=-P^{IK}\wedge P_{\ \,K}^J
\end{eqnarray}
 and
the vanishing torsion equation is
\begin{equation}
\nabla P^{IJ}\equiv
dP^{IJ}+P^{IK}\wedge\o_{1K}^{\,\,\,\,\,\,J}+\o^I_{2\,\,K}\wedge
P^{KJ}=0\\\label{dq}
\end{equation}


\subsection{Geometry of the $\s$--model in presence of $n$ $D3$--branes}
We now introduce additional $n$ Yang--Mills multiplets
$(A^i_{\m},\,\l_A^i,\,\l^{Ai},\,q_I^i)$, $I=1\dots 6$,\\ $i=1\dots
n$.

The isometry group is now ${\rm SL}(2,\mathbb{R})\times {\rm
SO}(6,6+n)$ and the coset representative $\mathbb{L}$ factorizes
in the product of the $\frac{{\rm SO}(6,6+n)}{{\rm SO}(6)\times
{\rm SO}(6+n)}$ coset representative $L$ and the $\frac{{\rm
SL}(2,\mathbb{R})}{{\rm SO}(2)}$ coset representative $S$: \be
\mathbb{L} = S\,L
\end{equation}
In the following we shall characterize the matrix form of the
various ${\rm SO}(6,6+n)$ generators in the ${\bf 12+n}$ and
define the embedding of ${\rm SU}(1,1)\otimes {\rm SO}(6,6+n)$
inside $Sp(24+2n,\mathbb{R})$.\\ With respect to the subgroup
${\rm SL}(6,\mathbb{R})\times {\rm SO}(1,1)\times {\rm SO}(n)$ the
${\rm SO}(6,6+n)$ generators decompose as follows:
\begin{eqnarray} \mathfrak{ so}(6,6+n)&=&\mathfrak{
sl}(6,\mathbb{R})^0+\mathfrak{ so}(1,1)^0+ \mathfrak{
so}(n)^0+{\bf (15^\prime,1)}^{+2}+{\bf
(15,1)}^{-2}+\nonumber\\&&+{\bf (6^\prime,n)}^{+1}+{\bf
(6,n)}^{-1} \label{dec}
\end{eqnarray}
where the superscript refers to the $\mathfrak{  so}(1,1)$
grading. Let us choose for the ${\bf 12+n}$ invariant metric
$\eta$ the following matrix : \begin{equation}\eta=\begin{pmatrix}
{\matrix{0_{6\times 6} & \bfone_{6\times 6} &0_{6\times n}\cr
\bfone_{6\times 6} &0_{6\times 6}&0_{6\times n}\cr 0_{n\times
6}&0_{n\times 6}&-\bfone_{n\times n}}}\end{pmatrix}
\end{equation}\label{eta} where the blocks are defined by the
decomposition of the ${\bf 12+m}$ into ${\bf 6}+{\bf 6}+{\bf m}$.
The generators in in the right hand side of (\ref{dec}) have the
following form: \be  \mathfrak{ sl}(6,\mathbb{R}):
\left(\matrix{A & 0 & 0\cr 0 & - A^T &0\cr 0& 0 & 0
}\right)\,\,\,\quad\qquad\qquad\,\,\,;\,\,\,\mathfrak{
so}(1,1)\,:\,\left(\matrix{\bfone & 0 & 0\cr 0 & -\bfone &0\cr 0&
0 & 0 }\right)\end{equation}
 \be {\bf (15^\prime,1)}^{+2}:
T_{[\Lambda\Sigma]}=\left(\matrix{0 & t_{[\Lambda\Sigma]} & 0\cr 0
& 0 &0\cr 0& 0 & 0} \right)\,\,\qquad\,;\,\,\,{\bf
(15,1)}^{-2}\,:\,(T_{[\Lambda\Sigma]})^T \end{equation}
 \be {\bf(6^\prime,n)}^{+1}:T_{(\Lambda i)}=\left(\matrix{0 & 0 &
t_{(\Lambda i)}\cr 0 & 0 &0\cr 0& (t_{(\Lambda i)})^T & 0
}\right)\,\,;\,\,\,{\bf (6,n)}^{-1}\,:\,(T_{(\Lambda i)})^T
\end{equation} where we have used the following notation: \be
t_{[\Lambda\Sigma]}{}^{\Gamma\Delta} =
\delta^{\Gamma\Delta}_{\Lambda\Sigma}\,\,\,\,\,;\,\,\,\,\,t_{(\Lambda
i)}{}^{\Sigma k}\,=\,\delta_\Lambda^\Sigma
\delta_{i}^k\nonumber\\\nonumber\\\quad \Lambda,\Sigma =1,\dots
,6\,\,\,;\,\,\,i,k=1,\dots ,n \end{equation} As in the preceding
case, we split the scalar fields into those which span the $
\frac{{\rm GL}(6,\mathbb{R})}{{\rm SO}(6)}$ submanifold  and which
parametrize the corresponding coset representative $L_{{\rm
GL}(6,\mathbb{R})}$ from the axions parametrizing the $({\bf
15',1})^{+2}$ translations. We indicate them as before
respectively with $E=E^T\equiv E^I_{\L}$,  $E^{-1}\equiv
(E^{-1})_I^{\L}$ and $B=-B^T\equiv B^{\Lambda\Sigma}$. In
presence of $D3$--branes we have in addition the generators in
the  ${\bf (6^\prime,n)}^{+1}$ that we parametrize with  the
$6\times n$ matrices $a\equiv a^{\Lambda} _i$ (in the following
we will also use the notation $q^I_i\equiv E^I_\L a^\L_i$ ). The
coset representatives $L_{{\rm GL}(6,\mathbb{R})}$ and $L$ can
thus be constructed as follows:
\begin{eqnarray}
L=\exp{\left(-B^{\Lambda\Sigma}T_{[\Lambda\Sigma]}+ a^{\Lambda
i}T_{(\Lambda i)}\right) }L_{{\rm GL}(6,\mathbb{R})}&=&
\left(\matrix{E^{-1} & -C E & a\cr 0 & E &0 \cr 0 & a^T E & \bfone
}\right)\nonumber\\&&\nonumber\\L_{{\rm
GL}(6,\mathbb{R})}&=&\left(\matrix{E^{-1} & 0 &  0\cr 0 & E
&0 \cr 0 & 0 & \bfone}\right)\nonumber\\&&\nonumber\\
C&=& B-\frac{1}{2}\,aa^T \label{cosetto}
\end{eqnarray}
where the sum over repeated indices is understood and $a\equiv a^{\L i}$ are $6\times n$ matrices.\\
As in the preceding case, $E\equiv E^I_{\L}$ and $E^{-1}\equiv
(E^{-1})_I^{\L}$ are. Note that the coset representative $L$ is
orthogonal respect the metric $\eta$.

The left invariant 1-form $\G=L^{-1}dL$ turns out to be:
\begin{equation}\G
=\begin{pmatrix}{E\,dE^{-1}&-E[dB-\frac{1}{2}(da\,a^T-a\,da^T)]E&E\,da\cr{\bf
0}_{6\times 6} &E^{-1}dE&{\bf 0}_{6\times n}\cr{\bf 0}_{n\times
6}&da^T\,E& \bfone_{n\times n}}\end{pmatrix}
\end{equation}
Proceeding as before we can extract, from the left invariant
1-form, the connection and the vielbein \eq{decomp} in the basis
where we take the diagonal subgroup ${\rm SO}(6)_{d}$ inside
${\rm SO}(6)\times {\rm SO}(6+n)$, where now $T_H$ are the
generators of ${\rm SO}(6)_1\times {\rm SO}(6)_2\times {\rm
SO}(n)$. It is sufficient to take the antisymmetric and symmetric
part of $\G$ corresponding to the connection and the vielbein
respectively. We find:
\begin{equation}\label{omega3}\O=
\begin{pmatrix}{\o^{IJ}&-P^{[IJ]}&P^{Ii}\cr
-P^{[IJ]}&\o^{IJ}&-P^{Ii}\cr
-P^{iI}&P^{iI}&0}\end{pmatrix};\quad\quad\mathcal{P}=
\begin{pmatrix}{P^{(IJ)}&-P^{[IJ]}&P^{Ii}\cr
P^{[IJ]}&-P^{(IJ)}&P^{Ii}\cr
P^{iI}&P^{iI}&0}\end{pmatrix}\end{equation}
\nin where \begin{eqnarray} &&\o^{IJ}=\frac{1}{2}(EdE^{-1}-dE^{-1}\,E)^{IJ}\\
\label{p1}&&P^{(IJ)}=\frac{1}{2}(EdE^{-1}+dE^{-1}\,E)^{IJ}\\
\label{p2}&&P^{[IJ]}=\frac{1}{2}\{E[dB-\frac{1}{2}(da\,a^T-a\,da^T)]E\}^{IJ}\\
\label{p3}&&P^{Ii}=\frac{1}{2}E_{\L}^Ida^{\L i}\end{eqnarray}
>From the Maurer--Cartan equations
\begin{equation}
d\G+\G\wedge\G=0
\end{equation}
we derive the expression of the curvatures and the equations
expressing the absence of torsion in the diagonal basis:
\begin{eqnarray}
&&\nabla^{(d)}P^{[IJ]}=-P^{(IK)}\wedge P^{[KJ]}+P^{[IK]}\wedge P^{(KJ)}+2P^{I}_{\ i}\wedge P^{iJ}\\
&&\nabla^{(d)}P^{iJ}=P^{i}_{\ I}\wedge P^{IJ}
\end{eqnarray}
while equations \eq{mauca1}, \eq{mauca3} remain unchanged.\\
Note that, as it is apparent from equation \eq{omega3}, the
connection of ${\rm SO}(n)$ is zero in this gauge: $\o^{ij}=0$. To
retrieve the form of the connection in the Cartan basis it is
sufficient to rotate $\O$ and $\mathcal{P}$, given in equation
\eq{omega3}, by the generalized $D$ matrix \eq{matD}
\begin{equation}D=\frac{1}{\sqrt{2}}
\begin{pmatrix}
{1&1&0\cr 1&-1&0\cr 0&0&1\cr}
\end{pmatrix}
\end{equation}
We find
\begin{equation}
\O=\left(\begin{array}{c|c c}\o^{IJ}-P^{[IJ]}&{\bf 0}_{6 \times
6} &{\bf 0}_{6 \times n}\cr\hline {\bf 0}_{6 \times 6}&
\o^{IJ}+P^{[IJ]}&P^{Ii}\cr {\bf 0}_{n \times 6}& -P^{Ii} &{\bf
0}_{n \times n}\end{array}\right)
\end{equation}
\begin{equation}
\mathcal{P}=\left(\begin{array}{c|c c}{\bf 0}_{6 \times 6} &
P^{(IJ)}+P^{[IJ]}& P^{Ii}\cr\hline P^{(IJ)}-P^{[IJ]}& {\bf 0}_{6
\times 6}& {\bf 0}_{6 \times n}\cr P^{iI} &{\bf 0}_{6 \times
n}&{\bf 0}_{n \times n}\end{array}\right)
\end{equation}

\begin{eqnarray}
&&R_1^{IJ}\equiv d\o_1^{IJ}+\o_1^{IK}\wedge\o_{1\,K}^{\ \ \ \ \,J}=-P^{KI}\wedge P^{\ \ \,J}_K-2P^{Ii}\wedge P^{iJ}\\
&&R_2^{IJ}\equiv d\o_2^{IJ}+\o_2^{IK}\wedge\o_{2\,K}^{\ \ \ \
\,J}=-P^{IK}\wedge P_{\ \,K}^J+2P^{Ii}\wedge P^{iJ}
\end{eqnarray}
 and
the vanishing torsion equation is
\begin{eqnarray}
&&\nabla P^{IJ}\equiv dP^{IJ}+P^{IK}\wedge\o_{1K}^{\ \
J}+\o_2^{IK}\wedge P_K^{\ \ J}+2P^{Ii}\wedge P_i^{J}=0\\
&&dP^{Ii}+\o_1^{IJ}\wedge P_{J}^{\ i}+P^{(IJ)}\wedge P_{J}^{\ i}=0
\end{eqnarray}

\section{The symplectic embedding and duality rotations}
\setcounter {equation}{0} \addtocounter{section}{0} Let us now
discuss the embedding  of the isometry group ${\rm
SL}(2,\mathbb{R})\times {\rm SO}(6,6+n)$ inside ${\rm
Sp}(24+2n,\mathbb{R})$. We start from the embedding in which the
${\rm SO}(6,6+n)$ is diagonal\footnote{The signs in the embedding
$\iota$ of ${\rm SL}(2,\mathbb{R})$ have been chosen in such a
way that the action on the doublet charges in the final embedding
$\iota^\prime$ were the same as $S$.}:
\begin{eqnarray} {\rm SO}(6,6+n)&\stackrel{\iota}{\hookrightarrow}&{\rm
Sp}(24+2n,\mathbb{R})\nonumber\\
g &\in & {\rm
SO}(6,6+n)\stackrel{\iota}{\longrightarrow}\,\iota(g)=\left(\matrix{g
&0\cr 0 & (g^{-1})^T }\right)\in
{\rm Sp}(24+2n,\mathbb{R})\nonumber\\
S&=&\left(\matrix{\a & \b\cr \g & \d}\right)\,\in\, {\rm
SL}(2,\mathbb{R})\stackrel{\iota}{\longrightarrow}\,\iota(S)=\left(\matrix{\a\bfone
& -\b\eta \cr -\g \eta & \d\bfone}\right)\in
{\rm Sp}(24+2n,\mathbb{R})\nonumber\\
\a\d-\b\g&=&1 \label{basis1}
\end{eqnarray}
where each block of the symplectic matrices is a $(12+n)\times
(12+n)$ matrix. In this embedding a generic symplectic section has
the following grading structure with respect to
$\mathfrak{so}(1,1)$:
\begin{eqnarray} V_{{\rm Sp}}&=& \left(\matrix{v^{(+1)}\cr
v^{(-1)} \cr v^{(0)}\cr u^{(-1)}\cr u^{(+1)} \cr u^{(0)} }\right)
\end{eqnarray}
where $v^{(\pm 1)}$ and $u^{(\pm 1)}$ are six dimensional vectors
while $v^{(0)}$ and $u^{(0)}$ have dimension $n$. Identifying the
$v$'s with the electric field strengths and the $u$'s with their
magnetic dual, we note that the embedding $\iota$ \eq{basis1}
corresponds to the standard embedding where ${\rm
SL}(2,\mathbb{R})$ acts as electric--magnetic duality while ${\rm
SO}(6,6+n)$ is purely
electric.\\
We are interested in defining an embedding $\iota^\prime$ in
which the generators in the ${\bf (15^\prime , 1)}^{+2}$ act as
nilpotent off diagonal matrices or Peccei--Quinn generators and
the ${\rm SL}(2,\mathbb{R})$ group has a block diagonal action on
the $v^{(\pm 1)}$ and $u^{(\pm 1)}$ components and an off
diagonal action on the $v^{(0)}$ and $u^{(0)}$ components.\\
Indeed, our aim is to gauge (at most) twelve of the fifteen
translation generators in the representation ${\bf (15^\prime ,
1)}^{+2}$ and a suitable subgroup $G\subset {\rm SO}(n)$.\\
The symplectic transformation $\mathcal{O}$ which realizes this
embedding starting from the one in (\ref{basis1}) is easily found
by noticing that $(v^{(+ 1)},\,u^{(+1)})$ and $(v^{(-
1)},\,u^{(-1)})$ transform in the ${\bf (6^\prime , 2)}^{+1}$ and
${\bf (6, 2)}^{-1}$ with respect to ${\rm GL}(6,\mathbb{R})\times
{\rm SL}(2,\mathbb{R})$ respectively. Therefore we define the new
embedding:
\begin{eqnarray}
\iota^\prime &=& \mathcal{O}\iota \mathcal{O}^{-1}\nonumber\\
\mathcal{O}&=& \left(\matrix{0 &0&0&\bfone_{6\times 6}&0&0\cr 0
&\bfone_{6\times 6}&0&0&0&0\cr 0 &0&\bfone_{m\times m}&0&0&0\cr
-\bfone_{6\times 6}&0&0&0&0&0\cr 0&0&0&0&\bfone_{6\times 6}&0\cr
0&0&0&0&0& \bfone_{m\times m}}\right)
\end{eqnarray}
In this embedding the generic ${\rm SL}(2,\mathbb{R}) $ element
$S$ has the following form: \begin{eqnarray} \iota^\prime
(S)&=&\left(\matrix{\d\,\bfone_{6\times 6} &-\g\,\bfone_{6\times
6}&0&0&0&0\cr -\b\,\bfone_{6\times 6} &\a\,\bfone_{6\times
6}&0&0&0&0\cr 0 &0&\a\,\bfone_{m\times m}&0&0&\b\,\bfone_{m\times
m}\cr 0&0&0&\a\,\bfone_{6\times 6}&\b\,\bfone_{6\times 6}&0\cr
0&0&0&\g\,\bfone_{6\times 6}&\d\,\bfone_{6\times 6}&0\cr
0&0&\g\,\bfone_{m\times m}&0&0& \d\,\bfone_{m\times m}}\right)
\end{eqnarray}
while the generic element of ${\rm SO}(6,6+n)/{\rm SO}(6)\times
{\rm SO}(6+n)$ takes the form \begin{eqnarray} \iota^\prime (L)=
\begin{pmatrix}{E&0&0&0&0&0\cr 0&E&0&0&0&0\cr
0&a^TE&\bfone&0&0&0\cr 0&CE&-a&E^{-1}&0&0\cr
-CE&0&0&0&E^{-1}&-a\cr
-a^TE&0&0&0&0&\bfone}\end{pmatrix}\end{eqnarray} The product
$\S=\iota^\prime (L)\iota^\prime (S)$ of these two matrices gives
the desired embedding in ${\rm Sp}(24+2n,\mathbb{R})$ of the
relevant coset. If we write the ${\rm Sp}(24+2n,\mathbb{R})$
matrix in the form
\begin{equation}\S=\begin{pmatrix}{A&B\cr
C&D\cr}\end{pmatrix}\end{equation} and define \be
f=\frac{1}{\sqrt{2}}(A-iB);\quad\quad
h=\frac{1}{\sqrt{2}}(C-iD)\end{equation} we obtain
\cite{Andrianopoli:1996ve} \be f=\frac{1}{2}
\begin{pmatrix}{\d E&-\g E&0\cr -\b E&\a E&0\cr -\b a^TE&\a
a^TE&\a-i\b\cr}\end{pmatrix}\end{equation}

\be h=\frac{1}{2}\begin{pmatrix}{-\b CE-i\a E^{-1}&\a CE-i\b
E^{-1}&-(\a-i\b)a\cr -\d CE-i\g E^{-1}&\g CE-i\d
E^{-1}&-(\g-i\d)a\cr -\d a^TE& \g
a^TE&\g-i\d\cr}\end{pmatrix}\end{equation}  The kinetic matrix of
the vectors is defined as \cite{Gaillard:1981rj},
\cite{Andrianopoli:1996ve} $\mathcal{N}=h \cdot f^{-1}$ and we
find
\begin{equation}\mathcal{N}\!=\!\!
\begin{pmatrix}{-2iL^1\oL^1E^{-1}E^{-1}&\frac{1}{2}aa^T\!\!-\!2iL^{(1}\oL^{2)}\!E^{-1}\!E^{-1}\!\!-\!2iB
L^{[1}\oL^{2]}&-a\cr\frac{1}{2}aa^T\!\!-\!2iL^{(2}\oL^{1)}\!E^{-1}\!E^{-1}\!\!-\!2iB
L^{[2}\oL^{1]}&-2iL^2\oL^2E^{-1}E^{-1}+\frac{L^2}{L^1}aa^T&-\frac{L^2}{L^1}a\cr
-a^T&-\frac{L^2}{L^1}a^T&\frac{L^2}{L^1}}\end{pmatrix}\end{equation}
or in components

\begin{eqnarray}&&\mathcal{N}^{\L\a\S\b}=-2iL^{(\a}\oL^{\b)}(E^{-1})^{\L}_{\,\,I}(E^{-1})^{I\S}+B^{\L\S}\e^{\a\b}
-i(aa^T)^{\L\S}\,L^{(\a}\left(\oL^{\b)}-L^{\b)}\frac{\oL^1}{L^1}\right)\nn\\
&&\mathcal{N}^{\L\a\,i}=-a^{\L i}\frac{L^{\a}}{L^1}\nn\\
&&\mathcal{N}^{ij}=\frac{L^2}{L^1}\d^{ij}\end{eqnarray} where we
have used the relation \eq{procione}.

\section{The Gauging}
\setcounter {equation}{0} \addtocounter{section}{0} Our aim is to
gauge a group of the following form:
\begin{equation}
T_{12}\times G\subset {\rm SO}(6,6+n)
\end{equation}
where $T_{12}$ denote 12 of the ${\bf (15^\prime ,1)}^{+2}$
Peccei--Quinn translations $T_{[\Lambda\Sigma]}$ in ${\rm
SO}(6,6)$ and the group $G$ is in general a compact semisimple
subgroup of ${\rm SO}(n)$ of dimension $n$. In particular if
$G={\rm U}(N)$ we must have $N^2=n$. The gauge group is a
subgroup of the global symmetry group of the ungauged action
whose algebra, for the choice of the symplectic embedding defined
in the previous section, is:
\begin{eqnarray}
&&\mathfrak{sl}(6,\mathbb{R})^0+\mathfrak{ so}(1,1)^0+ \mathfrak{
so}(n)^0+ ({\bf 15}^\prime,{\bf 1})^{+2}+ ({\bf 6}^\prime,{\bf
n})^{+1}
\end{eqnarray}
We note that the maximal translation group $T_{12}$ which can be
gauged is of dimension twelve since the corresponding gauge vector
fields are $A_{\Lambda\alpha}$ belong to the $({\bf 6},{\bf
2})^{-1}$ of ${\rm GL}(6,\mathbb{R})\times
 {\rm SL}(2,\mathbb{R})$. Let us denote the gauge generators of the
$T_{12}$ factor by $T^{\Lambda \alpha}$, corresponding to the
gauge vectors $A_{\L \alpha}$ and by $T^i$ ($i=1,\dots, n$) those
of the $G$ factor associated with the vectors $A_i$. These two
sets of generators are expressed in terms of the $ ({\bf
15}^\prime ,{\bf 1})^{+2}$ generators $T_{[\Lambda\Sigma]}$ and of
the ${\rm SO}(n)$ generators $T_{[ij]}$ respectively by means of
suitable {\it embedding matrices} $f^{
\Gamma\Sigma\Lambda\alpha}$ and $c^{kij}$:
\begin{eqnarray}
T^{\Lambda \alpha}&=& f^{ \Gamma\Sigma\Lambda\alpha}\,T_{[\Gamma\Sigma]}\nonumber\\
T^k &=& c^{kij}\,T_{[ij]}
\end{eqnarray}
where $c^{ijk}$ are the structure constants of $G$, with $i\, j\,
k$ completely antisymmetric. The constants $f^{
\Gamma\Sigma\Lambda\alpha}$ are totally antisymmetric in
$\Gamma\Sigma\Lambda$ as a consequence both of supersymmetry and
gauge invariance or, in our approach, of the closure of the
Bianchi identities. They transform therefore with respect to
${\rm SL}(6,\mathbb{R})\times {\rm SO}(1,1)\times {\rm
SL}(2,\mathbb{R})$ in the $ ({\bf 20}^\prime,{\bf 2})^{+3}$. Note
that $f^{ \Gamma\Sigma\Lambda\alpha}$ are the remnants in $D=4$
of the
fluxes of the Type IIB three--forms.\\
We may identify the scalar fields of the theory with the elements
of the coset representative $L$ of ${\rm SO}(6,6+n)/{\rm
SO}(6)\times {\rm SO}(6+n)$ namely, $E^{\L}_I$, $B^{\L\S}$,
$a^\L_i$. The scalar field associated with the coset ${\rm
SL}(2,\mathbb{R})/{\rm SO}(2)\simeq {\rm SU}(1,1)/{\rm U}(1)$ is
instead represented by the complex 2-vector
$L^{\a}$ satisfying the constraint \eq{procione}.\\
The gauging can be performed in the usual way replacing the
coordinates differentials with the gauge covariant differentials
$\nabla_{(g)}$:
\begin{eqnarray}
\label{coorgau1}&&dL^{\a}\longrightarrow\nabla_{(g)} L^{\a}\equiv
dL^{\a}\\
&&dE^\L_I\longrightarrow \nabla_{(g)} E^\L_I\equiv dE^\L_I\\
&&dB^{\L\S} \longrightarrow \nabla_{(g)} B^{\L\S}=dB^{\L\S}+f^{\L\S\G\a}A_{\G\a}\\
\label{coorgau3}&&da^\L_i\longrightarrow \nabla_{(g)}
a^\L_i=da^\L_i+ c_{i}^{\ j k}A_ja^{\L}_k
\end{eqnarray}
Note that $f^{\L\S\G\a}$ are the constant components of the
translational Killing vectors in the chosen coordinate system,
namely $k^{\L\S|\G\a}=f^{\L\S\G\a}$ \cite{D'Auria:2002tc}, where
the couple $\L\S$ are coordinate indices while $\G\a$ are indices
in the adjoint representation of the gauge subgroup $T_{12}$; in
the same way the Killing vectors of the compact gauge subgroup
$G$ are given by $k^{\L\,|j}_i=c_{i}^{\ jk} a^\L_k$ where the
couple $\L i$ are coordinate indices, while $jk$ are in the
adjoint representation
of $G$.\\
>From equations \eq{coorgau1}--\eq{coorgau3} we can derive the
structure of the gauged left--invariant 1--form $\hat{\G}$
\begin{equation}\label{gaucon0}
\hat{\G}=\G+\d_{(T_{12})}\G+\d_{(G)}\G
\end{equation}
where $\d_{(T_{12})}\G$ and $\d_{(G)}\G$ are the shifts of $\G$
due to the gauging of $T_{12}$ and $G$ respectively. From these
we can compute the shifts of the vielbein and of the connections.
We obtain:
\begin{eqnarray}
&&\hat P^{IJ} = P^{IJ} + \d_{(T_{12})}P^{IJ}+\d_{G}P^{IJ}\\
&&\hat P^{Ii} = P^{Ii} + \d_{(T_{12})}P^{Ii}+\d_{G}P^{Ii}\\
&&\hat \o_{1,2}^{IJ} = \o_{1,2}+ \d_{(T_{12})}\o_{1,2}^{IJ}+
\d_{G}\o_{1,2}^{IJ}
\end{eqnarray}
where
\begin{eqnarray}
&&\d_{(T_{12})}P^{IJ}=\frac{1}{2}E^I_\L
f^{\L\S\G\a}A_{\G\a}E^J_\S\\
&&\d_{G}P^{IJ}=\frac{1}{2}c^{ijk}E^I_\L a^\L_iA_ja_k^\S E_\S^J\\
&&\d_{(T_{12})}P^{Ii}=0\\
&&\d_{G}P^{Ii}=\frac{1}{2}c^{ijk}E^I_\L a^\L_kA_j\\
&&\d_{(T_{12})}\o_1^{IJ}=-\d_{(T_{12})}\o_2^{IJ}=-\frac{1}{2}E^I_\L
f^{\L\S\G\a}A_{\G\a}E^J_\S\\
&&\d_{G}\o_1^{IJ}=-\d_{G}\o_2^{IJ}=-\frac{1}{2}c^{ijk}E^I_\L
a^\L_iA_ja_k^\S E_\S^J
\end{eqnarray}
Note that only the antisymmetric part of $P^{IJ}$ is shifted,
while the diagonal connection $\o_d =\o_1+\o_2$ remains untouched.
An important issue of the gauging is the computation of the
"fermion shifts", that is of the extra pieces appearing in the
supersymmetry transformation laws of the fermions when the gauging
is turned on. Indeed the scalar potential can be computed from the
supersymmetry of the Lagrangian as a quadratic form in the
fermion shifts. The shifts have been computed using the (gauged)
Bianchi identities in superspace as it is explained in Appendix A.
We have:
\begin{eqnarray}
\d\p_{A\m}^{(shift)}&=&S_{AB}\g_{\m}\ve^B=-\frac{i}{48}
(\overline{F}^{IJK-}+\overline{C}^{IJK-})(\G_{IJK})_{AB}\g_{\m}\e^B\\
\d\c^{A\,(shift)}&=&N^{AB}\e_B=-\frac{1}{48}(\overline{F}^{IJK+}+\overline{C}^{IJK+})(\G_{IJK})^{AB}\e_B\\
\d\l^{I\,(shift)}_{A}&=&Z^{I\,B}_A\e_B=
\frac{1}{8}(F^{IJK}+C^{IJK})(\G_{JK})_{A}^{\,\,B}\e_B\\
\d\l^{(shift)}_{iA}&=&W^{\,\,\,B}_{iA}\e_B=\frac{1}{8}L_2\,E^J_\L
E^K_\S a^{\L j}a^{\S k}\,c_{ijk}(\G_{JK})_{A}^{\,\,B}\e_B
\end{eqnarray}
where we have used the selfduality relation (see Appendix D for
conventions)
$(\G_{IJK})_{AB}=\frac{i}{3!}\e_{IJKLMN}\G^{LMN}_{AB}$ and
introduced the quantities \be
\label{self}F^{\pm\,IJK}=\frac{1}{2}\left(F^{IJK}\pm
i\,^*F^{IJK}\right)\ee \be
\label{self2}C^{\pm\,IJK}=\frac{1}{2}\left(C^{IJK}\pm
i\,^*C^{IJK}\right)\ee where

\begin{equation}
F^{IJK}=L^{\a}f_{\a}^{IJK},\quad\quad
f^{IJK\a}=f^{\L\S\G\a}E^I_{\,\,\L}E^J_{\,\,\S}E^K_{\,\,\G},
\quad\quad\overline{F}^{IJK}=\overline{L}^{\a}f_{\a}^{IJK}\end{equation}
and $C^{IJK}$ are the boosted structure constants defined as \be
C^{IJK}=L_2E^I_\L E^J_\S E^K_\G a^{\L i}a^{\S j}a^{\G
k}\,c_{ijk},\quad\quad\overline{C}^{IJK}=\oL_2E^I_\L E^J_\S
E^K_\G a^{\L i}a^{\S j}a^{\G k}\,c_{ijk} \ee while the complex
conjugates of the self--dual and antiself--dual components are
\be (F^{\pm\,IJK})^*=\overline{F}^{\mp IJK},\quad\quad
(C^{\pm\,IJK})^*=\overline{C }^{\mp IJK}\ee

For the purpose of the study of the potential, it is convenient
to decompose the 24 dimensional representation of ${\rm
SU}(4)_{(d)}\subset {\rm SU}(4)_1\times {\rm SU}(4)_2$ to which
$\l_A^I$ belongs, into its irreducible parts, namely
$\bf{24}=\bf{\overline{20}}+\bf{\overline{4}}$. Setting:
\be\l_A^I=\l_A^{I\,\,(\overline{20})}-\frac{1}{6}(\G^{I})_{AB}\l^{B(\overline{4})}
\ee where \be
\l^{A(\overline{4})}=(\G_{I})^{AB}\l_B^I;\quad\quad(\G_{I})^{AB}\l_B^{I\,\,(\overline{20})}=0\ee
we get \ba \label{deco1}
\d\l^{A(\overline{4})}&=&Z^{AB(\overline{4})}\e_B=\frac{1}{8}(F^{+IJK}+C^{+IJK})(\G_{IJK})^{AB}\e_B\\
\label{shgauI204}\d\l_A^{I\,\,(\overline{20})}&=&Z_A^{I(\overline{20})\,\,B}\e_B=\frac{1}{8}(F^{-IJK}+C^{-IJK})(\G_{JK})_A^{\,\,B}\e_B\ea

\section{Space--time Lagrangian}
\setcounter {equation}{0} \addtocounter{section}{0}

The space--time Lagrangian and the associated supersymmetry
transformation laws, have been computed using the geometric
approach in superspace. We give in the Appendices A and B a
complete
derivation of the main results of this section.\\
In the following, in order to simplify the notation, we have
suppressed the "hats" to the gauged covariant quantities:
$\hat{\nabla}\rightarrow\nabla;\ \hat{P}\rightarrow P;\ \hat\o_{1,2}\rightarrow \o_{1,2}$.\\
In particular, the gauged covariant derivatives on the spinors of
the gravitational multiplet and of the Yang--Mills multiplets are
defined as follows:
\begin{eqnarray}
&&\nabla\p_A=\mathcal{D}\p_A+\frac{1}{2}q\p_A-\frac{1}{4}(\G_{IJ})_A^{\,\,\,B}\,\o_1^{IJ}\p_B\\
&&\nabla\c^A=\mathcal{D}\c^A+\frac{3}{2}q\c^A-\frac{1}{4}(\G_{IJ})^A_{\,\,\,B}\,\o_1^{IJ}\c^B\\
&&\nabla\l_{IA}=\mathcal{D}\l_{IA}-\frac{1}{2}q\l_{IA}-\frac{1}{4}(\G_{IJ})_A^{\,\,\,B}\,\o_1^{IJ}\l_{IB}+\o_2^{IJ}\l_{JA}\\
&&\nabla\l_{iA}=\mathcal{D}\l_{iA}-\frac{1}{2}q\l_{iA}-\frac{1}{4}(\G_{IJ})_A^{\,\,\,B}\,\o_1^{IJ}\l_{iB}\\
\end{eqnarray}
$\nabla$ is the gauged covariant derivative with respect to all
the connections that act on the field, while $\mathcal{D}$ is the
Lorentz covariant derivative acting on a generic spinor $\theta$
as follows
\begin{equation}\mathcal{D}\theta\equiv
d\theta-\frac{1}{4}\o^{ab}\g_{ab}\theta\end{equation} The action
of the ${\rm U}(1)$ connection $q$ \eq{q0} appearing in the
covariant derivative $\nabla$ is defined as a consequence of the
different ${\rm U}(1)$ weights of the fields \eq{pesos}.

 The complete action is:
\begin{equation}\label{action}
S=\int\sqrt{-g}\mathcal{L}d^4x
\end{equation}
where
\begin{equation}
\mathcal{L}=\mathcal{L}_{(kin)}+\mathcal{L}_{(Pauli)}+\mathcal{L}_{(mass)}-\mathcal{L}_{(potential)}
\end{equation}
where
\begin{eqnarray}
\mathcal{L}_{(kin)}&=&-\frac{1}{2} R
-i\left(\mathcal{N}^{\L\a\S\b}\mathcal{F}^{+\m\n}_{\L\a}\mathcal{F}^+_{\S\b\m\n}-\overline{\mathcal{N}}^{\L\a\S\b}
\mathcal{F}^{-\m\n}_{\L\a}\mathcal{F}^-_{\S\b\m\n}\right)+\nn\\
&&-2i\left(\mathcal{N}^{i\S\b}\mathcal{F}^{+\m\n}_i\mathcal{F}^+_{\S\b\m\n}-\overline{\mathcal{N}}^{i\S\b}
\mathcal{F}^{-\m\n}_i\mathcal{F}^-_{\S\b\m\n}\right)+\nn\\
&&-i\left(\mathcal{N}^{ij}\mathcal{F}^{+\m\n}_i\mathcal{F}^+_{j\m\n}
-\overline{\mathcal{N}}^{ij}\mathcal{F}^{-\m\n}_i\mathcal{F}^-_{j\m\n}\right)+\nn\\
&&+\frac{2}{3}f^{\L\S\G\g}\e^{\a\b}A_{\G\g\,\m}A_{\S\b\,\n}F_{\L\a\,\r\s}\e^{\m\n\r\s}+\overline{p}_\m
p^\m+\frac{1}{2}P^{IJ}_\m P_{IJ}^\m+P^{Ii}_\m P_{Ii}^\m+\nn\\
&&+\frac{\ve^{\m\n\r\s}}{\sqrt{-g}}\left(\pb^A_{\m}\g_{\n}\nabla_{\r}\p_{A\s}-\pb_{A\m}\g_{\n}\nabla_{\r}\p^A_{\s}\right)
-2i\left(\cb^A\g^\m\nabla_\m\c_A+\cb_A\g^\m\nabla_\m\c^A\right)+\nn\\
&&\label{lkin}-i\left(\lb^A_I\g^\m\nabla_\m\l_A^I+\lb_{IA}\g^\m\nabla_\m\l^{IA}\right)
-2i\left(\lb^A_i\g^\m\nabla_\m\l_A^i+\lb_{iA}\g^\m\nabla_\m\l^{iA}\right)
\end{eqnarray}
where, using equations \eq{p0}, \eq{p1}, \eq{p2}, \eq{p3} we have:
\begin{eqnarray}
&&p_{\m}\overline{p}^{\m}=L_{\a}\oL_{\b}\partial_{\m}L^{\a}\partial^{\m}\oL^{\b}\\
&&P^{IJ}_\m P_{IJ}^\m=P^{(IJ)}_\m P_{(IJ)}^\m+P^{[IJ]}_\m P_{[IJ]}^\m\nn\\
&&P^{(IJ)}_\m
P_{(IJ)}^\m=-4\partial_{\m}E^I_{\L}\partial^{\m}(E^{-1})_{I}^{\L}=4g_{\L\S}\partial_{\m}(E^{-1})_{I}^{\L}
\partial^{\m}(E^{-1})^{I\S}\nn \\
&&P^{[IJ]}_\m
P_{[IJ]}^\m=\frac{1}{4}g_{\L\G}g_{\S\D}\left(\nabla_{(g)\mu}
B^{\L\S}-\frac{1}{2}
a^{\L}_i\!\stackrel{\leftrightarrow}{\nabla}_{(g)\mu}\!a^{i\S}\right)\left(\nabla_{(g)}^\mu
B^{\G\D}-\frac {1}{2} {a^{\G}_j\!
\stackrel{\leftrightarrow}\nabla}_{(g)}^\mu \!a^{j\D}\right)\nonumber \\
&&P^{Ij}_\m
P_{Ij}^\m=g_{\L\S}\partial_{\m}a^{\L}_i\partial^{\m}a^{\S i}
\end{eqnarray}
and we have defined\quad $g_{\L\S}\equiv E_{I\L}E^I_{\S}$,\quad
$\mathcal{F}^{\pm}=\frac{1}{2}(\mathcal{F}\pm i\,^*\mathcal{F})$

\begin{eqnarray}
\mathcal{L}_{(Pauli)}&=&-2p_\m\cb^A\g^\n\g^\m\p_{A\n}-P^{IJ}_\m
\G_I^{AB}\lb_{JA}\g^\n\g^\m\p_{B\n}-2P^{Ii}_\m(\G_I)^{AB}\lb_{iA}\g^\n\g^\m\p_{B\n}+\nn\\
&&-2Im\left(\mathcal{N}\right)^{\L\a\S\b}\left[\mathcal{F}^{+\m\n}_{\L\a}\left(L_\b{E_\S}^I(\G_{I})^{AB}\pb_{A\m}\p_{B\n}+
2i\oL_\b{E_\S}^I(\G_{I})^{AB}\cb_A\g_\n\p_{B\m}+\right.\right.\nn \\
&&\left.\left.+2i\oL_\b{E_\S}^I\lb^{I}_{A}\g_\n\p^{A}_{\m}+\frac{1}{4}L_{\b}E_\S^I(\G_I)_{AB}\lb^A_J\g_{\m\n}\l^{JB}+\oL_\b E_\S^I\lb_I^A\g_{\m\n}\c_A\right.\right.+\nn\\
&&\left.\left.+\frac{1}{2}L_{\b}E_\S^I(\G_I)_{AB}\lb^{iA}\g_{\m\n}\l_i^B\right)\right]+\nn\\
&&-2Im\left(\mathcal{N}\right)^{i\S\b}\left[\mathcal{F}^{+\m\n}_i\left(L_\b{E_\S}^I(\G_I)^{AB}\pb_{A\m}\p_{B\n}
+2i\oL_\b{E_\S}^I(\G_I)^{AB}\cb_A\g_\n\p_{B\m}+\right.\right.\nn\\
&&\left.\left.+2i\oL_{\b}E_{\S}^I\lb^I_{A}\g_\n\p^{A}_{\m}+\frac{1}{4}L_{\b}E_{\S}^I(\G_I)_{AB}\lb^A_J\g_{\m\n}\l^{JB}+\oL_{\b}E_{\S}^I\lb_I^A\g_{\m\n}\c_A+\right.\right.\nn\\
&&\left.\left.+\frac{1}{2}L_{\b}E_\S^I(\G_I)_{AB}\lb^{iA}\g_{\m\n}\l_i^B\right)\right.+\nn\\
&&+\mathcal{F}^{+\m\n}_{\S\b}\left(E_\L^Ia_i^\L
L_2(\G_I)^{AB}\pb_{A\m}\p_{B\n}
+2iE_\L^Ia_i^\L\oL_2(\G_I)^{AB}\cb_A\g_\n\p_{B\m}+\right.\nn\\
&&\left.\left.+2iE_\L^Ia_i^\L\oL_2\lb_{IA}\g_\n\p^{A}_{\m}+4i\oL_2\lb_{iA}\g_\n\p^{A}_{\m}+\frac{1}{4}L_2E^I_\S a^\S_i(\G_I)_{AB}\lb^A_J\g_{\m\n}\l^{JB}+\right.\right.\nn\\
&&\left.\left.+\oL_2E^I_\S a^\S_i\lb^A_I\g_{\m\n}\c_A+\frac{1}{2}L_2E^I_\S a^\S_i(\G_I)_{AB}\lb^A_j\g_{\m\n}\l^{jB}+2\oL_2\lb_i^A\g_{\m\n}\c_A\right)\right]+\nn\\
&&-2Im\left(\mathcal{N}\right)^{ij}\left[\mathcal{F}^{+\m\n}_i\left(E_\L^Ia_j^\L
L_2(\G_I)^{AB}\pb_{A\m}\p_{B\n} +2iE_\L^Ia_j^\L \oL_2
(\G_I)^{AB}\cb_A\g_\n\p_{B\m}\right.\right.+\nn\\
&&\left.\left.+2iE_\L^Ia_j^\L
\oL_2\lb^{I}_{A}\g_\n\p^{A}_{\m}+4i\oL_2\lb_{jA}\g_\n\p^{A}_{\m}+\frac{1}{4}L_2E^I_\S a^\S_i(\G_I)_{AB}\lb^A_J\g_{\m\n}\l^{JB}\right.\right.\nn\\
&&\left.\left.+\oL_2E^I_\S
a^\S_i\lb^A_I\g_{\m\n}\c_A+\frac{1}{2}L_2E^I_\S
a^\S_i(\G_I)_{AB}\lb^A_j\g_{\m\n}\l^{jB}+2\oL_2\lb_i^A\g_{\m\n}\c_A\right)\right]+\nn\\&&+c.c.
\end{eqnarray}

\begin{eqnarray}
\label{lagmass}\mathcal{L}_{(mass)}&=&\left[+2iS^{AB}\pb_{A\m}\g^{\m\n}\p_{B\n}-4iN^{AB}\pb_{\m
A}\g^\m\c_B-2i{Z_{IB}}^A\pb_{A\m}\g^\m\l^{IB}
-4i{W_{iB}}^A\pb_{A\m}\g^\m\l^{iB}\right.\nn\\
&&+6\left.\left({Q^A_{I\,B}}\lb^{I}_{A}\c^B+{R^A_{i\,B}}\lb^{i}_{A}\c^B+T_{IJ}^{AB}\lb^{I}_{A}\l^{J}_{B}+U_{ij}^{AB}\lb^{i}_{A}
\l^{j}_{B}+V_{Ij}^{AB}\lb^{I}_{A}\l^{j}_{B}\right)\right]+\\
&&+c.c.
\end{eqnarray}

\begin{eqnarray}
\label{lagpot}\mathcal{L}_{(potential)}&=&\frac{1}{4}\left(-12S^{AB}S_{AB}+4N^{AB}N_{AB}+2{Z_{IB}}^A{Z^{IB}}_A+4{W_{iB}}^A{W^{iB}}_A\right)
\end{eqnarray}

The structures appearing in $\mathcal{L}_{(mass)}$ and
$\mathcal{L}_{(potential)}$ are given by
\begin{eqnarray}
\label{shgrav}S_{AB}&=&-\frac{i}{48} (\overline{F}^{IJK-}+\overline{C}^{IJK-})(\G_{IJK})_{AB}\\
\label{shdil}N^{AB}&=&-\frac{1}{48}(\overline{F}^{IJK+}+\overline{C}^{IJK+})(\G_{IJK})^{AB}\\
\label{shgauI}Z^{I\,B}_{A}&=&\frac{1}{8}(F^{IJK}+C^{IJK})(\G_{JK})_A^{\,\,B}\\
\label{shgaui}W^{\,\,\,B}_{iA}&=&\frac{1}{8}L_2\,q^{Jj}q^{Kk}\,c_{ijk}(\G_{JK})_{A}^{\,\,B}\\
Q^{IA}_{\phantom{A}B}&=&-\frac{1}{12}(F^{IJK}+C^{IJK})(\G_{JK})^A_{\,\,B}\\
\label{masschilambda}R^{iA}_{\phantom{A}B}&=&-\frac{1}{24}L_2\,q^{Jj}q^{Kk}\,c_{ijk}(\G_{JK})^{A}_{\,\,B}\\
T^{IJAB}&=&-\frac{1}{3}\d^{IJ}N^{AB}+\frac{1}{12}(\overline{F}^{IJK}+\overline{C}^{IJK})(\G_K)^{AB}\\
\label{massii}U^{ijAB}&=&-\frac{2}{3}\d^{ij}N^{AB}-\frac{1}{3}\oL_2 q^{I}_kc^{ijk}(\G_I)^{AB}\\
\label{massgaux}V^{IiAB}&=&-\frac{1}{3}c^{ijk}\oL_2 q_j^I
q_k^K(\G_K)^{AB}
\end{eqnarray}

The Lagrangian is invariant under the following supersymmetry
transformation laws:

\begin{eqnarray}
\label{susytra}\d V^a_{\mu}&=&-i\pb_{A\m}\g^a\ve^A +c.c.\nn\\
\d A_{\L\a\m}&=&-L_{\a}E_{\L}^{\,I}(\G_I)^{AB}\pb_{A\m}\ve_B +
iL_{\a}E_{\L}^{\,\,I}(\G_I)_{AB}\cb^A\g_{\m}\ve^B+\nn\\
&&+iL_{\a}E_{\L\,I}\lb^{IA}\g_{\m}\ve_A +c.c.\nn\\
\d A_{i\m}&=&-L_2 E^I_\L a^\L_i(\G_I)^{AB}\pb_{A\m}\ve_B+
iL_2E^I_\L a^\L_i(\G_I)_{AB}\cb^A\g_{\m}\ve^B+\nn\\
&&+iL_2E_{\L I} a^\L_i\lb^{IA}\g_{\m}\ve_A+
2iL_2\lb^A_i\g_{\m}\ve_A +c.c.\nn\\
p_{\b}\d L^{\b}&\equiv &=2\cb_A\ve^A\Longrightarrow\d L^{\a}=2\overline{L}^{\a}\cb_A\ve^A\nn\\
\d^{(I}_K E^{J)}_{\L} \d (E^{-1})^{\L K}&=&-(\G^{(I})^{AB}\lb^{J)}_{A}\ve_{B}+c.c.\nn\\
\frac{1}{2} E^I_\L E^J_\S(\d B^{\L\S}-\d a^{[\L}_i a^{\S]i})&=&
(\G^{[I})^{AB}\lb^{J]}_{A}\ve_{B}+c.c.\nn\\
\frac{1}{2}E^I_\L\delta a^{\L i}&=&(\G^I)^{AB}\lb^i_{A}\ve_{B}+c.c.\nn\\
\d\p_{\m
A}&=&\mathcal{D}_{\m}\ve_A-\overline{L}^{\a}(E^{-1})_{I}^{\L}(\Gamma^{I})_{AB}\mathcal{F}^-_{\L\a\mid\m\n}\g^{\n}\ve^B
+S_{AB}\g_\m\ve^B\nn\\
\d\c^A&=&\frac{i}{2}\overline{p}_{\m}\g^{\m}\ve^A+\frac{i}{4}\overline{L}^{\a}(E^{-1})_{I}^{\,\L}(\G^I)^{AB}
\mathcal{F}^-_{\L\a\mid\m\n}\g^{\m\n}\ve_B+N^{AB}\ve_B\nn\\
\d\l_{IA}&=&\frac{i}{2}(\G^J)_{AB}P_{JI\mid\,\m}\g^{\m}\ve^B-\frac{i}{2}L^{\a}(E^{-1})_{I}^{\,\L}
\mathcal{F}^-_{\L\a\mid\m\n}\g^{\m\n}\ve_A+Z^{\,\,\,B}_{IA}\ve_B\nn\\
\d\l_{iA}&=&\frac{i}{2}(\G^J)_{AB}P_{Ji\mid\,\m}\g^{\m}\ve^B+\frac{1}{4}\frac{1}{\oL_2}\mathcal{F}^-_{i\mid\m\n}\g^{\m\n}\ve_A+\nn\\
&&-\frac{1}{4}\frac{1}{\oL_2}a_i^\L\mathcal{F}^-_{2\L\mid\m\n}\g^{\m\n}\ve_A+W^{\,\,\,B}_{iA}\ve_B
\end{eqnarray}
where we have defined\be\label{qu} q^I_i=E^I_\L a^\L_I\ee
\section{The scalar potential and its extrema}
\setcounter {equation}{0} \addtocounter{section}{0} From the
expression of the potential given in the Lagrangian \eq{lagpot}
and using the fermionic shifts given in equations
\eq{shgrav}--\eq{shgaui}, one obtains that the potential is given
by a sum of two terms, each being a square modulus, namely: \be
V=\frac{1}{12}|F^{IJK-}+C^{IJK-}|^2+\frac{1}{8}|L_2c_{ijk}q^{jJ}q^{kK}|^2\ee
where $F^{IJK-}$ and $C^{IJK-}$ were defined in equations \eq{self} and \eq{self2} and
$q^I_i$ in equation \eq{qu}.\\
In the first term $F^{IJK-}$ represents the contribution to the
potential of the bulk fields, while $C^{IJK-}$ is the
contribution from the D-branes sector. On the other hand the
second term is the generalization of the potential already
present in the super Yang--Mills theory \cite
{D'Auria:2002th}.\\  We note that using the decomposition given
in equations \eq{deco1},\eq{shgauI204}, namely:\be
Z^{I\,B}_{A}=Z_A^{I\,\,B(\overline{20})}-\frac{1}{6}(\G^I)_A^{\ C}
Z_{CB}^{(\overline{4})}\ee in equation \eq{lagpot}, the
contribution of the gravitino shift $S_{AB}$ cancels exactly
against the contribution $Z_{AB}^{(4)}$ of the representation
$\overline{\bf{4}}$ of the gaugino; furthermore, since
$Z_A^{I\,\,B(\overline{20})}$ is proportional to $N_{AB}$, the
bulk part of the
potential is proportional to $N_{AB}N^{BA}$.\\
We now discuss the extrema of this potential. Since $V$ is
positive semidefinite, its extrema are given by the solution of
$V=0$, that is \be L_2c_{ijk}q^{jJ}q^{kK}=0,\quad
F^{IJK-}+C^{IJK-}=0\ee  In absence of fluxes $ F^{IJK}=0$, the
only solution is that the $q^{jJ}$ belong to the
 Cartan subalgebra of $G$, then $C^{IJK}=0$, but all the moduli of the orientifold are not
  stabilized as well as the $L_{\a}$. In this case the moduli space is
\be\frac{{\rm SU}(1,1)}{{\rm U}(1)}\times\frac{{\rm
SO}(6,6+N)}{{\rm SO}(6)\times {\rm SO}(6+N)}\ee where $N$ is the
 dimension of the Cartan subalgebra of $G$.\\
In presence of fluxes, $ F^{IJK}\neq 0$ we have, besides $q^{jJ}$
belonging to the Cartan subalgebra of $G$, also  $ F^{IJK-}=0$.
We now show that this condition freezes the dilaton field $S$ and
several moduli of ${\rm GL}(6)/{\rm SO}(6)_d$. We note that the
equation $ F^{IJK-}=0$ can be rewritten as:
\begin{equation}\label{cond1}
L^1f_1^{IJK-} + L^2f_2^{IJK-}=0 \Longrightarrow \frac
{f_1^{IJK-}}{f_2^{IJK-}}=- \frac {L^2}{L^1}\equiv S
\end{equation}
so that $S$ must be a constant. We set: \be\label{vacdil} S=i\a
\rightarrow \frac{L^2}{L^1}=-i\a
\Longrightarrow\frac{\phi_2}{\phi_1}=\frac{1-\a}{1+\a} \ee where
$\a$ is a complex constant and $\Re\a \equiv e^{\varphi_0}>0$
Equation \eq{cond1} becomes: \be \label{cond2}
f_1^{\L\S\D}=+\Re\a\,\,^*f_2^{\L\S\D}-\Im\a f_2^{\L\S\D}\ee We
rewrite this equation using $f_{1,2}^{IJK -
}=\frac{1}{2}\left(f_{1,2}^{IJK}- i\,^*f_{1,2}^{IJK}\right)$ and
by replacing the ${\rm SO(6)}$ indices $I,J,K$ with ${\rm GL(6)}$
indices via the coset representatives $E^I_{\L}$; we find:
\begin{equation}\label{dual}
f_1^{\L\S\G}+\Im \a f_2^{\L\S\G}=\frac {1}{3!}\Re \a \,\,
detE^{-1}\e^{\L\S\G\D\Pi\O}g_{\D\D'}g_{\Pi
\Pi'}g_{\O\O'}f_2^{\D'\Pi'\O'}
\end{equation}
where $g^{\L\S}\equiv E^{\L}_I E^{\S I}$ is the (inverse) moduli
metric of $T^6$ in ${\rm GL}(6)/{\rm SO}(6)_d$.\\ It is convenient
to analyze equation \eq{dual} using the complex basis defined in
Appendix C. In this basis only 4 fluxes (together with their
complex conjugates) corresponding to the eigenvalues of the
gravitino mass matrix, are different from zero.\\ Going to the
complex basis where each Greek index decomposes as
$\L=(i,\overline{\i})$, $i=1,2,3$ \cite{D'Auria:2002tc} and
lowering the indices on both sides, one obtains an equation
relating a $(p,q)$ form on the l.h.s $(p,q=0,1,2,3;\,p+q=3)$ to a
combination of $(p',q')$ forms on the r.h.s. Requiring that all
the terms with $p'\neq p,\,q'\neq q$ are zero and that the r.h.s.
of equation \eq{dual} be a constant, on is led to fix different
subsets of the $g^{\L\S}$ moduli, depending on the
residual degree of supersymmetry.\\
Suppose now that we have $N=3$ unbroken supersymmetry, that is
$m_1=m_2=m_3=0$, $m_4\propto |f^{123}|\neq 0$ (see Appendix C).
The previous argument, concerning $(p,q)$--forms, allows us to
conclude that all the components $g_{i,j}$ and
$g_{\overline{\i}\overline{\jmath}}$ are zero, so that at the
$N=3$ minimum we have: \be
g_{\L\S}\longrightarrow\begin{pmatrix}{g_{i\overline{\jmath}}&0\cr
0&g_{\overline{\i}j}}\end{pmatrix}\ee In the $N=2$ case we have
$m_2=m_3=0$, $m_1\propto |f^{1\overline{2}\overline{3}}|\neq 0$,
$m_4\propto |f^{123}|\neq 0$ and a careful analysis of equation
\eq{dual} shows that, besides the previous frozen moduli, also the
$g_{1\overline{2}},\,g_{1\overline{3}}$ components are frozen.\\
Finally, in the $N=1$ case we set one of the masses equal to
zero, say $m_2=0$, and $m_1\propto
|f^{1\overline{2}\overline{3}}|\neq 0$, $m_3\propto
|f^{12\overline{3}}|\neq 0$ $m_4\propto |f^{123}|\neq 0$; in this
case the only surviving moduli are the diagonal ones, namely
$g_{i\overline{\i}}$, so that, using the results given in
Appendix C, also the real components of $g_{\L\S}$ are diagonal
\be g_{\L\S}\longrightarrow{\rm diag}\{
e^{2\varphi_1},\,e^{2\varphi_2},\,e^{2\varphi_3},\,e^{2\varphi_1},\,e^{2\varphi_2},\,e^{2\varphi_3}\}\ee
the exponentials representing the radii of the manifold
$T_{(14)}^2\times T_{(25)}^2\times T_{(36)}^2$. In terms of the
vielbein $E^I_{\L}$ we have: \be \label{quella}E^I_{\L}={\rm diag}
( e^{\varphi_1}, e^{\varphi_2}, e^{\varphi_3}, e^{\varphi_1},
e^{\varphi_2}, e^{\varphi_3} )\ee Finally, when all the masses
are different from zero ($N=0$), no further condition on the
moduli is obtained.\\
Note that in every case equation \eq{dual} reduces to the ${\rm
SL(2,\mathbb R)}\times {\rm GL(6,\mathbb R)}$ non covariant
constraint among the fluxes:
\begin{equation}\label{dual1}
f_1^{\L\S\G}+\Im \a f_2^{\L\S\G}=\frac {1}{3!}\Re \a \,\,
\e^{\L\S\G\D\Pi\O}f_2^{\D\Pi\O}
\end{equation}
In the particular case $\a=1$, which implies $\varphi=C=0$, from
equations \eq{cond1}, \eq{vacdil}, we obtain:
\begin{equation}
\label{dualiz}f_1^{-\L\S\D}=if_2^{-\L\S\D}
\end{equation}
In this case the minimum of the scalar potential is given by \be
\label{scalvac}\phi_2=0\Longrightarrow|\phi_1|=1\ee or, in terms
of the $L^{\a}$ fields,
$L^1=\frac{1}{\sqrt{2}},\,\,\,L^2=-\frac{i}{\sqrt{2}}$ (up to an
arbitrary phase) . Furthermore \eq{cond2} reduces to the duality
relation: \be\label{rel}f_{1}^{\ \L\S\D} =\frac
{1}{3!}\e^{\L\S\D\G\Pi\O}f_2^{\G\Pi\O}\ee which is the  ${\rm
SL(2,\mathbb R)}\times {\rm GL(6)}$ non covariant constraint
imposed in \cite{Tsokur:1994gr}.\\
Let us now discuss the residual moduli space in each case. For
this purpose we introduce, beside the metric $g^{\L\S}$, also the
15 axions $B^{\L\S}$ with enlarge ${\rm GL}(6)/{\rm SO}(6)_d$ to
${\rm SO}(6,6)/{\rm SO}(6)\times {\rm SO}(6)$ (using complex
coordinates $B^{\L\S}=(B^{i\overline{\jmath}},\
B^{ij}=-B^{ji})$.\\
Since we have seen that for $N=1,\,0$, the frozen moduli from
$F^{IJK-}=0$ are all the $g^{i\overline{\jmath}}$ and $g^{ij}$
except the diagonal ones $g^{i\overline{\i}}$, correspondingly,
in the $B^{\L\S}$ sector, all $B^{ij}$ and
$B^{i\overline{\jmath}}$ are frozen, except the diagonal
$B^{i\overline{\i}}$, and they are eaten by the 12 bosons through
the Higgs mechanism. Indeed the three diagonal
$B^{i\overline{\i}}$ are inert under gauge transformation (see
Appendix C). The metric moduli space is $({\rm O}(1,1))^3$ which,
adding the axions, enlarges to the coset space $\left({\rm
U}(1,1)/{\rm U}(1)\times {\rm U}(1)\right)^3$. Adding the
Yang-Mills moduli in the $D3$-brane sector, the full moduli space
of a generic vacuum with completely broken supersymmetry, or
$N=1$ supersymmetry contains $6+6N$ moduli which parametrize
three copies
of ${\rm U}(1,1+N)/{\rm U}(1)\times {\rm U}(1+N)$.\\
Let us consider now the situation of partial supersymmetry
breaking ( for a more detailed discussion see Appendix C). For
$N=3$ supersymmetry the equation $ F^{IJK-}=0$ freezes all
$g^{ij}$ moduli but none of the $g^{i\overline{\jmath}}$. The
relevant moduli space of metric $g^{i\overline{\jmath}}$ is nine
dimensional and given by ${\rm GL}(3,\mathbb{C})/{\rm U}(3)$.
Correspondingly there are six massive vectors whose longitudinal
components are the $B^{ij}$ axions.  Adding the nine uneaten
$B^{i\overline{\jmath}}$ axions the total moduli space is ${\rm
U}(3,3)/{\rm U}(3)\times {\rm U}(3)$. Further adding the $6N$
Cartan moduli,
the complete moduli space is ${\rm U}(3,3+N)/{\rm U}(3)\times {\rm U}(3+N)$.\\
For $N=2$ unbroken supersymmetry the equation $ F^{IJK-}=0$ fixes
all $g^{ij}$ and $g^{i\overline{\jmath}}$ except the diagonal
ones and $g^{2\overline{3}}$. The moduli space of the metric is
${\rm SO}(1,1)\times {\rm GL}(2,\mathbb C)/{\rm U}(2)$. There are
10 massive vectors which eat all $B^{ij}$ moduli and
$B^{i\overline{\jmath}}$ except the diagonal ones and
$B^{2\overline{3}}$; the complete moduli space enlarges to ${\rm
SU}(1,1)/{\rm U}(1)\times {\rm SU}(2,2)/{\rm SU}(2)\times {\rm
SU}(2)\times {\rm U}(1)$. This space is the product of the
one--dimensional K\"ahler manifold and the two--dimensional
quaternionic manifold as required by $N=2$ supergravity. By
further adding the $6N$ Cartan moduli,
the moduli space enlarges to  ${\rm SU}(1,1+N)/{\rm U}(1)\times {\rm SU}(1+N)\times {\rm SU}(2,2+N)/{\rm SU}(2)\times {\rm SU}(2+N)\times {\rm U}(1)$.\\
Finally in the case of $N=1$ unbroken supersymmetry, the frozen
moduli are the same as in the $N=0$ case, and the moduli space is
indeed the product of three copies of K\"ahler--Hodge manifolds,
as appropriate to chiral multiplets.

\section{The mass spectrum}
\setcounter {equation}{0} \addtocounter{section}{0} The spectrum
of this theory contains 128 states (64 bosons and 64 fermions)
coming from the bulk states of IIB supergravity and $16N^2$
states coming from the $n$ $D3$--branes. The brane sector is
$N=4$ supersymmetric. Setting $\a=1$, the bulk part has a mass
spectrum which has a surprisingly simple form.\\
In units of the overall factor $\frac{\sqrt2}{24}e^{\frac{K}{2}}$,
$K={2\varphi_1+2\varphi_2+2\varphi_3}$ being the K\"ahler
potential of the moduli space $\left(\frac{{\rm SU}(1,1)}{{\rm
U}(1)}\right)^3$, one finds\footnote{Note that we have corrected a
mistake in the spin 1 mass formula (5.21)--(5.23) as given in
reference \cite{D'Auria:2002tc}}:
\begin{center}\large{\bf Fermions}\end{center}
\ban
(4)&spin\,\frac{3}{2}&\,\quad\quad |m_i|\quad\quad\quad\  i=1,2,3,4\\
2(4)&spin\,\frac{1}{2}&\quad\quad |m_i|\\
(16)&spin\,\frac{1}{2}&|m_i\pm m_j\pm m_k|\quad i<j<k \ean

\begin{center}\large{\bf Bosons}\end{center}
\ban
(12)&spin\, 1&\,\quad |m_i\pm m_j|\quad\quad\quad  i<j\\
(6)&spin\, 0&\quad\quad m=0\\
(12)&spin\, 0&\,\quad |m_i\pm m_j|\\
(8)&spin\, 0&|m_1\pm m_2\pm m_3\pm m_4| \ean where $m_i$,
$i=1,\dots 4$ is the
modulus of the complex eigenvalues of the matrix $f_1^{IJK-}(\G_{IJK})_{AB}$ evaluated at the minimum.\\
Note that in the case $\a\neq 1$ all the masses $m_i$ acquire an
$\a$--dependent extra factor due to the relation
\be\overline{F}^{IJK-}=-\sqrt{2}(\Re\a)^{\frac{1}{2}}f_2^{IJK-}=i\sqrt{2}\frac{(\Re\a)^{\frac{1}{2}}}{\a}f_1^{IJK-}\ee
so that all the spectrum is rescaled by a factor
$g(\a)\equiv\sqrt{2}\frac{(\Re\a)^{\frac{1}{2}}}{\a}$.\\
This spectrum is identical (in suitable units) to a truncation (to
half of the states) of the mass spectrum of the $N=8$
spontaneously
broken supergravity \`a la Scherk--Schwarz \cite{Scherk:1979zr}--\cite{Sezgin:ac}.\\
The justification of this statement in given in the next
section.\\
We now note some properties of the spectrum. For arbitrary values
of $m_i$ the spectrum satisfies the quadratic and quartic mass
relations \be\sum_J (2J+1)(-1)^{2J}m_J^{2k}=0\quad k=1,2\ee Note
that, in proving the above relations, the mixed terms for $k=1$
are of the form $m_im_j$ and they separately cancel for bosons
and fermions, due to the symmetry $m_i\longrightarrow -m_i$ of
the spectrum. On the other hand, for $k=2$
 the mixed terms $m_i^2m_j^2$ are even in $m_i$ and thus cancel between bosons and fermions.\\
If we set some of the $m_i=0$ we recover the spectrum of $N=3,\,2,\,1,\,0$ supersymmetric phases.\\
If we set $|m_i|=|m_j|$ for some $i,\,j$ we recover some unbroken
gauge symmetries. This is impossible with the $N=3$ phase (when
$m_2=m_3=m_4=0$) but it is possible in the $N\leq 2$ phases. For
instance in the $N=2$ phase, for $m_3=m_4=0$ and $|m_1|=|m_2|$
there is an additional massless vector multiplet, while in the
$N=1$ phase, for $m_4=0$, $|m_1|=|m_2|=|m_3|$ there are three
massless vector multiplets and finally for the $N=0$ phase and
all $|m_i|$ equal,
there are six massless vectors.\\
The spectrum of the $D3$--brane sector has an enhanced $(N=4)$
supersymmetry, so when the gauge group is spontaneously broken to
its Cartan subalgebra ${\rm U}(N)\longrightarrow {\rm U}(1)^N$,
$N(N-1)$ charged gauge bosons become massive and they are
$\frac{1}{2}$ BPS saturated multiplets of the $N=4$ superalgebra
with central charges (the fermionic sector for the brane gaugini
is discussed at the end of Section 8).
The residual $N$ Cartan multiplets remain massless and their scalar partners
complete the $6+6N$ dimensional moduli space of the theory, that is classically
given by three copies of $\frac{{\rm SU}(1,1+N)}{{\rm U}(1)\times {\rm SU}(1,1+N)}$.\\
Adding all these facts together we may say that the spectrum is
classified by the following quantum numbers $(q,e_i)$, where $q$
are ``charges'' of the bulk gauge group, namely $|m_i|$, $|m_i\pm
m_j|$, $|m_i\pm m_j\pm m_k|$, $|m_i\pm m_j\pm m_k\pm m_{\ell}|$
and $e_i$ are the $N-1$ charges of the ${\rm SU}(N)$
root--lattice. In the supergravity spectrum there is a sector of
the type $(q,0)$ (the 128 states coming from the bulk) and a
sector of the type $(0,e_i)$ (the sector coming from the
$D3$--brane).

\section{Embedding of the $N=4$ model with six matter multiplets in the $N=8$}
\setcounter {equation}{0} \addtocounter{section}{0} There are two
inequivalent ways of embedding the $N=4$ model with an action
which is invariant under global ${\rm SL}(2,\mathbb{R})\times
{\rm GL}(6,\mathbb{R})$, within the $N=8$ theory. They correspond
to the two different embeddings of the ${\rm
SL}(2,\mathbb{R})\times {\rm GL}(6,\mathbb{R})$ symmetry of the
$N=4$ action inside ${\rm E}_{7(7)}$ which is the global symmetry
group of the $N=8$ field equations and Bianchi identities.\par
\begin{figure}[htbp]
\centering \epsfig{file=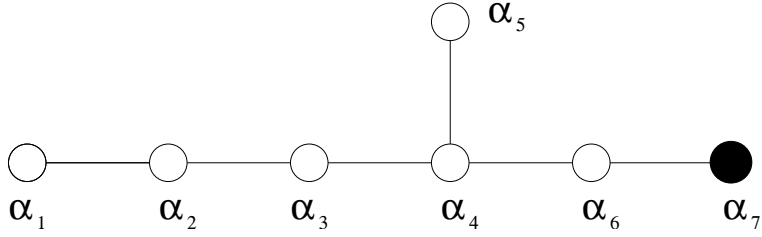, width=10cm} \caption{ ${\rm
E}_{7(7)}$ Dynkin diagram. The empty circles denote ${\rm
SO}(6,6)_T$ roots, while the filled circle denotes the  ${\rm
SO}(6,6)_T$ spinorial weight.}\label{e77}
\end{figure}

The $N=8$ model describes the low energy limit of Type II
superstring theory compactified on a six torus $T^6$. As shown in
\cite{Andrianopoli:1996zg, Andrianopoli:1996bq, Bertolini:1999uz}
the ten dimensional origin of the 70 scalar fields of the model
can be characterized group theoretically once the embeddings of
the isometry group ${\rm SO}(6,6)_T$ of the moduli space of $T^6$
and of the duality groups of higher dimensional maximal
supergravities are specified within ${\rm E}_{7(7)}$. This
analysis makes use of the {\em solvable Lie algebra
representation} which consists in describing the scalar manifold
as a solvable group manifold generated by a solvable Lie algebra
of which the scalar fields are the parameters. The solvable Lie
algebra associated with ${\rm E}_{7(7)}$ is defined by its
Iwasawa decomposition and is generated by the seven Cartan
generators and by the 63 shift generators corresponding to all
the positive roots. In this representation the Cartan subalgebra
is parametrized by the scalars coming from the diagonal entries
of the internal metric while all the other scalar fields are in
one to one correspondence with the ${\rm E}_{7(7)}$ positive
roots. We shall represent the  ${\rm E}_{7(7)}$ Dynkin diagram as
in figure \ref{e77}.
\begin{figure}[htbp]
\centering \epsfig{file=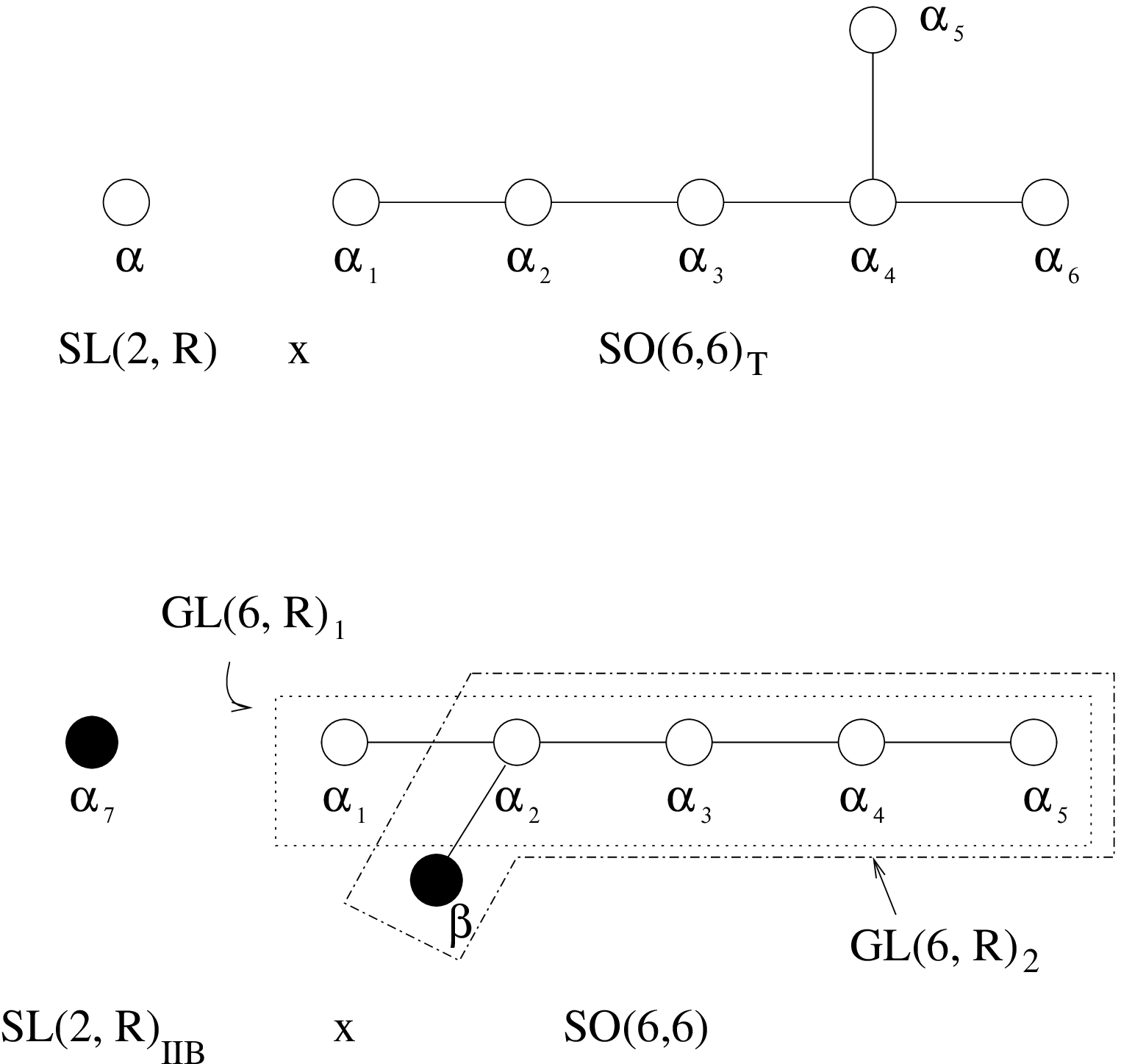, width=10cm} \caption{${\rm
SL}(2,\mathbb{R})\times {\rm SO}(6,6)_T$ and ${\rm
SL}(2,\mathbb{R})\times {\rm SO}(6,6)$ Dynkin diagrams. The root
$\alpha$ is the ${\rm E}_{7(7)}$ highest root while $\beta$ is
$\alpha_3+2\alpha_4+\alpha_5+2\alpha_6+\alpha_7$. The group ${\rm
SL}(2,\mathbb{R})_{IIB}$ is the symmetry group of the ten
dimensional type IIB theory.}\label{embe}
\end{figure}
The positive roots are expressed as combinations
$\alpha=\sum_{i=1}^7 n^i \alpha_i$ of the simple roots in which
the positive integers $n^i$ define the {\em grading} of the root
$\alpha$ with respect to $\alpha_i$. The isometry group of the
$T^6$ moduli space ${\rm SO}(6,6)_T$ is defined by the sub--Dynkin
diagram $\{\alpha_i\}_{i=1,\dots , 6}$ while the Dynkin diagram
of the duality group $E_{11-D(11-D)}$ of the maximal supergravity
in dimension $D\,>\,4$ is obtained from the ${\rm E}_{7(7)}$
Dynkin diagram by deleting the simple roots $\{\alpha_1,\dots
,\alpha_{D-4} \}$. Using these conventions in Table 1
\cite{Bertolini:1999uz} the correspondence between the 63 non
dilatonic scalar fields deriving from dimensional reduction of
type IIB theories and positive roots of ${\rm E}_{7(7)}$ is
illustrated.
\begin{table}[ht]
\vskip -25pt \label{tableroots}\caption{Correspondence between
the 63 non dilatonic scalar fields from type IIB string theory on
$T^6$ ($C^{(0)}$, $C^{(2)}\equiv C_{ij}$ and $C^{(4)}\equiv
C_{ijkl}$) and positive roots of ${\rm E}_{7(7)}$ according to the
solvable Lie algebra formalism. The $N=4$ Peccei--Quinn scalars
correspond to roots with grading 1 with respect to $\beta$,
namely those with $n^6=2$ and $n^7=1$ which are marked by an
arrow in the table.}
\begin{center}
{\scriptsize
\begin{tabular}{|c|c|c|c|}
\hline {\small IIB} &  {\small $\epsilon_i$--components}
& {\small $n^i$ gradings} \\
\hline \hline
 $C^{(0)}$ & $\frac{1}{2}(-1,-1,-1,-1,-1, {-} 1,\sqrt{2})$ &
$(0,0,0,0,0,0,1)$\\
\hline \hline
 $\mathcal{B}_{5\,6}$ & $(0,0,0,0,1,1,0)$ & $(0,0,0,0,0,1,0)$\\
\hline
 $g_{5\,6}$ &  $(0,0,0,0,1,-1,0)$ & $(0,0,0,0,1,0,0)$\\
\hline
 $C_{5\, 6}$ & $\frac{1}{2}(-1,-1,-1,-1,1,  1,\sqrt{2})$ &
$(0,0,0,0,0,1,1)$\\
\hline \hline
 $\mathcal{B}_{4\,5}$ & $(0,0,0,1,1,0,0)$ & $(0,0,0,1,1,1,0)$\\
\hline
 $g_{4\,5}$ &  $(0,0,0,1,-1,0,0)$ & $(0,0,0,1,0,0,0)$\\
\hline
 $\mathcal{B}_{4\,6}$ &  $(0,0,0,1,0,1,0)$ & $(0,0,0,1,0,1,0)$\\
\hline
 $g_{4\,6}$ & $(0,0,0,1,0,-1,0)$ & $(0,0,0,1,1,0,0)$\\
\hline
 $C_{4\, 5}$ &  $\frac{1}{2}(-1,-1,-1,1,1, {-} 1,\sqrt{2})$ &
$(0,0,0,1,1,1,1)$\\
\hline $C_{4\, 6}$ &  $\frac{1}{2}(-1,-1,-1,1,-1, 1,\sqrt{2})$ &
$(0,0,0,1,0,1,1)$\\
\hline \hline
 $\mathcal{B}_{3\,4}$ &  $(0,0,1,1,0,0,0)$ & $(0,0,1,2,1,1,0)$\\
\hline
 $g_{3\,4}$ &  $(0,0,1,-1,0,0,0)$ & $(0,0,1,0,0,0,0)$\\
\hline
 $\mathcal{B}_{3\,5}$ &  $(0,0,1,0,1,0,0)$ & $(0,0,1,1,1,1,0)$\\
\hline
 $g_{3\,5}$ &  $(0,0,1,0,-1,0,0)$ & $(0,0,1,1,0,0,0)$\\
\hline
 $\mathcal{B}_{3\,6}$ &  $(0,0,1,0,0,1,0)$ & $(0,0,1,1,0,1,0)$\\
\hline
 $g_{3\,6}$ & $(0,0,1,0,0,-1,0)$ & $(0,0,1,1,1,0,0)$\\
\hline
 $C_{3\, 4}$ &  $\frac{1}{2}(-1,-1,1,1,-1, {-} 1,\sqrt{2})$ &
$(0,0,1,2,1,1,1)$\\
\hline $C_{3\, 5}$ &  $\frac{1}{2}(-1,-1,1,-1,1,{-} 1,\sqrt{2})$ &
$(0,0,1,1,1,1,1)$\\
\hline
 $C_{3\, 6}$ &  $\frac{1}{2}(-1,-1,1,-1,-1, 1,\sqrt{2})$ &
$(0,0,1,1,0,1,1)$\\
\hline
 $C_{3\, 4\,5\,6}$ &  $\frac{1}{2}(-1,-1,1,1,1,
  1,\sqrt{2})$ &
$(0,0,1,2,1,2,1)$  $\leftarrow$\\
\hline \hline
$\mathcal{B}_{2\,3}$ & $(0,1,1,0,0,0,0)$ & $(0,1,2,2,1,1,0)$\\
\hline
$g_{2\,3}$ & $(0,1,-1,0,0,0,0)$ & $(0,1,0,0,0,0,0)$\\
\hline
 $\mathcal{B}_{2\,4}$ & $(0,1,0,1,0,0,0)$ & $(0,1,1,2,1,1,0)$\\
\hline
 $g_{2\,4}$ & $(0,1,0,-1,0,0,0)$ & $(0,0,1,0,0,0,0)$\\
\hline
$\mathcal{B}_{2\,5}$ &  $(0,1,0,0,1,0,0)$ & $(0,1,1,1,1,1,0)$\\
\hline
 $g_{2\,5}$ &  $(0,1,0,0,-1,0,0)$ & $(0,0,0,1,0,0,0)$\\
\hline
 $\mathcal{B}_{2\,6}$ &  $(0,1,0,0,0,1,0)$ & $(0,1,1,1,0,1,0)$\\
\hline
 $g_{2\,6}$ &  $(0,1,0,0,0,-1,0)$ & $(0,0,0,0,1,0,0)$\\
\hline
 $C_{2\, 3}$ & $\frac{1}{2}(-1,1,1,-1,-1,{-} 1,\sqrt{2})$ &
$(0,1,2,2,1,1,1)$\\
\hline $C_{2\, 4}$ &  $\frac{1}{2}(-1,1,-1,1,-1, {-} 1,\sqrt{2})$
&
$(0,1,1,2,1,1,1)$\\
\hline
 $C_{2\, 5}$ &  $\frac{1}{2}(-1,1,-1,-1,1, {-} 1,\sqrt{2})$ &
$(0,1,1,1,1,1,1)$\\
\hline
 $C_{2\, 6}$ &  $\frac{1}{2}(-1,1,-1,-1,-1, 1,\sqrt{2})$ &
$(0,1,1,1,0,1,1)$\\
\hline
 $C_{2\, 4\,5\,6}$ &  $\frac{1}{2}(-1,1,-1,1,1,  1,\sqrt{2})$ &
$(0,1,1,2,1,2,1)$ $\,\,\leftarrow$\\
\hline
 $C_{2\,3\,5\, 6}$ &  $\frac{1}{2}(-1,1,1,-1,1,
 1,\sqrt{2})$ &
$(0,1,2,2,1,2,1)$ $\,\,\leftarrow$\\
\hline
 $C_{2\,3\,4\, 6}$ & $\frac{1}{2}(-1,1,1,1,-1,
 1,\sqrt{2})$ &
$(0,1,2,3,1,2,1)$ $\,\,\leftarrow$\\
\hline
 $C_{2\,3\,4\, 5}$ &  $\frac{1}{2}(-1,1,1,1,1,{-} 1,\sqrt{2})$ &
$(0,1,2,3,1,2,1)$ $\,\,\leftarrow$\\
\hline \hline
 $\mathcal{B}_{1\,2}$ &  $(1,1,0,0,0,0,0)$ & $(1,2,2,2,1,1,0)$\\
\hline
 $g_{1\,2}$ &  $(1,-1,0,0,0,0,0)$ & $(1,0,0,0,0,0,0)$\\
\hline
 $\mathcal{B}_{1\,3}$ &  $(1,0,1,0,0,0,0)$ & $(1,1,2,2,1,1,0)$\\
\hline
 $g_{1\,3}$ &  $(1,0,-1,0,0,0,0)$ & $(1,1,0,0,0,0,0)$\\
\hline
$\mathcal{B}_{1\,4}$ &  $(1,0,0,1,0,0,0)$ & $(1,1,1,2,1,1,0)$\\
\hline
$g_{1\,4}$ & $(1,0,0,-1,0,0,0)$ & $(1,1,1,0,0,0,0)$\\
\hline
 $\mathcal{B}_{1\,5}$ &  $(1,0,0,0,1,0,0)$ & $(1,1,1,1,1,1,0)$\\
\hline
$g_{1\,5}$ &  $(1,0,0,0,-1,0,0)$ & $(1,1,1,1,0,0,0)$\\
\hline
$\mathcal{B}_{1\,6}$ &  $(1,0,0,0,0,1,0)$ & $(1,1,1,1,0,1,0)$\\
\hline
 $g_{1\,6}$ &  $(1,0,0,0,0,-1,0)$ & $(1,1,1,1,1,0,0)$\\
\hline $\mathcal{B}_{\mu\nu}$ &  $(0,0,0,0,0,0,\sqrt{2})$ &
$(1,2,3,4,2,3,2)$\\
\hline
 $C_{\mu\nu}$ &  $\frac{1}{2}(1,1,1,1,1, 1,\sqrt{2})$ &
$(1,2,3,4,2,3,1)$\\
\hline
 $C_{1\,2}$ &  $\frac{1}{2}(1,1,-1,-1,-1,{-} 1,\sqrt{2})$ &
$(1,2,2,2,1,1,1)$\\
\hline $C_{1\,3}$ &  $\frac{1}{2}(1,-1,1,-1,-1,{-} 1,\sqrt{2})$ &
$(1,1,2,2,1,1,1)$\\
\hline $C_{1\,4}$ &  $\frac{1}{2}(1,-1,-1,1,-1,{-} 1,\sqrt{2})$ &
$(1,1,1,2,1,1,1)$\\
\hline $C_{1\,5}$ &  $\frac{1}{2}(1,-1,-1,-1,1,{-} 1,\sqrt{2})$ &
$(1,1,1,1,1,1,1)$\\
\hline
 $C_{1\,6}$ &$\frac{1}{2}(1,-1,-1,-1,-1, 1,\sqrt{2})$ &
$(1,1,1,1,0,1,1)$\\
\hline $C_{1\,4\,5\,6}$ & $\frac{1}{2}(1,-1,-1,1,1, 1,\sqrt{2})$
& $(1,1,1,2,1,2,1)$
$\,\,\leftarrow$\\
\hline
 $C_{1\,3\,5\,6}$ &
$\frac{1}{2}(1,-1,1,-1,1, 1,\sqrt{2})$ & $(1,1,2,2,1,2,1)$
$\,\,\leftarrow$\\
\hline $C_{1\,3\,4\,6}$ & $\frac{1}{2}(1,-1,1,1,-1, 1,\sqrt{2})$
& $(1,1,2,3,1,2,1)$
$\,\,\leftarrow$\\
\hline
 $C_{1\,3\,4\,5}$ &
$\frac{1}{2}(1,-1,1,1,1,{-} 1,\sqrt{2})$ & $(1,1,2,3,2,2,1)$
$\,\,\leftarrow$\\
\hline
 $C_{1\,2\,5\,6}$ &
$\frac{1}{2}(1,1,-1,-1,1, 1,\sqrt{2})$ & $(1,2,2,2,1,2,1)$
$\,\,\leftarrow$\\
\hline $C_{1\,2\,4\,6}$ & $\frac{1}{2}(1,1,-1,1,-1, 1,\sqrt{2})$
& $(1,2,2,3,1,2,1)$
$\,\,\leftarrow$\\
\hline $C_{1\,2\,4\,5}$ & $\frac{1}{2}(1,1,-1,1,1,{-}
1,\sqrt{2})$ & $(1,2,2,3,2,2,1)$
$\,\,\leftarrow$\\
\hline
 $C_{1\,2\,3\,6}$ &
$\frac{1}{2}(1,1,1,-1,-1, 1,\sqrt{2})$ & $(1,2,3,3,1,2,1)$
$\,\,\leftarrow$\\
\hline $C_{1\,2\,3\,5}$ & $\frac{1}{2}(1,1,1,-1,1,{-}
1,\sqrt{2})$ & $(1,2,3,3,2,2,1)$
$\,\,\leftarrow$\\
\hline
 $C_{1\,2\,3\,4}$ &
$\frac{1}{2}(1,1,1,1,-1,{-} 1,\sqrt{2})$ & $(1,2,3,4,2,2,1)$
$\,\,\leftarrow$\\
\hline
\end{tabular}}
\end{center}
\end{table}
In this framework the R--R scalars, for instance, are defined by
the positive roots which are spinorial with respect to ${\rm
SO}(6,6)_T$, i.e. which have grading $n^7=1$ with respect to the
spinorial simple root $\alpha_7$. On the contrary the NS--NS
scalars are defined by the roots with $n^7=0,\,2$.\par Let us
first discuss the embedding of the ${\rm SL}(2,\mathbb{R})\times
{\rm SO}(6,6)$ duality group of our model within ${\rm E}_{7(7)}$.
 In the solvable Lie algebra language the
Peccei--Quinn scalars parametrize the maximal abelian ideal of the
solvable Lie algebra generating the scalar manifold. As far as the
manifold ${\rm SO}(6,6)/{\rm SO}(6)\times {\rm SO}(6)$ is
concerned, this abelian ideal is 15 dimensional and is generated
by the shift operators corresponding to positive ${\rm SO}(6,6)$
roots with grading one with respect to the simple root placed at
one of the two symmetric ends of the corresponding Dynkin diagram
${\rm D}_6$. Since in our model the Peccei--Quinn scalars are of
R--R type, the ${\rm SO}(6,6)$ duality group embedded in ${\rm
E}_{7(7)}$ does not coincide with ${\rm SO}(6,6)_T$. Indeed one of
its symmetric ends should be a spinorial root of ${\rm
SO}(6,6)_T$. Moreover the ${\rm SL}(2,\mathbb{R})$ group
commuting with ${\rm SO}(6,6)$ should coincide with the ${\rm
SL}(2,\mathbb{R})_{IIB}$ symmetry group of the ten dimensional
type IIB theory, whose Dynkin diagram consists in our formalism
of the simple root $\alpha_7$. This latter condition uniquely
determines the embedding of ${\rm SO}(6,6)$ to be the one defined
by ${\rm
D}_6=\{\alpha_1,\alpha_2,\alpha_3,\alpha_4,\alpha_5,\beta\}$,
where $\beta=\alpha_3+2\alpha_4+\alpha_5+2\alpha_6+\alpha_7$ is
the spinorial root (see figure \ref{embe}).  On the other hand
the 20 scalar fields parametrizing the manifold ${\rm
SL}(6,\mathbb{R})/{\rm SO}(6)$ are all of NS--NS type (they come
from the components of the $T^6$ metric). This fixes the
embedding of ${\rm SL}(6,\mathbb{R})$ within ${\rm E}_{7(7)}$
which we shall denote by ${\rm SL}(6,\mathbb{R})_1$: its Dynkin
diagram is $\{\alpha_1,\alpha_2,\alpha_3,\alpha_4,\alpha_5\}$. The
Peccei--Quinn scalars are then defined by the positive roots with
grading one with respect to the spinorial end $\beta$ of ${\rm
D}_6$  which is not contained in ${\rm SL}(6,\mathbb{R})_1$. In
table 1 the scalar fields in ${\rm SO}(1,1)\times {\rm
SL}(6,\mathbb{R})_1/{\rm SO}(6)$ which are not dilatonic (i.e. do
not correspond to diagonal entries of the $T^6$ metric)
correspond to the ${\rm SL}(6,\mathbb{R})_1$ positive roots which
are characterized by $n^6=n^7=0$ and are the off--diagonal
entries of the internal metric. The Peccei--Quinn scalars on the
other hand correspond to the roots with grading one with respect
to $\beta$, which in Table 1 are those with $n^6=2,\,n^7=1$ and
indeed, as expected, are identified with the internal components
of the type IIB four form.
\par The above analysis
based on the microscopic nature of the scalars present in our
model has led us to select one out of two inequivalent embeddings
of the ${\rm SL}(6,\mathbb{R})$ group within ${\rm E}_{7(7)}$
which we shall denote by ${\rm SL}(6,\mathbb{R})_1$ and ${\rm
SL}(6,\mathbb{R})_2$. The former corresponds to the $A_5$ Dynkin
diagram running from $\alpha_1$ to $\alpha_5$ while the latter to
the $A_5$ diagram running from $\beta$ to $\alpha_5$. The ${\rm
SL}(6,\mathbb{R})_1$ symmetry group of our $N=4$ Lagrangian is
uniquely defined as part of the maximal subgroup ${\rm
SL}(3,\mathbb{R})\times {\rm SL}(6,\mathbb{R})_1$ of ${\rm
E}_{7(7)}$ (in which ${\rm SL}(3,\mathbb{R})$ represents an
enhancement of ${\rm SL}(2,\mathbb{R})_{IIB}$
\cite{Andrianopoli:1996zg}) with respect to which the relevant
${\rm E}_{7(7)}$ representations branch as follows:
\begin{eqnarray}
\label{56e7}{\bf 56}&\rightarrow & {\bf (1,20)}+ {\bf (3^\prime,6)}+ {\bf (3,6^\prime)}\\
{\bf 133}&\rightarrow &  {\bf (8,1)}+ {\bf (3,15)}+ {\bf
(3^\prime,15^\prime)}+ {\bf (1,35)}
\end{eqnarray}
Moreover with respect to the ${\rm SO}(3)\times {\rm SO}(6)$
subgroup of ${\rm SL}(3,\mathbb{R})\times {\rm
SL}(6,\mathbb{R})_1$ the relevant ${\rm SU}(8)$ representations
branch in the following way:
\begin{eqnarray}
{\bf 8}&\rightarrow & {\bf (2,4)}\nonumber\\
{\bf 56}&\rightarrow & {\bf (2,20)}+{\bf (4,\overline{4})}\nonumber\\
{\bf 63}&\rightarrow & {\bf (1,15)}+{\bf (3,1)}+{\bf (3,15)}\nonumber\\
{\bf 70}&\rightarrow &  {\bf (1,20)}+ {\bf (3,15)}+ {\bf (5,1)}
\end{eqnarray}
The group ${\rm GL}(6,\mathbb{R})_2$ on the other hand is
contained inside both ${\rm SL}(8,\mathbb{R})$ and ${\rm
E}_{6(6)}\times {\rm O}(1,1)$ as opposite to $ {\rm
GL}(6,\mathbb{R})_1$. As a consequence of this it is possible in
the $N=8$ theory to choose electric field strengths and their
duals in such a way that ${\rm SL}(2,\mathbb{R})\times {\rm
GL}(6,\mathbb{R})_2$ is contained in the global symmetry group of
the action while this is not the case for the group ${\rm
SL}(2,\mathbb{R})\times {\rm GL}(6,\mathbb{R})_1\subset {\rm
SL}(3,\mathbb{R})\times {\rm SL}(6,\mathbb{R})_1$. Indeed as it
is apparent from eq. (\ref{56e7}) the electric/magnetic charges
in the ${\bf 56}$ of ${\rm E}_{7(7)}$ do not branch  with respect
to ${\rm SL}(6,\mathbb{R})_1$ into two 28 dimensional reducible
representations as it would be required in order for ${\rm
SL}(6,\mathbb{R})_1$ to be contained in the symmetry group of the
Lagrangian. On the other hand the group ${\rm
SL}(2,\mathbb{R})\times {\rm O}(1,1)\times {\rm SO}(6)\subset {\rm
SL}(3,\mathbb{R})\times {\rm SL}(6,\mathbb{R})_1$ can be
contained in the symmetry group of the $N=8$ action for a certain
choice of the electric and magnetic field
strengths\footnote{Actually this property holds for the whole
${\rm SL}(3,\mathbb{R})\times{\rm SO}(6)$}, since with respect to
it the ${\bf 56}$ branches as follows (the grading as usual
refers to the ${\rm O}(1,1)$ factor):
\begin{eqnarray}
\label{56e7n}{\bf 56}&\rightarrow & {\bf (1,10)}_0+ {\bf
(1,\overline{10})}_0+{\bf (1,6)}_{+2}+{\bf (1,6)}_{-2}+{\bf
(2,6)}_{+1}+{\bf (2,6)}_{-1}
\end{eqnarray}
In truncating to the $N=4$ model the charges in the ${\bf
(1,10)}_0+ {\bf (1,\overline{10})}_0+{\bf (1,6)}_{+2}+{\bf
(1,6)}_{-2}$ are projected out and the symmetry group of the
Lagrangian is enhanced to ${\rm SL}(2,\mathbb{R})\times {\rm
GL}(6,\mathbb{R})_1$.

\subsection{The masses in the $N=4$ theory with
gauged Peccei--Quinn isometries and ${\rm USp}(8)$ weights} As we
have seen, in the $N=4$ theory with gauged Peccei--Quinn
isometries, the parameters of the effective action at the origin
of the scalar manifold are encoded in the tensor
$f_\alpha{}^{IJK}$. The condition for the origin to be an
extremum of the potential, when $\a=1$, constrains the fluxes in
the following way:
\begin{eqnarray}
f_1{}^{-IJK}-{\rm i} f_2{}^{-IJK}&=&0
\end{eqnarray}
therefore all the independent gauge parameters will be contained
in the combination $f_1{}^{-IJK}+{\rm i} f_2{}^{-IJK}$
transforming in the ${\bf 10}^{+1}$ with respect to ${\rm U}(4)$
and in its complex conjugate which belongs to the $\overline{{\bf
10}}^{-1}$. Using the gamma matrices each of these two tensors can
be mapped into $4\times 4$ symmetric complex matrices:
\begin{eqnarray}
B_{AB}&=&\left(f_1{}^{-IJK}+{\rm i} f_2{}^{-IJK}\right)\Gamma^{IJK}{}_{AB}\,\,\,\in\,\,\,{\bf 10}^{+1}\nonumber\\
\overline{B}^{AB}&=&\left(f_1{}^{+IJK}-{\rm i}
f_2{}^{+IJK}\right)\Gamma_{IJK}{}^{AB}\,\,\,\in\,\,\,\overline{{\bf
10}}^{-1}
\end{eqnarray}
where the matrix $B_{AB}$ is proportional to the gravitino mass
matrix $S_{AB}$. If we denote by $A_A{}^B$ a generic generator of
$\mathfrak{u}(4)$ we may formally build the representation of a
generic  $\mathfrak{usp}(8)$ generator in the ${\bf 8}$:
\begin{eqnarray}
\left(\matrix{A_A{}^B & B_{AC}\cr -\overline{B}^{DB} &
-A_C{}^D}\right)&\in &{\mathfrak{ usp}}(8)\label{usp8}
\end{eqnarray}
The ${\rm U}(1)$ group in ${\rm U}(4)$ is generated by
$A_A{}^B={\rm i}\,\delta_A{}^B$. Under a ${\rm U}(4)$
transformation $\mathcal{A}$ the matrix B transforms as follows:
\begin{eqnarray}
B &\rightarrow & \mathcal{A}B\mathcal{A}^t
\end{eqnarray}
Therefore using ${\rm U}(4)$ transformations the off diagonal
generators in the $\mathfrak{usp}(8)/\mathfrak{u}(4)$ can be
brought to the following form
\begin{eqnarray}
\left(\matrix{{\bf 0} & B^{(d)}\cr -B^{(d)} & {\bf 0}}\right)&\equiv &  m_i\, H_i\nonumber\\
B^{(d)}&=&{\rm diag}(m_1,m_2,m_3,m_4)\,\,\,\,\,\,m_i>0
\label{usp8cart}
\end{eqnarray}
where the phases and thus the signs of the $m_i$ were fixed using
the ${\rm U}(1)^4$ transformations inside ${\rm U}(4)$ and $H_i$
denote a basis of generators of the $\mathfrak{usp}(8)$ Cartan
subalgebra. The gravitino mass matrix represents just the upper
off diagonal block of the $\mathfrak{usp}(8)$ Cartan generators
in the ${\bf 8}$.\par As far as the vectors are concerned we may
build the $\mathfrak{usp}(8)$ generators in the ${\bf 27}$ in
much the same way as we did for the gravitini case, by using the
$\mathfrak{u}(4)$ generators in the ${\bf 15}$ and in the ${\bf
6}+{\bf 6}$ to form the diagonal $15\times 15$ and $12\times 12$
blocks of a $27\times 27$ matrix.
\begin{eqnarray}
\left(\matrix{A_{15\times 15} & K_{15\times 12}\cr K_{12\times
15} & A_{12\times 12}}\right)&\in & \mathfrak{usp}(8)
\end{eqnarray}
Here $A_{15\times 15}\equiv
A^{\L\S}_{\phantom{a}\phantom{b}\G\D}$, $A_{12\times 12}\equiv
A^{\L\a}_{\phantom{a}\phantom{b}\G\b}$ while $K_{15\times
12}\equiv K^{\L\S|\G\a}=f^{\L\S\G\a}$ and $K_{12\times
15}=-K_{15\times 12}^T$.\\
The vector mass matrix is:
\begin{eqnarray}
M_{{\rm (vector)}}^2 &\propto &K_{15\times 12}^t K_{15\times 12}
\label{vec}
\end{eqnarray}
 By acting by means of ${\rm U}(4)$ on the rectangular matrix $K_{15\times 12}$ it is possible to reduce it to the upper off--diagonal part of a generic element of the
$\mathfrak{usp}(8)$ Cartan subalgebra:
\begin{eqnarray}
K_{15\times 12}&=&\left(\matrix{a_1&0&\dots&\dots &0\cr 0&a_2 &0&\dots &0\cr\vdots & &\ddots & &\vdots\cr\vdots & & &\ddots  &\vdots\cr 0 &\dots & \dots &0& a_{12}\cr 0 &0&\dots &\dots& 0\cr 0 &0&\dots &\dots& 0\cr  0 &0&\dots &\dots& 0}\right)\nonumber\\&&\nonumber\\
a_\ell &=& |m_i\pm m_j|\,\,\,\,\,\, 1\le i<j \le
4\,\,\,;\,\,\,m_i\ge 0
\end{eqnarray}
Using equation (\ref{vec}) we may read the mass eigenvalues for
the vectors which are just $a_\ell$.\par The above argument may
be extended also to the gaugini and the scalars as discussed in
the next Section.
\subsection{Duality with a truncation of the spontaneously broken $N=8$ theory from Scherk--Schwarz reduction
}As discussed in the previous sections the microscopic
interpretation of the fields in our $N=4$ model is achieved by
its identification, at the ungauged level, with a truncation of
the $N=8$ theory describing the field theory limit of IIB string
theory on $T^6$. To this end the symmetry group of the $N=4$
action is interpreted as the ${\rm SL}(2,\mathbb{R})\times {\rm
GL}(6,\mathbb{R})_1$ inside the ${\rm SL}(3,\mathbb{R})\times {\rm
SL}(6,\mathbb{R})_1$ maximal subgroup of ${\rm E}_{7(7)}$, which
is the natural group to consider when interpreting the four
dimensional theory from the type IIB point of view, since the
${\rm SL}(3,\mathbb{R})$ factor represents an enhancement of the
type IIB symmetry group ${\rm SL}(2,\mathbb{R})_{IIB}\times {\rm
SO}(1,1)$, where ${\rm SO}(1,1)$ is associated to the $T^6$
volume, while ${\rm SL}(6,\mathbb{R})_1$ is the group acting on
the moduli of the $T^6$ metric. A different microscopic
interpretation of the ungauged $N=4$ theory would follow from the
identification of its symmetry group with the group ${\rm
SL}(2,\mathbb{R})\times {\rm GL}(6,\mathbb{R})_2$ contained in
both ${\rm E}_{6(6)}\times {\rm O}(1,1)$ and ${\rm
SL}(8,\mathbb{R})$ subgroups of ${\rm E}_{7(7)}$, where, although
the ${\rm SL}(2,\mathbb{R})$ factor is still ${\rm
SL}(2,\mathbb{R})_{IIB}$, the fields are naturally interpreted in
terms of dimensionally reduced M--theory since ${\rm
GL}(6,\mathbb{R})_2$ this time is the group acting on the moduli
of the $T^6$ torus from $D=11$ to $D=5$. At the level of the
$N=4$ theory the ${\rm SL}(6,\mathbb{R})_1$ and the ${\rm
SL}(6,\mathbb{R})_2$ are equivalent, while their embedding in
${\rm E}_{7(7)}$ is different and so is the microscopic
interpretation of the fields in the corresponding theories. Our
gauged model is obtained by introducing in the model with ${\rm
SL}(2,\mathbb{R})\times {\rm GL}(6,\mathbb{R})_1$ manifest
symmetry  a gauge group characterized by a flux tensor
transforming in the ${\bf (2,20)}^{+3}$. It is interesting to
notice that if the symmetry of the ungauged action were
identified with ${\rm SL}(2,\mathbb{R})\times {\rm
GL}(6,\mathbb{R})_2$ formally we would have the same $N=4$ gauged
model, but, as we are going to show, this time we could interpret
it as a truncation to $N=4$ of the spontaneously broken $N=8$
theory deriving from a Scherk--Schwarz reduction from $D=5$. The
latter, as mentioned in the introduction, is a gauged $N=8$
theory which is completely defined once we specify the gauge
generator $T_0\in \mathfrak{e}_{6,6}$ to be gauged by the
graviphoton arising from the five dimensional metric. The gauging
(couplings, masses etc...) is therefore characterized by the
${\bf 27}$ representation of $T_0$, namely by the flux matrix
$f_r^s{}_0$ ($r,s=1,\dots, 27$), element of ${\rm
Adj}(\mathfrak{e}_{6,6})={\bf 78}$ \cite{deWit:2002vt}.
Decomposing this representation with respect to ${\rm
SL}(2,\mathbb{R})\times {\rm SL}(6,\mathbb{R})_2$ we have:
\begin{eqnarray}
{\bf 78}&\rightarrow & {\bf (3,1)}+{\bf (1,35)}+{\bf (2,20)}
\end{eqnarray}
The representation ${\bf (2,20)}$ defines the gaugings in which
we choose:
\begin{eqnarray}
T_0 &\in& \frac{\mathfrak{e}_{6,6}}{\mathfrak{sl}(2,\mathbb{R})+
\mathfrak{sl}(6,\mathbb{R})_2}
\end{eqnarray}
 These generators can be either compact or non--compact. However, it is known that only for compact $T_0$ the gauged $N=8$ theory is a ``no--scale'' model with  a Minkowski vacuum at the origin of the moduli space (flat gaugings). Let us consider the relevant branchings of ${\rm E}_{7(7)}$
  representations with respect to ${\rm SL}(2,\mathbb{R})\times {\rm SL}(6,\mathbb{R})_2$:
\begin{eqnarray*}
{\bf 56}\!&\!\rightarrow\! &\! {\bf (2,6^\prime)}_{+1} +{\bf (2,6)}_{-1}+{\bf (1,15^\prime)}_{-1} +{\bf (1,15)}_{+1}+{\bf (1,1)}_{-3} +{\bf (1,1)}_{+3}\nonumber\\
{\bf 133}\!&\!\rightarrow\! &\!  {\bf (3,1)}_0\!+{\bf
(1,1)}_0\!+{\bf (1,35)}_0\!+{\bf (2,20)}_0\!+{\bf
(2,6)}_{+2}\!+{\bf (2,6^\prime)}_{-2}\!+{\bf (1,15^\prime)}_{+2}
+{\bf (1,15)}_{-2}
\end{eqnarray*}
where the ${\bf (2,6)}_{-1}+{\bf (1,15^\prime)}_{-1}$ in the
first branching denote the vectors deriving from five dimensional
vectors while ${\bf (1,1)}_{-3}$ is the graviphoton. The
truncation to $N=4$ is achieved at the bosonic level by
projecting the ${\bf 56}$ into ${\bf (2,6^\prime)}_{+1} +{\bf
(2,6)}_{-1}$ and the ${\bf 133}$ into the adjoint of ${\rm
SL}(2,\mathbb{R})\times {\rm SO}(6,6)$, namely ${\bf (3,1)}_0+{\bf
(1,1)}_0+{\bf (1,35)}_0+{\bf (1,15^\prime)}_{+2}+{\bf
(1,15)}_{-2}$.\par If we chose $T_0$ within ${\bf (2,20)}$  as a
$27\times 27$ generator it has only non vanishing entries
$f^{\Lambda\Sigma\Gamma\alpha}$ in the blocks ${\bf
(1,15^\prime)}\times {\bf (2,6^\prime)}$ and ${\bf
(2,6^\prime)}\times {\bf (1,15^\prime)}$ and inspection into the
couplings of these theories shows that the truncation to $N=4$ is
indeed consistent and that we formally get the $N=4$ gauged
theory  considered in this paper with six matter multiplets.
Moreover the extremality condition $f_1{}^{-IJK}-{\rm i}
f_2{}^{-IJK}=0$ discussed in the previous section coincides with
the condition on $T_0$ to be compact:
\begin{eqnarray}
f_1{}^{-IJK}-{\rm i} f_2{}^{-IJK}&=&0
\,\,\,\,\,\Leftrightarrow\,\,\,\,
T_0\,\in\,\frac{\mathfrak{usp}(8)}{\mathfrak{so}(2)+
\mathfrak{so}(6)}\,\,\,\,\mbox{($N=8$ flat gauging)}\nonumber
\end{eqnarray}
After restricting the ${\bf (2,20)}$ generators $T_0$ to
$\mathfrak{usp}(8)$ they will transform in the ${\bf 10}^{+1}+{\bf
\overline{10}}^{\ -1}$ with respect to ${\rm SO}(2)\times {\rm
SO}(6)$, ${\bf 10}^{+1}$ being the same representation as the
gravitino mass matrix.  In the previous section the itinerary
just described from the $N=8$ to the $N=4$ theory was followed
backwards: we have reconstructed the $27\times 27$
$\mathfrak{usp}(8)$ matrix $T_0$ starting from the symmetry
$\mathfrak{u}(4)$ of the ungauged $N=4$ action and the fluxes
$f_\alpha{}^{IJK}$ defining the
gauging.\\

As far as the fermions are concerned, we note that in the $N=8$
theory, the gravitini in the $\bf{8}$ of ${\rm USp}(8)$ decompose
under ${\rm SO}(2)\times {\rm SO}(6)_2\subset {\rm USp}(8)$ as
\be{\bf 8}\longrightarrow {\bf 4}^{+\frac{1}{2}}+\overline{\bf
4}^{-\frac{1}{2}}\ee so that a vector in the ${\bf 8}$ can be
written as
\begin{equation}V^a=\begin{pmatrix}{v_A^{(+\frac{1}{2})}\cr v^{A(-\frac{1}{2})}}\end{pmatrix}
\quad\quad A,B=1,\dots 8;\quad a,b=1,\dots 8\end{equation} From
equation \eq{usp8} we see that the off diagonal generators in the
coset $\frac{{\rm USp}(8)}{{\rm U}(4)}$ belong to the ${\rm
U}(4)$ representation ${\bf 10}^{+1}+\overline{\bf 10}^{\ -1}$
among which we find the symplectic invariant
\begin{equation}\mathbf{C}_{ab}=\begin{pmatrix}{{\bf 0}_{4\times 4}
& \bfone_{4\times 4}\cr -\bfone_{4\times 4}&{\bf 0}_{4\times 4}
}\end{pmatrix}\end{equation}


The basic quantities which define the fermionic masses and the
gradient flows equations of the $N=4$ model (in absence of
$D3$--brane couplings) are the symmetric matrices
\ba S_{AB}&=&-\frac{i}{48} (\overline{F}^{IJK-}+\overline{C}^{IJK-})(\G_{IJK})_{AB}\\
N_{AB}&=&-\frac{1}{48}(F^{IJK-}+C^{IJK-})(\G_{IJK})_{AB}\ea that
belong to the representations ${\bf 10}^{+1}$, ${\bf 10}^{-1}$ of
${\rm U}(4)$ respectively. Note that they have opposite ${\rm
U}(1)_R$ weight \be w[S_{AB}]=-w[N_{AB}]=1\ee If we indicate with
$\l_A^{(\overline{4})}$, $\l_{IA}^{(\overline{20})}$ the ${\bf
\overline{4} }$, ${\bf \overline{20}}$ irreducible
representations of the ${\bf 24}$ $\l^I_A$ bulk gaugini, the
weights of the left handed gravitini, dilatini and gaugini as
given in equation \eq{pesos} give in this case
\begin{equation}
\label{pesosx}w[\p_A]=\frac{1}{2};\quad w[\c^A]=\frac{3}{2};\quad
w[\l_{IA}^{(\overline{20})}]=-\frac{1}{2};\quad
w[\l^{A(\overline{4})}]=-\frac{1}{2};\quad
w[\l^i_{A}]=-\frac{1}{2}
\end{equation}
>From equations \eq{lagmass}, \eq{lagpot},
\eq{shgrav}--\eq{massgaux} it follows that, by suitable
projection on the irreducible representations ${\bf \overline{4}
}$, ${\bf \overline{20}}$, the following mass matrices associated
to the various bilinears either depend on the $S_{AB}$ or
$N_{AB}$ matrices, according to the following scheme\footnote{We
remind that $(S^{AB},\,N^{AB})$ have opposite ${\rm U}(1)$
weights, since they transform in the complex conjugate
representation with respect to $(S_{AB},\,N_{AB})$.}:
\begin{eqnarray}
\c^{A(\overline{4})}\c^{B(\overline{4})}&\longrightarrow& 0\\
\c^{A(\overline{4})}\l_{IB}^{(\overline{20})}&\longrightarrow& S^{AB}\\
\c^{A(\overline{4})}\l^{B(\overline{4})}&\longrightarrow& N_{AB}\\
\l_{IA}^{(\overline{20})}\l_{JB}^{(\overline{20})}&\longrightarrow&
S_{AB}\\
\l^{A(\overline{4})}\l^{B(\overline{4})}&\longrightarrow&
S_{AB}\\
\l^{A(\overline{4})}\l_{IB}^{(\overline{20})}&\longrightarrow&
N^{AB}\\
\p_A\c_B^{(4)}&\longrightarrow & N^{AB}\\
\p_A\l_B^{(4)}&\longrightarrow&S^{AB}\\
\p_A\l^{IB(20)}&\longrightarrow& N_{AB}\\
\p_A\p_B&\longrightarrow& S^{AB}\\
\l^i_A\,\l^j_B&\longrightarrow & N^{AB}
\end{eqnarray}
All these assignments come from the fact that $S_{AB}$, $N_{AB}$
are in the ${\bf 10}^{+1}$ ${\bf 10}^{-1}$ representations of
${\rm U}(4)$ and the mass matrices must have grading opposite to
the bilinear fermions, since the Lagrangian has zero grading.
Indeed, from the group theoretical decomposition we find, for
each of the listed bilinear fermions \ba \label{d1}\overline{\bf
4}^{\frac{3}{2}}\times\overline{\bf 4}^{\frac{3}{2}}&
\not\supset&{\bf 10}^{\pm1}\\
\label{d2}\overline{\bf 4}^{\frac{3}{2}}\times\overline{\bf 20}^{\ -\frac{1}{2}}& \supset&{\bf 10}^{+1}\\
\label{d3}\overline{\bf 4}^{\frac{3}{2}}\times\overline{\bf 4}^{\ -\frac{1}{2}}& \supset&\overline{\bf 10}^{\ +1}\\
\label{d4}\overline{\bf 20}^{\ -\frac{1}{2}}\times\overline{\bf
20}^{\ -\frac{1}{2}}& \supset&{\bf
10}^{-1}+\overline{\bf 10}^{\ -1}\\
\label{d5}\overline{\bf 4}^{\ -\frac{1}{2}}\times\overline{\bf 4}^{\ -\frac{1}{2}}& \supset&\overline{\bf 10}^{\ -1}\\
\label{d6}\overline{\bf 4}^{\ -\frac{1}{2}}\times\overline{\bf 20}^{\ -\frac{1}{2}}& \supset&{\bf 10}^{-1}\\
\label{d7}{\bf 4}^{\frac{1}{2}}\times {\bf 4}^{-\frac{3}{2}}& \supset&{\bf 10}^{-1}\\
\label{d8}{\bf 4}^{\frac{1}{2}}\times {\bf 4}^{\frac{1}{2}}& \supset&{\bf 10}^{+1}\\
\label{d9}{\bf 4}^{\frac{1}{2}}\times {\bf 20}^{\frac{1}{2}}& \supset&\overline{\bf 10}^{\ +1}\\
\label{d10}{\bf 4}^{\frac{1}{2}}\times {\bf 4}^{\frac{1}{2}}&
\supset&{\bf 10}^{\ +1}\\
\label{d11}{\bf 4}^{-\frac{1}{2}}\times {\bf 4}^{-\frac{1}{2}}&
\supset&{\bf 10}^{-1}
 \ea The decomposition of the
$\overline{\bf 20}^{\ -\frac{1}{2}}\times\overline{\bf 20}^{\
-\frac{1}{2}}$ implies that in principle we have both $S_{AB}$
and $N^{AB}$ appearing in the
$\l_{IA}^{(\overline{20})}\l_{JB}^{(\overline{20})}$ mass term.
However an explicit calculation shows that the
representation ${\bf 10}^{-1}$, corresponding to $N_{AB}$ is missing.\\
The above assignments are consistent with the Scherk--Schwarz
truncation of $N=8$ supergravity \cite{Sezgin:ac}, where the two
matrices $Q_{5ab}$, $P_{5abcd}$ contain the ${\bf 10},
\overline{\bf 10}$ of ${\rm SU}(4)$ of the $N=4$ theory. More
explicitly: \ban Q_{5ab}&\longrightarrow& (S_{AB}, S^{AB})\\
P_{5abcd}&\longrightarrow& (N_{AB}, N^{AB})\ean which is
consistent with the fact that, on the vacuum, $P_{5abcd}=0$ in
the Scherk--Schwarz $N=8$ model and $N_{AB}=0$ in our $N=4$
orientifold model.\\
Let us consider now the decomposition of the dilatino in the
${\bf 48}$ of ${\rm USp}(8)$ under ${\rm U}(4)$. We get: \be
\c_{abc}\longrightarrow
\c_{ABC}\oplus\c^{AB}_{\phantom{AB}C}+h.c.\ee corresponding to
\be{\bf 48}\longrightarrow\overline{\bf 4}+\overline{\bf 20}+{\bf
4}+{\bf 20} \ee We may then identify the chiral dilatino and
gaugino as
follows:\be\c^A=\epsilon^{ABCD}\c_{BCD};\quad\quad\l^{I(\overline{20})}_C=(\G^I)_{AB}\c^{AB}_{\phantom{AB}C}\ee
Moreover the decomposition ${\bf 8}\longrightarrow {\bf
4}^{+\frac{1}{2}}+\overline{\bf 4}^{\ -\frac{1}{2}}$ identifies
${\bf 4}^{+\frac{1}{2}}$ with $\l^{I(4)}_A$ and $\overline{\bf
4}^{\ -\frac{1}{2}}$ with $\l^{IA(\overline{4})}$ as they come
from the $\mathbf{C}$--trace part or the threefold antisymmetric
product ${\bf 8}\times{\bf 8}\times{\bf 8}$.\\
These results are consistent with the explicit reduction
appearing in reference \cite{Sezgin:ac}. Indeed the mass term of
reference \cite{Sezgin:ac} are of the following form \footnote{
Note that the terms \eq{2}, \eq{2x} do not appear explicitly in
the Lagrangian of reference \cite{Sezgin:ac} but they would appear
after diagonalization of the fermionic kinetic terms.} \ba
&&\label{1}Q_{5ab}\pb^{'a}_{\m}\g^{\m\r}\p^{'b}_{\r}\\
&&\label{2}Q_{5ab}\overline{\zeta}^{'a}\zeta^{'b}\\
&&\label{2x}Q_{5ab}\pb^{'a}_{\m}\g^{\m}\zeta^{'b}\\
&&\label{3}Q_{5a}^{\phantom{b}\ e}\cb^{'abc}\c{'}_{ebc}\\
&&\label{4}P_{5}^{abcd}\pb^{'}_{\m a}\g^{\m} \g^5 \c^{'}_{bcd}\\
&&\label{5}P_{5}^{abcd}\overline{\zeta}^{'}_a\g^5\c^{'}_{bcd}\ea
The term \eq{1} gives rise to the mass term of the gravitino
$S^{AB}\pb_{A\m}\g^{\m\n}\p_{B\n}$; the term
$N^{AB}\p_{A\n}\g^\m\c_B$ and the term
$Z^{I\phantom{A}B}_A\pb_{A\m}\g^\m\l^{IB}$ are obtained by
reduction of the structures \eq{2x}, \eq{4} via the decompositions
\eq{d7}, \eq{d8}. The mass term of the bulk gaugini
$T^{AB}_{IJ}\lb_A^I\l_B^J$ is reconstructed from the terms
\eq{2}, \eq{3}, \eq{5} through the decompositions \eq{d5},
\eq{d4}, \eq{d6}. Finally, the mass term
$Q^{I\phantom{A}B}_A\cb^A\l^I_B$ is obtained by reducing equation
\eq{3}, \eq{5} via the decomposition \eq{2}, \eq{3}.\\
In conclusion we see that our theory can be thought as a
truncation of the Scherk--Schwarz $N=8$ supergravity. Once the
Goldstino $\lb^{A(\overline{4})}$ is absorbed to give mass to the
gravitino $\p_{A\m}$ the spin $\frac{1}{2}$ mass matrix is given
by the entries
$\left(\c\c,\,\c\l^{(\overline{20})},\,\l^{(\overline{20})}\l^{(\overline{20})}\right)$.
Therefore the full spin $\frac{1}{2}$ mass spectrum is the
truncation of the Scherk--Schwarz $N=8$ spin $\frac{1}{2}$ to
this sectors.\\
This justifies the results for the mass spectrum given in Section
7. Analogous considerations can be done for the scalar sector.


We conclude by arguing that there is a duality between two
microscopically different theories:
\begin{eqnarray}
\left[\mbox{type IIB on an orientifold with
fluxes}\right]&\!\!\!\leftrightarrow\!\!\!
&\left[\matrix{\mbox{$N=4$ truncation of $N=8$ theories }\cr
\mbox{spontaneously broken \`a la
Scherk--Schwarz}}\right]\nonumber
\end{eqnarray}
since they are described by the same $N=4$ four dimensional
effective field theory.

Finally we consider the fermionic bilinear involving $D3$--brane
gaugini $\l^i_A$. From the structure of the matrices
$W_{iA}^{\phantom{a}B}$, $R_{iA}^{\phantom{a}B}$, $V_{AB}^{Ii}$,
equations \eq{shgaui}, \eq{masschilambda}, \eq{massgaux}, we
notice that they vanish when the $D3$--brane coordinates commute
(i.e. the scalars $q^I_i$ are in the Cartan subalgebra of $G$).\\
The diagonal mass $U_{AB}^{ij}$ \eq{massii} has a gravitational
part $\d^{ij}N_{AB}$ which vanishes on the vacuum while the
second term is non vanishing for those gaugini which are not in
the Cartan subalgebra of $G$. Indeed, for $G={\rm SU}(N)$, there
are exactly $N(N-1)$ ($\frac{1}{2}$ BPS) vector multiplets which
become massive when ${\rm SU}(N)$ is spontaneously broken to
${\rm U}(1)^{N-1}$.

\section*{Acknowledgements}

S. F. would like to thank the Dipartimento di Fisica, Politecnico
di Torino, R. D. and S. V. would like to thank the Th. Division of
Cern
for their kind hospitality during the  completion of this work.\\
M.T. would like to thank Henning Samtleben for useful
discussions.\\ Work supported in part by the European Community's
Human Potential Program under contract HPRN-CT-2000-00131 Quantum
Space-Time, in which  R. D'A. and S. V. are associated to Torino
University. The work of M. T. is supported by a European
Community Marie Curie Fellowship under contract
HPMF-CT-2001-01276. The work of S. F. has also been supported by
the D.O.E. grant DE-FG03-91ER40662, Task C. \vskip 5cm
\appendix
\section*{Appendix A: The solution of the Bianchi Identities and the supersymmetry transformation laws. }\label{appendiceA}
\setcounter {equation}{0} \addtocounter{section}{1}

\par
In this Appendix we describe the geometric approach for the
derivation of the $N=4$ supersymmetry transformation laws of the
physical fields.

The first step to perform is to extend the physical fields to
superfields in $N=4$ superspace: that means that the space--time
1-forms $\omega^{a\,b}$, $V^a$,$\psi^A$, $\psi_A$, $A_{\L\a}$,
$A_i $ and the space--time zero--forms $\c^A$, $\c_A$,
$\lambda_{A}^{I}$, $\lambda^{IA}$, $\lambda_{A}^{i}$,
$\lambda^{iA}$, $L^{\a}$, $E_{\L I}$, $B^{\L\S}$, $a^\L_i$ are
promoted to one--superforms and zero--superforms in $N=4$
superspace, respectively.\\
As a consequence  the superforms  must depend on the superspace
coordinates $\{x^\m;\theta^\a_A\}$ (where $x^\m$, $\m=1,2,3,4$
are the ordinary space--time coordinates and $\theta^\a_A$,
$\a=1,2,3,4$, $A=1,2,3,4$ are anticommuting fermionic coordinates
) in such a way that projected on the space--time submanifold
 (i.e. setting $\theta^\a_A=0, d\theta^\a_A=0$) they correspond to the ordinary physical fields.\\
 A basis of one--forms on the superspace is  given by $\{V^a,\p^\a_A\}$, $a=1,2,3,4$; here $V^a$ are the vierbein, and $ \p^\a_A$ are the fermionic vielbein identified with the
gravitini
 fields.

 The appropriate definition for the
super--curvatures (or super--field strengths)of the superfield
p--forms in the $N=4$ superspace is\footnote{Here and in the
following by "curvatures" we mean not only two--forms, but also
the one--forms defined as covariant differentials of the
zero--form superfields} as follows (we omit for simplicity the
sign of wedge product):

\begin{eqnarray}
\label{Lorentz}R^{ab}&=&d\o^{ab}-\o^a_c\o^{cb}\\
\label{torsion}T^a&=&\mathcal{D}V^a-i\pb_A\g^a\p^A=0\\
\label{feffe}F_{\L\a}&=&dA_{\L\a}-\frac{1}{2}L_{\a}E_{\L}^I(\G_I)^{AB}\pb_A\p_B-\frac{1}{2}\oL_{\a}E_{\L}^I(\G_I)_{AB}\pb^A\p^B\\
F_{i}&=&dA_{i}-\frac{1}{2}L_{2}q_{i}^I(\G_I)^{AB}\pb_A\p_B-\frac{1}{2}\oL_{2}q_{i}^I(\G_I)_{AB}\pb^A\p^B\\
\r_A&=&\mathcal{D}\p_A+\frac{1}{2}q\p_A-2Q_A^{\,\,B}\p_B\\
\nabla\c^A&=&\mathcal{D}\c^A-\frac{3}{2}q\c^A-2Q^A_{\,\,B}\c^B\\
\nabla\l_{IA}&=&\mathcal{D}\l_{IA}-\frac{1}{2}q\l_{IA}-2Q_A^{\,\,B}\l_{IB}+\o_2^{IJ}\l_{JA}\\
\nabla\l_{iA}&=&\mathcal{D}\l_{iA}-\frac{1}{2}q\l_{iA}-2Q_A^{\,\,B}\l_{iB}\\
p&=&-i\e_{\a\b}L^{\a}dL^{\b}\\
P^{IJ}&=&-\frac{1}{2}(EdE^{-1}+dE^{-1}\,E)^{IJ}+\frac{1}{2}\{E[\nabla B-\frac{1}{2}(\nabla a\,a^T-a\,\nabla a^T)]E\}^{IJ}\\
\label{delam}P^{Ii}&=&\frac{1}{2}E_{\L}^I\nabla a^{\L i}
\end{eqnarray}

$\nabla$ is the covariant derivative with respect to all the
connections that act on the field, including the gauge
contribution, while $\mathcal{D}$ is the Lorentz covariant
derivative acting on a generic vector $A^a$ and a generic spinor
$\theta$ respectively as follows
\begin{equation}\mathcal{D}A^a\equiv
dA^a-\o^{ab}A_b;\quad\quad\mathcal{D}\theta\equiv
d\theta-\frac{1}{4}\o^{ab}\g_{ab}\theta
\end{equation}
The coefficients appearing in front of the ${\rm U}(1)$
connection $q$
correspond to the different ${\rm U}(1)$ weights of the fields as shown in equation \eq{pesos}.\\
$Q^A_{\,\,B}$ is the R--symmetry ${\rm SU}(4)_1$ connection, that
in terms of the gauged ${\rm SO}(6)_1$ connection $\o_1^{IJ}$
reads as
$Q^A_{\,\,B}=\frac{1}{8}(\G_{IJ})_A^{\phantom{B}B}\o_1^{IJ}$
(see Appendix D for details).\\
Equation \eq{torsion} is a superspace constraint imposing the absence of supertorsion, on the $N=4$ superspace.\\
Note that the definition of the "curvatures" has been chosen in
such a way that in absence of vector multiplets the equations by
setting $R^{ab}=T^a=\rho^A=\rho _A=F^I=0,\,I=1,\dots,6$ give the
Maurer--Cartan equations of the $N=4$ Poincar\'e superalgebra
dual to the $N=4$ superalgebra of (anti)--commutators, (the
1-forms $\omega^{ab}, V^a, \psi ^A,\psi _A, A^I$ are dual to the
corresponding generators of the supergroup).\\
By $d$--differentiating the supercurvatures definition
\eq{Lorentz}--\eq{delam}, one obtains the Bianchi identities that
are their integrability conditions:
\begin{eqnarray}
&&\mathcal{R}^{ab}V_b+i\pb_A\g^a\r^A+i\pb^A\g^a\r_A=0\nn\\
&&\mathcal{D}R^{ab}=0\nn\\
&&\nabla F_{\L\a}-L_{\a}E_{\L}^I(\G_I)^{AB}\pb_A\r_B-\oL_{\a}E_{\L}^I(\G_I)_{AB}\pb^A\r^B+\frac{1}{2}\nabla L_{\a}E_{\L}^I(\G_I)^{AB}\pb_A\p_B+\nn\\&&+\frac{1}{2}\nabla\oL_{\a}E_{\L}^I(\G_I)_{AB}\pb^A\p^B\!+\!\frac{1}{2}L_{\a}\nabla E_{\L}^I(\G_I)^{AB}\pb_A\p_B\!+\!\frac{1}{2}\oL_{\a}\nabla E_{\L}^I(\G_I)_{AB}\pb^A\p^B=0\nn\\
&&\nabla F_{i}-L_{2}q_{i}^I(\G_I)^{AB}\pb_A\r_B-\oL_{2}q_{i}^I(\G_I)_{AB}\pb^A\r^B+\frac{1}{2}\nabla L_{2}q_{i}^I(\G_I)^{AB}\pb_A\p_B+\nn\\&&+\frac{1}{2}\nabla\oL_{2}q_{i}^I(\G_I)_{AB}\pb^A\p^B+\frac{1}{2}L_{2}\nabla q_{i}^I(\G_I)^{AB}\pb_A\p_B+\frac{1}{2}\oL_{2}\nabla q_{i}^I(\G_I)_{AB}\pb^A\p^B=0\nn\\
&&\nabla\r_A+\frac{1}{4}\mathcal{R}^{ab}\g_{ab}\p_A-\frac{1}{2}R\p_A+2R_A^{\,\,B}\p_B=0\nn\\
&&\nabla^2\c^A+\frac{1}{4}\mathcal{R}^{ab}\g_{ab}\c^A+\frac{3}{2}R\c^A+2R^A_{\,\,B}\c^B=0\nn\\
&&\nabla^2\l_{IA}+\frac{1}{4}\mathcal{R}^{ab}\g_{ab}\l_{IA}+\frac{1}{2}R\l_{IA}+2R_A^{\,\,B}\l_{IB}-R_{2IJ}\l_{A}^J=0\nn\\
&&\nabla^2\l_{iA}+\frac{1}{4}\mathcal{R}^{ab}\g_{ab}\l_{iA}+\frac{1}{2}R\l_{iA}+2R_A^{\,\,B}\l_{iB}=0\nn\\
&&\nabla p=0\nn\\
&&\nabla P^{IJ}+\frac{1}{2}E_\L^I(f^{\L\S\G\a}F_{\G\a}+c^{ijk}a^{\L}_ia^{\S}_k F_j)E_\S^J=0\nn\\
\label{bianchi}&&\nabla P^{Ii}+\frac{1}{2}a^\L_k
E_{\L}^Ic^{ijk}F_j=0
\end{eqnarray}

\nin Here
$R_A^{\,\,B}=\frac{1}{8}(\G_{IJ})_A^{\phantom{B}B}R_1^{IJ}$ is
the gauged ${\rm SU}(4)$ curvature with \be
R_1^{IJ}=d\o_1^{IJ}+\o^{IK}\wedge\o_K^{\phantom{K}J}-\frac{1}{2}E^I_\L
f^{\L\S\G\a}F_{\G\a}E^J_\S\ee and $R=dq$ is the ${\rm U}(1)$
curvature.\\ The solution can be obtained as follows: first of all
one requires that the expansion of the curvatures along the
intrinsic $p$--forms basis in superspace namely: $V^a, V^a\wedge
V^b, \psi , \psi \wedge V^b,\psi \wedge \psi$, is given in terms
only of the physical fields (rheonomy). This insures that no new
degree of freedom is introduced in the theory.
\\
Secondly one writes down such expansion in a form which is
compatible with all the symmetries of the theory, that is:
covariance under ${\rm U}(1)$ and ${\rm SO}_d(6) \otimes {\rm
SO}(n)$, Lorentz transformations and reparametrization of the
scalar manifold. Besides it is very useful to take into account
the invariance under the following rigid rescalings of the fields
(and their corresponding curvatures):
\begin{equation}
 (\omega^{ab}, A_{\Lambda \a}, E^I_\L, B^{\L\S},a^\L_i) \rightarrow
 (\omega^{ab}, A_{\Lambda \a}, E^I_\L, B^{\L\S},a^\L_i)
 \end{equation}
 \begin{equation}
 V^a \rightarrow \ell V^a
 \end{equation}
 \begin{equation}
 (\psi^A,\psi_A) \rightarrow \ell^{1 \over 2}(\psi^A,\psi_A)
 \end{equation}
 \begin{equation}
 (\lambda_{iA},\lambda^{I}_A, \c^A)
 \rightarrow  \ell^{-{1 \over 2}}(\lambda_{iA},\lambda^{I}_A, \c^A)
 \label{resc}
 \end{equation}
 Indeed the first three rescalings and the corresponding ones for the
 curvatures leave invariant the definitions of the curvatures and the
 Bianchi identities. The last one follows from the fact that in the solution for the $\s--$ model vielbeins $p$, $P^{IJ}$, $P^{Ii}$
 the spin $\frac{1}{2}$ fermions must appear contracted with the gravitino 1-form.\\
Performing all the steps one finds the final parametrizations of
the superspace curvatures, namely:

\begin{eqnarray}
\label{effe}F_{\L\a}&=&\mathcal{F}_{\L\a}^{ab}V_aV_b+i(\oL_{\a}E_{\L}^I(\G_I)^{AB}\cb_A\g_a\p_B+
L_{\a}E_{\L}^I(\G_I)_{AB}\cb^A\g_a\p^B+\nn\\&&+L_{\a}E_{\L}^I\lb_{I}^A\g_a\p_A+\oL_{\a}E_{\L}^I\lb_{IA}\g_a\p^A)V^a\\
F_{i}&=&\mathcal{F}_{i}^{ab}V_aV_b+i(\oL_{2}q_{i}^I(\G_I)^{AB}\cb_A\g_a\p_B+
L_{2}q_{i}^I(\G_I)_{AB}\cb^A\g_a\p^B+\nn\\&&+L_{2}q_{i}^I\lb_{I}^A\g_a\p_A+
\oL_{2}q_{i}^I\lb_{IA}\g_a\p^A+2L_{2}\lb_{i}^A\g_a\p_A+2\oL_{2}\lb_{iA}\g_a\p^A)V^a\\
\label{roferm}\r_A&=&\r_{Aab}V^aV^b-\oL^{\a}(E^{-1})^{\L}_I(\G^I)_{AB}\mathcal{F}_{\L\a}^{-ab}\g_b\p^BV_a+\nn\\
&&+\e_{ABCD}\c^B\pb^C\p^D+S_{AB}\g_a\p^BV^a\\
\nabla\c^A&=&\nabla\c^A_{a}V^a+\frac{i}{2}\overline{p}_a\g^a\p^A+
\frac{i}{4}\oL^{\a}(E^{-1})^{\L}_I(\G^I)^{AB}\mathcal{F}_{\L\a}^{-ab}\g_{ab}\p_B+N^{AB}\p_B\\
\nabla\l_{IA}&=&\nabla\l_{IAa}V^a+\frac{i}{2}(\G_J)_{AB}P^{JI}_a\g^a\p^B-
\frac{i}{2}L^{\a}(E^{-1})^{\L}_I\mathcal{F}_{\L\a}^{-ab}\g_{ab}\p_A+Z_{IA}^{\,\,\,B}\p_B\\
\nabla\l_{iA}&=&\nabla\l_{iAa}V^a+\frac{i}{2}(\G_I)_{AB}P^{iI}_a\g^a\p^B-
\frac{1}{4\oL_{2}}q_i^I(E^{-1})^{\L}_I\mathcal{F}_{\L2}^{-ab}\g_{ab}\p_A+\nn\\&&+
\frac{1}{4\oL_{2}}\mathcal{F}_{i}^{-ab}\g_{ab}\p_A+W_{iA}^{\,\,\,B}\p_B\\
p&=&p_aV^a+2\cb_A\p^A\\
P^{IJ}&=&P^{IJ}_aV^a+(\G^I)^{AB}\lb^J_A\p_B+(\G^I)_{AB}\lb^{JA}\p^B\\
P^{Ii}&=&P^{Ii}_aV^a+(\G^I)^{AB}\lb^i_A\p_B+(\G^I)_{AB}\lb^{iA}\p^B
\end{eqnarray}
The previous parametrizations are given up to three fermions
terms, except equation \eq{roferm} where the term with two
gravitini has been computed; in fact this term is in principle
involved in the computation of the gravitino shift but by
explicit computation its
contribution vanishes.\\
As promised the solution for the curvatures is given as an
expansion along the 2--form basis $ \left (V \wedge V\,,\,V
\wedge \psi\,,\,\psi\wedge \psi \right)$ or the 1--form basis
$\left(V\,,\,\psi\right)$ with coefficients given in terms of the
physical fields.
\par
It is important to stress that the components of the field
strengths along the bosonic vielbeins are not the space--time
field strengths since  $V^a=V^a_{\mu}dx^{\mu}\,+\, V^a_{\alpha}d
\theta^{\alpha}$ where $(V^a_{\mu}\,,\,V^a_{\alpha})$ is a
submatrix of the super--vielbein matrix $E^I \equiv\left(
V^a\,,\,\psi\right)$. The physical field strengths are given by
the expansion of the forms along the $ dx^{\mu}$-differentials
and by restricting the superfields to space--time ($\theta = 0$
component). For example, from the parametrization \eq{effe},
expanding along the $dx^{\mu}$--basis one finds:
\begin{eqnarray}
F^\Lambda_{\mu\nu} &=& F^\Lambda_{ab} V^a_{[\mu} V^b_{\nu]}+{\rm
i}\oL_{\a}E_{\L}^I(\G_I)^{AB}\cb_A\g_{[\m}\p_{\n]B}+ {\rm
i}L_{\a}E_{\L}^I(\G_I)_{AB}\cb^A\g_{[\m}\p^B_{\n]}\nn \\
&+&\!\!\!iL_{\a}E_{\L}^I\lb_{I}^A\g_{[\m}\p_{\n] A}+{\rm
i}\oL_{\a}E_{\L}^I\lb_{IA}\g_{[\m}\p^A_{\n]}
\end{eqnarray}
where $F^\Lambda_{\mu\nu}$ is defined by the expansion of eq.
\eq{feffe} along the $dx^\m$--differentials.
 When all
the superfields are restricted to space--time we may treat the
$V^a_{\mu}$ vielbein as the usual 4--dimensional invertible matrix
converting intrinsic indices in coordinate indices and we see
that the physical field--strength $\mathcal {F}_{\L\a\mu\nu}$
differs from $F_{\Lambda \a\, ab} V^a_{[\mu} V^b_{\nu]}\equiv
\tilde { F}_{\Lambda \a \mu\nu} $ by a set of spinor currents
($\tilde { F}_{\Lambda \a \mu\nu}$ is referred to as the
supercovariant field--strength).

Analougous considerations hold for the other field--strengths
components along the bosonic vielbeins.

Note that the solution of the Bianchi identities also implies a
set of differential constraints on the components along the
bosonic vielbeins which are to be identified, when the fields are
restricted to space--time only, with the equations of motion of
the theory. Indeed the analysis of the Bianchi identities for the
fermion fields give such equations (in the sector containing the
2--form basis $\bar\psi_A\gamma^a \psi^A$). Further the superspace
derivative along the $\psi_A\,\left(\psi^A\right)$ directions,
which amounts to a supersymmetry transformation, yields the
equations of motion of the bosonic fields. Indeed the closure of
the Bianchi identities is  equivalent to the closure of the $N=4$
supersymmetry algebra on the physical fields and we know that in
general such closure implies the equations of motion .

The determination of the superspace curvatures enables us to
write down the $N=4$ SUSY transformation laws. Indeed we recall
that from the superspace point of view a supersymmetry
transformation is a Lie derivative along the tangent vector:
\begin{equation}
\epsilon = \bar\epsilon^A\,\vec D_A\,+\,\bar\epsilon_A\,\vec D^A
\end{equation}
where the basis tangent vectors $\vec D_A\,,\,\vec D^A$ are dual
to the gravitino 1--forms:
\begin{equation}
\vec D_A \left(\psi^B \right) = \vec D^A \left(\psi_B \right)
=\bf 1
\end{equation}
and $\bf 1$ is the unit in spinor space.
\par
Denoting by $\mu^I$ and $R^I$ the set of one--forms $\Bigl (
V^a,\,\psi_A,\,\psi^A,\,A_{\Lambda \a},A_i \Bigr )$ and of
two--forms $\Bigl ( T^a =0,\,\rho_A,\,\rho^A,\,F_{\Lambda \a},F_i
\Bigr )$ respectively, one has:
\begin{equation}
\ell \mu^I = \left(i_{\epsilon}d\,+\,di_{\epsilon}\right)\mu^I
\equiv \left(D \epsilon\right)^I\,+\,i_{\epsilon}R^I
\end{equation}
where $D$ is the derivative covariant with respect to the $N=4$
Poincar\'e superalgebra and $i_{\epsilon}$ is the contraction
operator along the tangent vector $\epsilon$.
\par
In our case:
\begin{eqnarray}
\left(D \epsilon\right)^a &=& {-\rm
i}\left(\bar\psi_A\gamma^a\epsilon^A\,+\,\bar\psi^A\gamma^a\epsilon_A \right)\\
\left(D \epsilon\right)^{\alpha} &=& \nabla\epsilon^{\alpha}\\
\left(D \epsilon\right)_{\Lambda \a}= \left(D
\epsilon\right)_{i}&=& 0
\end{eqnarray}
(here $\alpha$ is a spinor index)\\
For the 0--forms which we denote shortly as $\nu^I$ we have the
simpler result:
\begin{equation}
\ell_{\epsilon} = i_{\epsilon}d \nu^I =
i_{\epsilon}\left(\nabla\nu^I\,-\,connection\,terms \right)
\end{equation}
Using the parametrizations given for $R^I$ and $\nabla\nu^I$ and
identifying $\delta_{\epsilon}$ with the restriction of
$\ell_{\epsilon}$ to space--time it is immediate to find the $N=4$
susy laws for all the fields. The explicit formulae are given by
the equations \eq{susytra}.
\section*{Appendix B: Derivation of the space time Lagrangian from
the geometric approach} \label{appendiceB}
\setcounter{equation}{0} \addtocounter{section}{1} In Appendix A
we have seen how to reconstruct the $N=4$ susy transformation
laws of the physical fields from the solution of the
Bianchi identities in superspace.\\
In principle, since the Bianchi identities imply the equations of
motion, the Lagrangian could also be completely determined.
However this would be a cumbersome procedure.
\par
In this Appendix we give a short account of the construction of
the Lagrangian on space--time from a geometrical Lagrangian in
superspace. Note that while the solution of the Bianchi
identities is completely equivalent to the ordinary "Superspace
approach" (apart from notations and a different emphasis on the
group--theoretical structure),the geometric approach for the
construction of the Lagrangian is completely
different from the usual superspace approach via integration in superspace\\
In the geometric (rheonomic) approach the superspace action is a
4--form in superspace integrated on a 4--dimensional (bosonic)
hypersurface ${\cal M}^{4}$ locally embedded in superspace
\begin{equation}
{\cal A} = \int_{{\cal M}^{4} \subset {\cal SM}} \, {\cal L}
\label{campodicalcio}
\end{equation}
where ${\cal SM}$ is the superspace manifold. Provided we do not
introduce the Hodge duality operator in the construction of $\cal
L$ the equations of motions derived from the generalized
variational principle $\delta \cal A = {\rm{0}}$ are 3--form or
4--form equations independent from the particular hypersurface
$\cal M^{\rm 4}$ on which we integrate and they are therefore
valid in all superspace. (Indeed in the variational principle we
have also to vary the hypersurface which can always compensated
by a diffeomorphism of the fields if the Lagrangian
is written olnly in terms of differential forms).\\
These superspace equations of motion can be analyzed along the
$3$--form basis. The components of the equations obtained along
bosonic vielbeins give the differential equations for the fields
which, identifying ${\cal M}^{  4}$ with space--time, are the
ordinary equations of motion of the theory. The components of the
same equations along $3$--forms containing at least one gravitino
("outer components") give instead algebraic relations which
identify the components of the various "supercurvatures" in the
outer directions in terms of the physical fields along the
bosonic vierbeins (rhenomy principle).
\par
 Actually if we have already solved
the Bianchi identities this requirement is equivalent to identify
the outer components of the curvatures obtained from the
variational principle with those obtained from the Bianchi
identities.
\par
There are simple rules which can be used in order to write down
the most general Lagrangian compatible with this requirement.\\
The implementation of these rules is described in detail in the
literature \cite{CDF} to which we refer the interested reader.
Actually one writes down the most general 4--form as a sum of
terms with indeterminate coefficients in such a way that $\cal L$
be a scalar with respect to all the symmetry transformations of
the theory (Lorentz invariance,{\rm U}(1), ${\rm SO}(6)_d \otimes
{\rm SO}(n)$ invariance, invariance under the rescaling \eq{resc}.
Varying the action and comparing the outer equations of motion
with the actual solution of the Bianchi identities one then fixes
all the undetermined coefficients.
\par
Let us perform the steps previously indicated. The most general
Lagrangian has the following form: (we will determine the complete
 Lagrangian up to four fermion terms):

\begin{eqnarray}
\mathcal{L}_{kinetic}\!&\!=\!&\!\mathcal{R}^{ab}V^cV^d\e_{abcd}+a_1(\pb_A\g_a\r^AV^a-\pb^A\g_a\r_AV^a)+[a_2(\cb^A\g^a\nabla\c_A
+\cb_A\g^a\nabla\c^A)+\nn\\&&+a_3(\lb^{IA}\g^a\nabla\l_{IA}+\lb_{IA}\g^a\nabla\l^{IA})+a_6(\lb^{iA}\g^a\nabla\l_{iA}+
\lb_{iA}\g^a\nabla\l^{iA})V^bV^cV^d\e_{abcd}+\nn\\&&+a_4[{\bf
p}^a(\overline{p}-2\cb^A\p_A)+
{\bf \overline{p}}^a(p-2\cb_A\p^A)-\frac{1}{4}{\bf \overline{p}}_f{\bf p}^fV^a]V^bV^cV^d\e_{abcd}+\nn\\
&&+a_5[{\bf
P}^{IJ}_a(P^{IJ}\!-\!(\G^I)^{AB}\lb^J_A\p_B\!-\!(\G^I)_{AB}\lb^{JA}\p^B)
-\frac{1}{8}{\bf P}^{IJ}_f{\bf P}^{IJ\,f}V^a]V^bV^cV^d\e_{abcd}+\nn\\
&&+a_7[{\bf
P}^{Ii}_a(P^{Ii}\!-\!(\G^I)^{AB}\lb^i_A\p_B\!-\!(\G^I)_{AB}\lb^{iA}\p^B)-
\frac{1}{8}{\bf P}^{Ii}_f{\bf P}^{Ii\,f}V^a]V^bV^cV^d\e_{abcd}+\nn\\
&&+a[\mathcal{N}^{\L\a\S\b}{\bf
F}^{+ab}_{\L\a}+\overline{\mathcal{N}}^{\L\a\S\b}{\bf
F}^{-ab}_{\L\a}+
\mathcal{N}^{i\S\b}{\bf F}^{+ab}_{i}+\overline{\mathcal{N}}^{i\S\b}{\bf F}^{-ab}_{i}]\cdot\nn\\
&&[F_{\S\b}-i(\oL_{\b}E_{\S}^I(\G_I)^{AB}\cb_A\g^d\p_B+L_{\b}E_{\S}^I(\G_I)_{AB}\cb^A\g^d\p^B+\nn\\&&+L_{\b}E_{\S}^I\lb_{I}^A\g^d\p_A+\oL_{\b}E_{\S}^I\lb_{IA}\g^d\p^A)V_d]V_aV_b+\nn\\
&&+a[\mathcal{N}^{\L\a j}{\bf F}^{+ab}_{\L\a}+\overline{\mathcal{N}}^{\L\a j}{\bf F}^{-ab}_{\L\a}+\mathcal{N}^{ij}{\bf F}^{+ab}_{i}+\overline{\mathcal{N}}^{ij}{\bf F}^{-ab}_{i}]\cdot\nn\\
&&[F_j-i(\oL_{2}q_{j}^I(\G_I)^{AB}\cb_A\g^d\p_B+L_{2}q_{j}^I(\G_I)_{AB}\cb^A\g^d\p^B+\nn\\&&+L_{2}q_{j}^I\lb_{I}^A\g^d\p_A+\oL_{2}q_{j}^I\lb_{IA}\g^d\p^A+L_{2}\lb_{j}^A\g^d\p_A+\oL_{2}\lb_{jA}\g^d\p^A)V_d]V_aV_b+\nn\\
&&-\frac{i}{24}a(\mathcal{N}^{\L\a\S\b}{\bf F}^{+fg}_{\L\a}{\bf F}^+_{fg\S\b}-\overline{\mathcal{N}}^{\L\a\S\b}{\bf F}^{-fg}_{\L\a}{\bf F}^-_{fg\S\b})V^aV^bV^cV^d\e_{abcd}+\nn\\
&&-\frac{i}{24}a(\mathcal{N}^{i\S\b}{\bf F}^{+lm}_{i}{\bf F}^+_{fg\S\b}-\overline{\mathcal{N}}^{i\S\b}{\bf F}^{-fg}_{i}{\bf F}^-_{fg\S\b})V^aV^bV^cV^d\e_{abcd}+\nn\\
&&-\frac{i}{24}a(\mathcal{N}^{ij}{\bf F}^{+fg}_{i}{\bf F}^+_{fg\,i}-\overline{\mathcal{N}}^{ij}{\bf F}^{-fg}_{i}{\bf F}^-_{fg\,j})V^aV^bV^cV^d\e_{abcd}+\nn\\
\end{eqnarray}

\begin{eqnarray}
\mathcal{L}_{Pauli}\!&\!=\!&\!b_1[p\cb^A\g_{ab}\p_A-\overline{p}\cb_A\g_{ab}\p^A]V^aV^b+\nn\\
&&+b_2P^{IJ}[(\G^I)^{AB}\lb^J_A\g_{ab}\p_B\!-\!(\G^I)_{AB}\lb^{JA}\g_{ab}\p^B)V^aV^b+\nn\\
&&+b_3P^{Ii}[(\G^I)^{AB}\lb^i_A\g_{ab}\p_B\!-\!(\G^I)_{AB}\lb^{iA}\g_{ab}\p^B)V^aV^b+\nn\\
&&+F_{\L\a}[c_1(\overline{\mathcal{N}}^{\L\a\S\b}L_{\b}E_{\S}^I(\G_I)^{AB}\pb_A\p_B+{\mathcal{N}}^{\L\a\S\b}\oL_{\b}E_{\S}^I(\G_I)_{AB}\pb^A\p^B)+\nn\\
&&+c_2(\mathcal{N}^{\L\a\S\b}\oL_{\b}E_{\S}^I(\G_I)^{AB}\cb_A\g_a\p_B+\overline{\mathcal{N}}^{\L\a\S\b}L_{\b}E_{\S}^I(\G_I)_{AB}\cb^A\g_a\p^B)V^a+\nn\\
&&+c_3(\mathcal{N}^{\L\a\S\b}L_{\b}E_{\S}^I\lb_I^A\g_a\p_A+\overline{\mathcal{N}}^{\L\a\S\b}\oL_{\b}\lb_{IA}\g_a\p^A)V^a]+\nn\\
&&+F_{\L\a}[c_4(\overline{\mathcal{N}}^{\L\a\,i}L_{2}q_{i}^I(\G_I)^{AB}\pb_A\p_B+\mathcal{N}^{\L\a\,i}\oL_{2}q_{i}^I(\G_I)_{AB}\pb^A\p^B)+\nn\\
&&+c_5(\mathcal{N}^{\L\a\,i}\oL_{2}q_{i}^I(\G_I)^{AB}\cb_A\g_a\p_B+\overline{\mathcal{N}}^{\L\a\,i}L_{2}q_{i}^I(\G_I)_{AB}\cb^A\g_a\p^B)V^a+\nn\\
&&+c_6(\mathcal{N}^{\L\a\,i}L_{2}q_{i}^I\lb_{I}^A\g_a\p_A+\overline{\mathcal{N}}^{\L\a\,i}\oL_{2}q_{i}^I\lb_{IA}\g_a\p^A)V^a+\nn\\
&&+c_7(\mathcal{N}^{\L\a\,i}L_{2}\lb_{i}^A\g_a\p_A+\overline{\mathcal{N}}^{\L\a\,i}\oL_{2}\lb_{iA}\g_a\p^A)V^a]+\nn\\
&&+F_i[c_1(\overline{\mathcal{N}}^{i\S\b}L_{\b}E_{\S}^I(\G_I)^{AB}\pb_A\p_B+{\mathcal{N}}^{i\S\b}\oL_{\b}E_{\S}^I(\G_I)_{AB}\pb^A\p^B)+\nn\\
&&+c_2(\mathcal{N}^{i\S\b}\oL_{\b}E_{\S}^I(\G_I)^{AB}\cb_A\g_a\p_B+\overline{\mathcal{N}}^{i\S\b}L_{\b}E_{\S}^I(\G_I)_{AB}\cb^A\g_a\p^B)V^a+\nn\\
&&+c_3(\mathcal{N}^{i\S\b}L_{\b}E_{\S}^I\lb_I^A\g_a\p_A+\overline{\mathcal{N}}^{i\S\b}\oL_{\b}\lb_{IA}\g_a\p^A)V^a]+\nn\\
&&+F_i[c_4(\overline{\mathcal{N}}^{ij}L_{2}q_{j}^I(\G_I)^{AB}\pb_A\p_B+\mathcal{N}^{ij}\oL_{2}q_{j}^I(\G_I)_{AB}\pb^A\p^B)+\nn\\
&&+c_5(\mathcal{N}^{ij}\oL_{2}q_{j}^I(\G_I)^{AB}\cb_A\g_a\p_B+\overline{\mathcal{N}}^{ij}L_{2}q_{j}^I(\G_I)_{AB}\cb^A\g_a\p^B)V^a+\nn\\
&&+c_6(\mathcal{N}^{ij}L_{2}q_{j}^I\lb_{I}^A\g_a\p_A+\overline{\mathcal{N}}^{ij}\oL_{2}q_{j}^I\lb_{IA}\g_a\p^A)V^a+\nn\\
\label{paolino}&&+c_7(\mathcal{N}^{ij}L_2\lb_j^A\g_a\p_A+\overline{\mathcal{N}}^{ij}\oL_2\lb_{jA}\g_a\p^A)V^a]+\,more\,terms
\end{eqnarray}

\begin{eqnarray}
\mathcal{L}_{gauge}\!&\!=\!&\!g_1(\pb_A\g_{ab}\p_BS^{AB}-\pb^A\g_{ab}\p^BS_{AB})V^aV^b+\nn\\
&&g_2(\pb_A\g^{a}\c_BN^{AB}+\pb^A\g^{a}\c^BN_{AB})V^bV^cV^d\e_{abcd}+\nn\\
&&+g_3(\pb_A\g^{a}\l^{IB}Z^{\,\,\,\,A}_{IB}+\pb^A\g^{a}\l^I_BZ_{I\,\,\,A}^B)V^bV^cV^d\e_{abcd}+\nn\\
&&+g_4(\pb_A\g^{a}\l^{iB}W^{\,\,\,\,A}_{iB}+\pb^A\g^{a}\l^i_BW_{i\,\,\,A}^B)V^bV^cV^d\e_{abcd}+\nn\\
&&+(\lb_I^A\c_BQ^{I\,\,\,B}_A+\lb_{IA}\c^BQ^{IA}_{\,\,\,\,\,B}+\lb_i^A\c_BM^{i\,\,\,B}_A+\lb_{iA}\c^BM^{iA}_{\,\,\,\,\,B})V^aV^bV^cV^d\e_{abcd}\!+\nn\\
&&+(\lb_{IA}\l_{JB}T^{IJAB}\!\!\!+\!\lb^{IA}\!\!\l^{JB}T_{IJAB}+\!\lb_{iA}\l_{jB}U^{ijAB}\!\!\!+\!\lb^{iA}\!\!\l^{jB}U_{ijAB})V^aV^bV^cV^d\e_{abcd}\!+\nn\\
&&-\frac{1}{24}(-\!12S_{AC}\overline{S}^{CA}\!\!+4\,N_{AC}\overline{N}^{CA}\!\!+2Z^{IC}_{\phantom{B}\,A}Z^{I\
A}_C\!\!+4W^{iC}_{\phantom{A}\,A}W^{i\
A}_C)V^aV^bV^cV^d\!\e_{abcd}
\end{eqnarray}

\begin{eqnarray}
\mathcal{L}_{torsion}\!&\!=\!&\!T_aV^aV^b(t_1\cb^A\g_b\c_a+t_2\lb_{IA}\g_b\l^{IA}+t_3\lb_{iA}\g_b\l^{iA})\\
&&\nn\quad\quad\quad\quad\quad\quad\quad\quad\quad\quad\quad\quad\quad\quad\quad\quad\quad\quad\quad\quad\quad\quad\quad\quad\quad\quad\quad
\quad\quad\quad\quad\quad
\end{eqnarray}

Note that in equation \eq{paolino} the statement "+ more terms"
means Pauli terms containing currents made out spin $\frac{1}{2}$
bilinears which can not be computed in this geometric approach
without knowledge of the four fermion couplings. However these
terms have been included in the space--time Lagrangian given in
Section 5 by imposing the invariance of the space--time
Lagrangian under supersymmetry transformations.

The introduction of the auxiliary 0--forms ${\bf p}^a,{\bf
P}^{IJ}_a,{\bf F}^{\pm ab}_{\L\a}, {\bf F}^{\pm ab}_{i}$ is a
trick which avoids the use of the Hodge operator for the
construction of the kinetic terms for the vectors and scalar
fields which otherwise would spoil the validity of the 3--form
equations of motion in all superspace; indeed the equation of of
motion of these auxiliary 0--forms identifies them with the
components of the physical field--strengths $ p^a, P^{IJ}_a,
F^{\pm ab}_{\L\a}, F^{\pm ab}_{i}$ along the bosonic vielbeins
$V^a$ thus reconstructing the usual kinetic terms on
space--time.\\ The $\mathcal{L}_{torsion}$--term has been
constructed in such a way as to give $T^a=0$.

Performing the variation of all the fields one fixes all the
undetermined coefficients, namely:
\begin{eqnarray}
&&a_1=4;\quad a_2=-\frac{4}{3}i;\quad a_3=-\frac{2}{3}i;\quad a_4=\frac{2}{3};\quad a_5=\frac{2}{3};\quad a_6=-\frac{4}{3}i;\nn\\
&&a_7=\frac{4}{3};\quad c_1=-2;\quad c_2=-4i;\quad c_3=-4i;\quad c_4=-2;\quad c_5=-4i;\nn\\
&& c_6=-4i;\quad c_7=-8i;\quad b_1=4i;\quad b_2=2i;\quad b_3=4i; \nn\\
&& g_1=-4;\quad g_2=-\frac{8}{3}i;\quad g_3=-\frac{4}{3}i;\quad g_4=-\frac{8}{3}i;\nn\\
&&t_1=-4;\quad t_2=-2;\quad t_3=-4;\quad a=-4
\end{eqnarray}
In order to obtain the space--time Lagrangian the last step to
perform is the restriction of the 4--form Lagrangian from
superspace to space--time. Namely we restrict all the terms to the
$\theta = 0\,,\,d \theta = 0$ hypersurface ${\cal M}^4$. In
practice one first goes to the second order formalism by
identifying the auxiliary 0--form fields as explained before.
Then one expands all the forms along the $dx^{\mu}$ differentials
and restricts the superfields to their lowest ($\theta = 0$)
component. Finally the coefficients of:
\begin{equation}
dx^{\mu}\wedge dx^{\nu}\wedge dx^{\rho}\wedge
dx^{\sigma}\,=\,{\epsilon^{\mu\nu\rho\sigma}\over \sqrt g}\left(
\sqrt g d^4x \right)
\end{equation}
give the Lagrangian density written in Section 5. The overall
normalization of the space--time action has been chosen such as to
be the standard one for the Einstein term. (To conform to the
usual definition of the Riemann tensor $R^{ab}_{\phantom{ab}cd}$
we have set $R^{ab}=-\frac{1}{2}R^{ab}_{\phantom{ab}cd}V^cV^d$).

\appendix
\section*{Appendix C: The moduli of $T^6$ in real and complex coordinates}
\setcounter {equation}{0} \addtocounter{section}{3} In this
appendix we give a more detailed discussion of the extrema of the
potential using a complex basis for the ${\rm GL}(6,\mathbb{R})$
indices for the moduli of the $T^6$ torus.\\
Let us consider the basis vectors $\{e_\L\}$, $(\L=1\dots 6)$ of
the fundamental representation of ${\rm GL}(6,\mathbb{R})$. We
introduce a complex basis $\{E_i,\overline{E}_i\}$ with
$i=1,\,2,\,3$ or, to avoid confusion on indices, $i=x,\,y,\,z$ in
the following way:
\begin{eqnarray}\label{complex} &&e_1+ie_4
=E_x;\,\,\,e_2+ie_5=E_y;\,\,\,e_3+ie_6=E_z
\\&&e_1-ie_4
=\overline{E}_x;\,\,\,e_2-ie_5=\overline{E_y};\,\,\,e_3-ie_6=\overline{E_z}
\end{eqnarray}
The axion fields and the (inverse) metric of $T^6$ can then be
written using (anti)holomorphic indices
$i,\,j,\,\overline{\i},\,\overline{\jmath}$ as follows: \ba
&&B^{\L\S}\longrightarrow
B^{ij},\,B^{i\overline{\jmath}},\,B^{\overline{\i}j},\,B^{\overline{\i}\,\overline{\jmath}}\\
&&g^{\L\S}\longrightarrow
g^{ij},\,g^{i\overline{\jmath}},\,g^{\overline{\i}j},\,g^{\overline{\i}\,\overline{\jmath}}\\
\ea In particular, the fluxes $f^{\L\S\G}_1\equiv f^{\L\S\G}$ are
given by
\begin{eqnarray}\label{Holof}
f^{xyz}&=&\frac{1}{8}\{f^{123}-f^{156}+f^{246}-f^{345}+ i\left(^*f^{123}-^*f^{156}+^*f^{246}-^*f^{345}\right)\}\\
f^{x\overline{y}\overline{z}}&=&\frac{1}{8}\{f^{123}-f^{156}-f^{246}+f^{345}
+i\left(^*f^{123}-^*f^{156}-^*f^{246}+^*f^{345}\right)\}\\
f^{xy\overline{z}}&=&\frac{1}{8}\{f^{123}+f^{156}-f^{246}-f^{345}+i\left(^*f^{123}+^*f^{156}-^*f^{246}-^*f^{345
}\right)\}\\
f^{x\overline{y}k}&=&\frac
{1}{8}\{f^{123}+f^{156}+f^{246}+f^{345}+
i\left(^*f^{123}+^*f^{156}+^*f^{246}+^*f^{345}\right)\}
\end{eqnarray}
while \be\label{tuttizero}
f^{x\overline{x}y}=f^{x\overline{x}z}=f^{y\overline{y}x}=f^{y\overline{y}z}=f^{z\overline{z}x}=f^{z\overline{z}y}=0\ee
and therefore, the twenty entries of $f_1^{\L\S\D}$ are reduced to
eight.\\
In this holomorphic basis the gravitino mass eigenvalues assume
the rather simple form:
\begin{eqnarray}\label{zeromass}
&&m_1\equiv|\m_1+i\m_1^{'}|=\frac{1}{6|L^2|}|f^{x\overline{y}\overline{z}}|\\
&&m_2\equiv|\m_2+i\m_2^{'}|=\frac{1}{6|L^2|}|f^{x\overline{y}z}|\\
&&m_3\equiv|\m_3+i\m_3^{'}|=\frac{1}{6|L^2|}|f^{xy\overline{z}}|\\
&&m_4\equiv|\m_4+i\m_4^{'}|=\frac{1}{6|L^2|}|f^{xyz}|
\end{eqnarray}
We note that the three axions
$B^{\L\S}=\{B_{14},\,B_{25},\,B_{36}\}\equiv\{-2iB_{x\overline{x}},\,-2iB_{y\overline{y}},\,-2iB_{z\overline{z}}\}$
are inert under $T_{12}$--gauge transformations, since we have 15
axions but only 12 bulk vectors.\\
When we consider the truncation to the $N=3$ theory we expect
that only 9 complex scalar fields become massless moduli
parametrizing ${\rm SU}(3,3)/{\rm SU}(3)\times {\rm SU}(3)\times
{\rm U}(1)$. Moreover, it is easy to see that if we set e.g.
$\m_1=\m_2=\m_3=0$ ( $\m_1^{'}=\m_2^{'}=\m_3^{'}=0$) which implies
$f^{345}=f^{156}=-f^{123}=- f^{246}$
($^*f^{345}=^*f^{156}=-^*f^{123}=- ^*f^{246}$) in the $N=3$
theory, we get that also the 6 fields
$B^{12}-B^{45},\,B^{13}-B^{46},\,B^{24}-B^{15},\,B^{34}-B^{16},\,B^{23}-B^{56},\,B^{35}-B^{26}$
are inert under gauge transformations.\\
In holomorphic coordinates, the translational gauging implies that
the differential of the axionic fields become covariant and they
are given by: \ba&&\nabla_{(g)}B^{ij}\equiv dB^{ij}+(\Re
f^{ijk1})A_{k 1}+(\Re f^{ij\overline{k}1})A_{\overline{k}1}+(\Im
f^{ijk2})A_{k 2}+
(\Im f^{ij\overline{k}2})A_{\overline{k}2}\\
&&\label{bij}\nabla_{(g)}B^{i\overline{\jmath}}\equiv
dB^{i\overline{\jmath}}+(\Re f^{i\overline{\jmath} k
1})A_{k1}+(\Re
f^{i\overline{\jmath}\overline{k}1})A_{\overline{k}1}+ (\Im
f^{i\overline{\jmath} k 2})A_{k2}+(\Im
f^{i\overline{\jmath}\overline{k}2})A_{\overline{k}2}\ea

Since in the $N=4\longrightarrow N=3$ truncation the only
surviving massless moduli fields are
$B_{i\overline{\jmath}}+ig_{i\overline{\jmath}}$, then the 3+3
axions $\{B_{ij},\,B_{\overline{\i}\,\overline{\jmath}}\}$ give
mass to 6 vectors, while $\d B_{i\overline{\jmath}}$ must be
zero. We see from equation \eq{bij} we see that we must put to
zero the components \be \label{n=3}
f^{i\overline{\jmath}k}=f^{i\overline{\jmath}\overline{k}}=f^{ij\overline{k}}=0\ee
while \be f^{ijk}\equiv f\e^{ijk}\neq 0\ee Looking at the
equations (\ref{zeromass}) we see that these relations are
exactly the same which set
$\m_1+i\m_1^{'}=\m_2+i\m_2^{'}=\m_3+i\m_3^{'}=0$ and
$\m_4+i\m_4^{'}\neq 0$, confirming that the chosen complex
structure corresponds to the $N=3$ theory. Note that the
corresponding $g^{i\overline{\jmath}}$ fields partners of
$B^{i\overline{\jmath}}$ in the chosen complex structure
parametrize the coset ${\rm O}(1,1)\times {\rm
SL}(3,\mathbb{C})/{\rm SU}(3)$. Actually the freezing of the
holomorphic $g^{ij}$ gives the following relations among the
components in the real basis of
$g^{\L\S}$: \ba&& g^{14}=g^{25}=g^{36}=0\\
&&\label{ziapina}g^{11}-g^{44}=0,\ \ g^{22}-g^{55}=0,\ \ g^{33}-g^{66}=0\\
&&g^{12}-g^{45}=0,\ \ g^{13}-g^{46}=0,\ \ g^{23}-g^{56}=0\\
&&g^{15}+g^{24}=0,\ \ g^{16}+g^{34}=0,\ \ g^{26}+g^{35}=0\ea The
 freezing of the axions  $B^{ij}$ in the holomorphic basis give the analogous equations:
\begin{eqnarray}
&&B^{12}-B^{45}=0,\ \ B^{13}-B^{46}=0,\ \ B^{23}-B^{56}=0\\
&&B^{15}+B^{42}=0,\ \ B^{16}+B^{43}=0,\ \ B^{26}+B^{53}=0 \\
&& B^{14}=B^{25}=B^{36}=0
\end{eqnarray}
The massless $g^{i\overline{\jmath}}$ and $B^{i\overline{\jmath}}$
are instead given by the following combinations:
\be\label{ziagina} g^{x\overline{x}}=\frac{1}{2}(g^{11}+g^{44}),\
\ g^{y\overline{y}}=\frac{1}{2}(g^{22}+g^{55}),\ \
g^{z\overline{z}}=\frac{1}{2}(g^{33}+g^{66})\ee \be
B^{x\overline{x}}=\frac{i}{2}B^{14},\ \
B^{y\overline{y}}=\frac{i}{2}B^{25},\ \
B^{z\overline{z}}=\frac{i}{2}B^{36}\ee

\ba\label{ziopino}
&&g^{x\overline{y}}=\frac{1}{2}(g^{12}+ig^{15}),\ \
g^{x\overline{z}}=\frac{1}{2}(g^{13}+ig^{16}),\ \
g^{y\overline{z}}=\frac{1}{2}(g^{23}+ig^{26})\\
&&B^{x\overline{y}}=\frac{1}{2}(B^{12}+iB^{15}),\ \
B^{x\overline{z}}=\frac{1}{2}(B^{13}+iB^{16}),\ \
B^{y\overline{z}}=\frac{1}{2}(B^{23}+iB^{26})\\
&&B^{xx}=B^{yy}=B^{zz}=0 \ea
\\Let us now
consider the reduction $N=4\longrightarrow N=2$ for which the
relevant moduli space is ${\rm SU}(2,2)/\left({\rm SU}(2)\times
{\rm SU}(2)\times {\rm U}(1)\right)\otimes {\rm SU}(1,1)/{\rm
U}(1)$ . Setting $\m_2+i\m_2^{'}=\m_3+i\m_3^{'}=0$ we find: \be
f^{x\overline{y}z}=f^{xy\overline{z}}=0\ee which , in real
components implies:\be f^{123}+f^{156}=0;\quad\quad
f^{246}+f^{345}=0 \ee and analogous equations for their Hodge
dual. This implies that in the $N=2$ phase two more axions are
gauge inert namely $B^{23}+B^{56}=2B^{23}$ and
$B^{26}+B^{35}=2B^{26}$ or, in holomorphic components,
$B^{y\overline z}$. The remaining fields are $g^{14},g^{25},
g^{36}, g^{23}, g^{26}$ and $B^{14},B^{25}, B^{36}, B^{23},
B^{26}$, the last ones
parametrize the coset $ {\rm SO}(1,1)\times {\rm SO}(2,2)/{\rm SO}(2)\times {\rm SO}(2)$. \\
If we now consider the truncation $N=4  \longrightarrow N=1$ the
relevant coset manifold is $({\rm SU}(1,1)/{\rm U}(1))^3$ which
contains 3 complex moduli. To obtain the corresponding complex
structure, it is sufficient to freeze
$g^{i\overline{\jmath}},\,B^{i\overline{\jmath}}$ with $i\neq j$.
In particular the  ${\rm SU}(1,1)^3$ can be decomposed into ${\rm
O}(1,1)^3\otimes^{s}T^3$ where  the three ${\rm O}(1,1)$ and the
three translations $T^3$ are parametrized by $g^{x\overline{x}}$,
$g^{y\overline{y}}$, $g^{z\overline{z}}$ and
$B^{x\overline{x}}$, $B^{y\overline{y}}$, $B^{z\overline{z}}$ respectively.\\
These axions are massless because of equation \eq{tuttizero} (Note
that the further truncation $N=1\longrightarrow N=0$ does not
alter the coset manifold ${\rm SU}(1,1)^3$ since we have no loss
of massless fields in this process). In this case we may easily
compute the moduli dependence of the gravitino masses. Indeed,
${\rm O}(1,1)^3$, using equations \eq{ziapina}, \eq{ziagina},
will have as coset representative the matrix

\begin{equation}\label{cosettone}E_{\L}^I=\begin{pmatrix}{e^{\varphi_1}&0&0&0&0&0\cr
0&e^{\varphi_2}&0&0&0&0\cr0&0&e^{\varphi_3}&0&0&0\cr0&0&0&e^{\varphi_1}&0&0\cr0&0&0&0&e^{\varphi_2}&0\cr
0&0&0&0&0&e^{\varphi_3}\cr}\end{pmatrix}\end{equation}

\nin where we have set
$g_{11}=e^{2\varphi_1},\,\,g_{22}=e^{2\varphi_2},\,\,g_{33}=e^{2\varphi_3}$,
the exponentials representing the radii of the manifold
$T_{(14)}^2\times T_{(25)}^2\times T_{(36)}^2$.\\ We see that in
the gravitino mass formula the  vielbein $E_{\L}^{I}$ reduces to
the diagonal components of the matrix \eq{cosettone} A
straightforward computation then gives: \be
S_{AB}\overline{S}^{AB}=\frac{2}{(48)^2}e^{(2\varphi_1+2\varphi_2+2\varphi_3)}
\begin{pmatrix}{m_1^2&0&0&0\cr0&m_2^2&0&0\cr0&0&m_3^2&0\cr0&0&0&m_4^2\cr}\end{pmatrix}\ee
We note that in the present formulation where we have used a
contravariant $B^{\L\S}$ as basic charged fields, the gravitino
mass depends on the $T^6$ volume. However if we made use of the
dual 4-form $C_{\L\S\G\D}$, as it comes from Type IIB string
theory, then the charge coupling would be given in terms of
$^*f^\a_{\L\S\G}$ and the gravitino mass matrix would be
trilinear in $E^\L_I$ instead of $E^I_\L$. Therefore all our
results can be translated in the new one by replacing
$R_i\rightarrow R^{-1}_i$.

\appendix
\section*{Appendix D: Conventions}
\setcounter {equation}{0} \label{appendiceD}
\addtocounter{section}{4}

We realize the isomorphism between the two fold antisymmetric
representation of ${\rm SU}(4)$ and the fundamental of ${\rm
SO}(6)$
using the $4\times 4$ $\G$--matrix $(\G^I)_{AB}=-(\G^I)_{BA}$.\\
We have used the following representation \vskip 1cm
\begin{eqnarray}
\G ^1 &=& \left( \begin{array}{cccc} 0&0&0&1 \\ 0&0&1&0 \\ 0&-1&0&0 \\
-1&0&0&0 \end{array} \right) \qquad \G ^4 = \left(
\begin{array}{cccc} 0&0&0& i \\ 0&0& -i& 0 \\ 0&
i &0&0 \\ -i &0&0&0 \end{array}
\right) \nonumber \\
\G ^2 &=& \left( \begin{array}{cccc} 0&0&-1&0 \\ 0&0&0&1 \\ 1&0&0&0 \\
0&-1& 0&0 \end{array} \right) \qquad  \G^5 = \left(
\begin{array}{cccc} 0&0&i&0 \\ 0&0&0&i \\ -i&0&0&0 \\ 0&-i&0&0
\end{array} \right)
\\
\G^3 &=& \left( \begin{array}{cccc} 0&1&0&0 \\ -1&0&0&0 \\ 0&0&0&1 \\
0&0&-1&0 \end{array} \right) \qquad \G^6 = \left(
\begin{array}{cccc} 0&-i&0&0 \\ i&0&0&0 \\
0&0&0&i \\ 0&0&-i&0
\end{array} \right)
\nonumber
\end{eqnarray}
Note that their anticommutator is
$\{\G_I,\overline{\G}_J\}=-\d_{IJ}$, where the complex conjugation
acts as
\be(\G^{I})^{AB}=(\overline{\G}^I)_{AB}=\frac{1}{2}\e^{ABCD}(\G^I)_{CD}\ee
We define
\ba&&(\G^{IJ})_A^{\phantom{B}B}=\frac{1}{2}\left[(\G^{[I})_{AC}(\G^{J]})^{CB}\right]\\
&&(\G^{IJK})_{AB}=\frac{1}{3!}\left[(\G^I)_{AC}(\G^J)^{CD}(\G^K)_{DB}+\,perm.\right]\ea
Here the matrices $(\Gamma ^{IJK})_{AB}$ are symmetric and satisfy
the relation
\begin{equation}
(\Gamma ^{IJK})_{AB} = \frac{i}{6}\varepsilon ^{IJKLMN} (\Gamma
_{LMN})_{AB} \label{c5}
\end{equation}
In this representation, the following matrices are diagonal:
\begin{eqnarray}
\Gamma ^{123} &=& \left( \begin{array}{cccc} 1&0&0&0 \\ 0&1&0&0
\\ 0&0&1&0
\\ 0&0&0&1 \end{array} \right) \qquad \Gamma ^{156} = \left(
\begin{array}{cccc} -1&0&0&0 \\ 0&1&0&0 \\ 0&0&1&0 \\ 0&0&0&-1
\end{array} \right) \nonumber \\
\Gamma ^{246} &=& \left( \begin{array}{cccc} -1&0&0&0 \\ 0&1&0&0 \\
0&0&-1&0
\\ 0&0&0&1 \end{array} \right) \qquad \Gamma ^{345} = \left(
\begin{array}{cccc} 1&0&0&0 \\ 0&1&0&0 \\ 0&0&-1&0 \\ 0&0&0&-1
\end{array}\right)
\end{eqnarray}as well as the matrices $\Gamma ^{456}$, $\Gamma ^{234}$, $\Gamma
^{135}$ and $\Gamma ^{126}$ related with them through the relation
(\ref{c5}).

\nin We define for a generic tensor
\begin{eqnarray}&& T_{\dots I\, J\dots}=\dots\frac{1}{2}(\G_I)^{AB}\frac{1}{2}(\G_J)^{CD}\dots T_{\dots [AB]\,[CD]\dots}\nn\\
&&\label{conversion} T_{\dots
[AB]\,[CD]\dots}=\dots\frac{1}{2}(\G^I)_{AB}\frac{1}{2}(\G^J)_{CD}\dots
T_{\dots I\, J\dots}\end{eqnarray} so that
\begin{eqnarray}&& E_{\L}^I(E^{-1})_I^{\S}= E_{\L}^{AB}(E^{-1})_{AB}^{\S}=\d_{\L}^{\S}\\
&& E^I_{\L}(E^{-1})^{\L}_J=\d_I^J\iff
E_{\L}^{AB}(E^{-1})_{CD}^{\L}=\d^{AB}_{CD}\end{eqnarray} In
particular we need to convert the ${\rm SO}(6)_1$ indices of
$\o_1^{IJ}$ into ${\rm SU}(4)$ $R$--symmetry indices as they
appear in the covariant derivative on spinors. For this purpose
we apply the previous definition \eq{conversion} to the connection
$\o_1^{IJ}$ defining \be \o^{AB}_{\quad
CD}\equiv\frac{1}{4}(\G_I)^{AB}(\G_J)_{CD}\,\o_1^{IJ}\end{equation}
then we introduce the connection $Q^A_{\,\,\,B}$ defined as \be
\o^{AB}_{\quad
CD}\equiv\d^{[A}_{\,\,[C}Q^{B]}_{\,\,\,D]}\end{equation} and thus
\be Q^A_{\,\,\,D}=-\frac{1}{2}\o^{AB}_{\quad
BD}=\frac{1}{8}(\G_{IJ})^A_{\,\,\,D}\,\o_1^{IJ}\end{equation} One
can easily realize that given the definition of the ${\rm
SO}(6)_1$ curvature as \be  R_1^{IJ}\equiv d\o^{IJ}_1+\o^I_{1\,\,
K}\wedge\o^{KJ}_1 \end{equation} one finds for consistence that
\be R^A_{\,\,\,B}\equiv dQ^A_{\,\,\,B}-2 Q^A_{\,\,\,C}\wedge
Q^C_{\,\,\,B}=\frac{1}{8}(\G_{IJ})^A_{\,\,\,B}\,R_1^{IJ}\end{equation}
As a consequence, the covariant derivative acting on spinors
turns out to be \be
D\theta_A=d\theta_A-2Q^A_{\,\,\,B}\theta_B=d\theta_A-\frac{1}{4}(\G_{IJ})^A_{\,\,\,B}\,\o_1^{IJ}\theta_B\end{equation}

\end{document}